\documentclass[aoas,preprint]{imsart}

\RequirePackage{amsthm,amsmath,amsfonts,amssymb}
\RequirePackage[authoryear]{natbib}
\RequirePackage[colorlinks,citecolor=blue,urlcolor=blue]{hyperref}
\RequirePackage{graphicx}

\usepackage{enumerate}
\usepackage{booktabs}
\usepackage{multirow}
\usepackage{caption}
\usepackage{subcaption}
\usepackage{float}
\usepackage{hyperref}
\hypersetup{
    colorlinks=true,
    allcolors=blue
}
\usepackage[percent]{overpic}
\usepackage{array}
\newcolumntype{?}{!{\vrule width 1pt}}

\usepackage{enumitem}

\usepackage{multibib}
\newcites{A}{Appendix References}

\startlocaldefs

\theoremstyle{remark}

\newtheorem{mydef}{Definition}
\newtheorem{prob}{Problem}
\newtheorem{prop}{Fact}
\theoremstyle{definition}

\endlocaldefs

\begin{document}

\begin{frontmatter}
\title{How Likely are Ride-share Drivers to Earn a Living Wage? \\ 
Large-Scale Spatiotemporal Density Smoothing with the Graph-fused Elastic Net}
\runtitle{Large-Scale Spatiotemporal Density Smoothing with the GFEN}

\begin{aug}
\author[A]{\fnms{Mauricio} \snm{Tec}\ead[label=e1]{mauriciogtec@utexas.edu}},
\author[C]{\fnms{Natalia} \snm{Zuniga-Garcia}\ead[label=e2,mark]{nzuniga@utexas.edu}},
\author[C]{\fnms{Randy B.} \snm{Machemehl}\ead[label=e3,mark]{rbm@mail.utexas.edu}}
\and
\author[A,B]{\fnms{James G.} \snm{ Scott}\ead[label=e4]{james.scott@mccombs.utexas.edu}}
\address[A]{Department of Statistics and Data Science,
The University of Texas at Austin,
\printead{e1}}

\address[B]{Department of Information, Risk, and Operations Management and Department of Statistics and Data Science,
The University of Texas at Austin,
\printead{e4}}

\address[C]{Department of Civil, Architectural and Environmental Engineering,
The University of Texas at Austin,
\printead{e2,e3}}
\end{aug}

\begin{abstract}
Ride-sourcing or transportation network companies (TNCs) provide on-demand transportation service for compensation, connecting drivers of personal vehicles with passengers through smartphone applications. In this study, we consider the problem of estimating a spatiotemporally varying probability distribution for the productivity of a TNC driver, using data on more than 1.2 million TNC trips in Austin, Texas. We propose a graph-based smoothing approach that allows for distinct spatial and temporal dynamics, including different degrees of smoothness, spatio-temporal interactions, and interpolation in regions with little or no data.  For such a goal, we introduce the Graph-fused Elastic Net (GFEN) and use it in combination with a dyadic tree decomposition for density estimation. In addition, we present an optimization-driven approach for fast point estimates scalable to massive graphs. Bayesian inference and uncertainty quantification with MCMC are also illustrated. The main results demonstrate that the optimization strategy is an effective exploration tool for selecting adequate regularization schemes using Bayesian optimization of the cross-validation loss. Two key empirical findings made possible by our method include: 1) the probability that a TNC driver can expect to earn a living wage in Austin exhibits high variability in space and time, from as low as 25\% to as high as 85\%; and 2) some drivers suffer considerable ``tail risk'', with the bottom 10\% of the earnings distribution falling below \$10 per hour---grossly below a living wage in Austin for a single adult---for specific times and locations.  All code and data for the paper are \href{https://github.com/mauriciogtec/gfen-reproduce}{publicly available}, as a \href{https://mauriciogtec.shinyapps.io/gfen/}{Shiny app for visualizing the results} and a \href{https://github.com/mauriciogtec/GraphFusedElasticNet.jl}{software package} in Julia for implementing the GFEN.
\end{abstract}


\end{frontmatter}

\section{Introduction} 
\label{sec:intro}

Ride-sourcing or transportation network companies (TNCs), such as Uber and Lyft, operate as a two-sided market that connects drivers with passengers through mobile applications. %
In recent years, TNCs have experienced rapid growth. For example, Uber saw 5.22 billion trips worldwide in 2018, up from 140 million trips in 2014 \citep{smithuber}.  
This growth has posed several challenges to transportation planners, policymakers, and researchers---for example, lack of infrastructure (e.g.~at airports), geographical variation in operating rules and regulations, and potential changes in travelers' behavior.  TNCs have also been the subject of controversy because of their aggressive business tactics and sometimes complex pricing systems \citep{li-2019}, whose effects on both rider and driver behavior are not well understood.    

This paper addresses a fundamental statistical question relevant to all these stakeholders: how best to quantify spatial and temporal variation in TNC driver earnings.  Due to various statistical challenges, which we articulate below, this variation is not well understood---nor are there suitable methods for estimating this variation reliably, at the scale and speed needed for analyzing millions of trips at high spatial and temporal resolution.  Our goal in this paper is to address this gap.  Specifically, we present a nonparametric method for estimating the probability density $f_{s,t}$, at location $s$ and time $t$, for the productivity of a TNC driver (defined roughly as profit per hour and explained in detail below).

We apply our method using data on more than 1.2 million ride-sourcing trips taken on RideAustin, a local non-profit TNC in Austin, Texas, during a period in 2016-17 when leading national TNCs left the city temporarily. %
The main results provide a number of interesting findings---many made possible only because our method yields a full probability distribution of driver earnings as a function of both space and time, giving us access to distributional features like quantiles and tail areas.  To give two examples: first, the probability that a TNC driver can expect to earn a living wage in Austin exhibits high variability with respect to space and time. For a parent of two children who works a typical Saturday late-night near downtown Austin, the probability of earning a living wage for the Austin area can exceed 90\%.  But at midday on a Monday far from the city center, this probability can fall below 40\%. Second, the bottom 10\% of earners among drivers accepting rides at the airport have productivity below $\$10$/hour in a typical Monday midday.  This figure is considerably lower than the living wage in Austin for a single adult with no children.

\subsection{Ride-Sourcing Productivity: Background \& Challenges}
\label{subsection:challenges}

Quantifying spatiotemporal variation in driver earnings is a relevant problem. This variation sits at the heart of many challenges faced by the designers of TNC pricing models and drivers themselves. For both the TNC platform and the drivers, a desirable property of a ride-sourcing platform is what \cite{zuniga-etal-2019} call \emph{destination invariance}, or ``the principle that two drivers dispatched on different trips from the same location at the same time do not envy each other’s expected future income.''  But in reality, some trip opportunities yield higher continuation payoffs than others, which implies that some trips are mispriced.  From the driver's perspective, this mispricing can result in needlessly high volatility in driver earnings, and therefore substantial variation in the likelihood that a driver will earn a living wage.  

Moreover, for both the driver and the TNC, such variation can also result in substantial market inefficiencies, potentially impacting service reliability at the whole network level. TNCs respond to this reality in multiple ways.  For example, Uber and Lyft tried to provide drivers with more flexibility by adding filters for desired destination drop-off  \citep{filter1}. However, this feature caused a negative impact on the platform by increasing riders' waiting time and other drivers' pick-up time. Strategic drivers used the filter to select trips with better-earning potential, leading Uber to limit this feature \citep{filter2}.

In this paper, we do not explicitly consider the question of how to design a better TNC pricing model.  Instead, we take the perspective that before one can develop such a pricing model that mitigates spatiotemporal variation in driver earnings, one must first quantify the extent of that variation---and we, therefore, seek to provide a scalable and reliable method for doing so.  %
Our approach encodes spatiotemporal structure using a graph. Each vertex corresponds to a traffic analysis zone (TAZ) at a specific hour of the week, and where edges are used to denote geographical and temporal adjacency. TAZs are contiguous geographic areas that divide a planning region into relatively similar land use and activity areas. %
This specification results in a graph of 225k vertices, one for each TAZ at a specific time. Our goal is to estimate a full probability distribution for a driver's productivity at each one of these vertices (we provide further details about the construction of the graph in Section \ref{section:graph}).  Owing to the size of the graph, we shall build upon the scalable spatial density smoothing framework of \citet{tansey-etal-2017}.

The first challenge comes from the question of how to smooth the raw data while still capturing important spatiotemporal effects. Space and time dimensions have different units and physical interpretations, suggesting that spatial versus temporal edges must be treated differently. Moreover, effects in space and time are likely to be a mix of smooth and non-smooth transitions. For example, the productivity of a driver may change drastically from one side to another of a highway or a river, but it would most likely be similar or even constant across highly interconnected regions with no obvious barriers.  In the time dimension, by contrast, effects are more likely to be smooth, and yet there may still be sudden transitions caused by specific events, such as the increased temporal density of airport arrivals.  The challenge here is to allow for separate but parsimonious spatial and temporal dynamics that incorporate a mix of both smooth and non-smooth features.

The second challenge arises from the fact that many vertices in our graph will have no data.  For the RideAustin dataset, discretized by the hour of the week, 45\% of the graph's vertices have no observations.  Every TAZ had at least one observation at some point of the week. But many do not have observations for every hour of the week.  The challenge here is to develop a method that can borrow information efficiently across spatial and temporal adjacencies, estimating a density at every location for every hour, even if no data was observed.

\subsection{Overview of Proposed Methodology}\label{sec:overview}

Our analysis of the RideAustin data is based on the following broad outline.  Given a graph $G$ with vertices $V$, we estimate a family of densities $\{f_v: v \in V\}$ with common support $\mathcal{X}$ using three steps: partition, smooth, and merge. First, we use a binary tree to recursively \textbf{partition} $\mathcal{X}$ into a series of half-spaces, each described by a conditional probability. Second, we \textbf{smooth} each half-space probability across the graph in a way that encourages similarity between adjacent nodes. Third, we \textbf{merge} the smoothed half-space probabilities to yield full density estimates $\{f_v\}$ at each node.

This process is broadly similar to the approach taken by \citet{tansey-etal-2017} for a ``spatial only'' graph; aside from the extension to a spatiotemporal setting, the other critical difference in our paper is that we handle the smoothing step in a fundamentally different way.   \citet{tansey-etal-2017} smooth using the Graph-fused Lasso (GFL) \citep{tibshirani-2015}. The GFL is an attractive approach to smoothing because of the anisotropic properties of its estimates and its algorithmic scalability, being able to provide point estimates for massive graphs at a low computational cost. However, the GFL has some limitations specific to our context.  First, the GFL does not distinguish between spatial and temporal edges on the graph.  Second, it is also not ideal for modeling a combination of smooth and non-smooth effects since the GFL produces estimates that are piece-wise constant across the graph. Finally, and most importantly, as we shall explain in Section \ref{section:methodology}, the GFL does not necessarily give sensible results in missing-data scenarios since its objective function will no longer be strictly convex. Ignoring this fact, or resolving it in a naive way, can lead to solutions with undesirable, counter-intuitive, or non-unique interpolation behavior.   

To address these limitations, we construct a spatial smoothing model that combines the traditional $\ell_1$ total variation penalty used by the GFL with an additional $\ell_2$ smoothing term to impose different penalties across spatial and temporal edges. This combination will have the effect of enabling both smooth and non-smooth transitions across both space and time.  The imposition of the $\ell_2$ penalty alone results in a solution equivalent to a MAP estimate under a specific Gaussian Markov Random Field prior \citep{cressie-1993}, which have been proposed as a fast kriging technique for large datasets \citep{hartman2008fast}.  But the combination of an $\ell_2$ penalty with the $\ell_1$ penalty yields a spatial smoother that is analogous to the elastic net regularization method for linear regression \citep{zou-2005}.  More importantly, this combination of penalties can model both smooth and non-smooth spatio-temporal variation while yielding a strictly convex optimization problem even in the case of missing data.  In summary, the anisotropic $\ell_1$ penalty provides sharp edges, relevant to model geographical boundaries. At the same time, urban discontinuities and the isotropic $\ell_2$ penalty interpolates missing data and encourages some smoothness, where warranted by the data. We call the resulting method the Graph-fused Elastic Net (GFEN).

\begin{figure}[t]
    \centering
    \includegraphics[width=13.5cm, height=4.6cm]{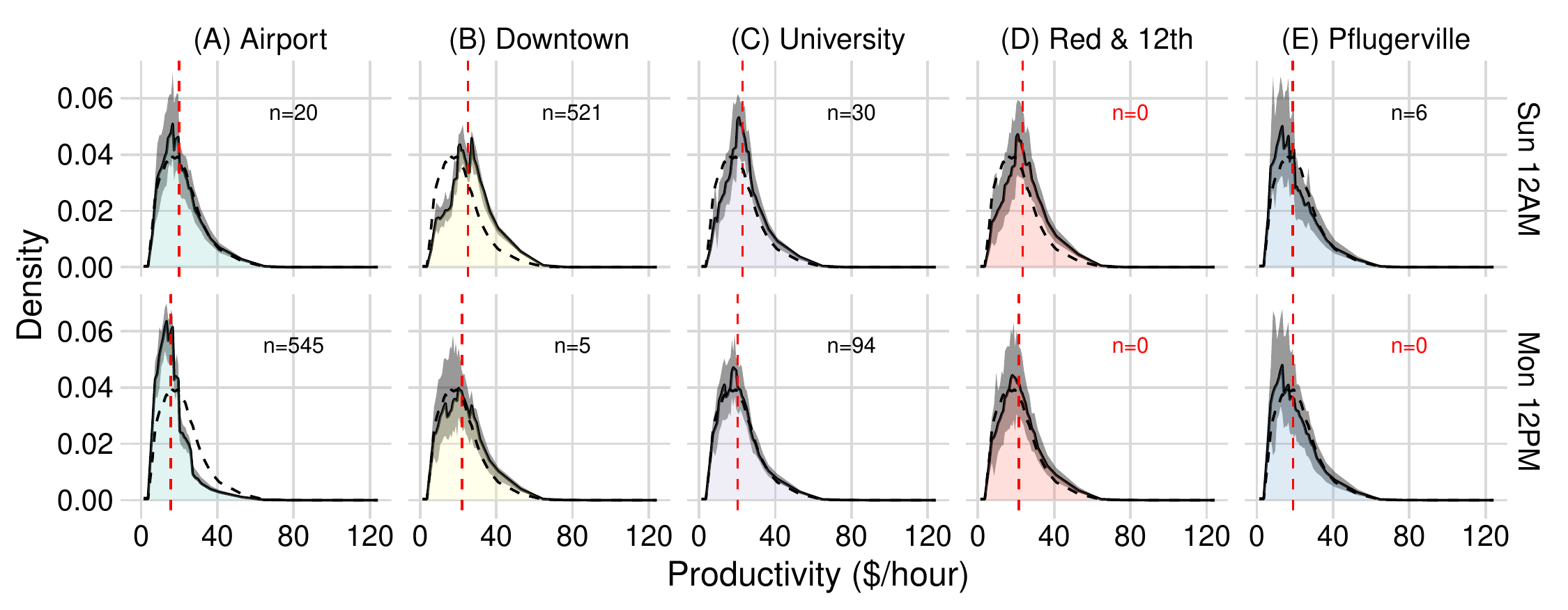}
    \caption{Examples of density estimation methodology with the GFEN for selected locations and hours. Solid curves are MAP estimates. Shaded regions are posterior credibility intervals of 90\% for each productivity level. The number of data points at each specific location and time is indicated with $n$ next to each density. Dotted curves are the ``global'' density reference estimate. The dotted vertical red line is the expected productivity for the MAP estimate.}
    \label{fig:densities-sample}
\end{figure}

Figure \ref{fig:densities-sample} provides a preview of results produced on the RideAustin data set using the GFEN and shows our estimated densities for five diverse locations in Austin:
\begin{enumerate}[label={(\Alph*)}]
    \item The Austin–Bergstrom International {Airport} (ABIA).
    \item {Downtown}, identified with the TAZ containing the intersection of Guadalupe \& 6th Streets, which has very high trip demand and small-sized TAZs.
    \item The University of Texas at Austin, located approximately 1.5km north from Downtown.
    \item The intersection of {Red River \& 12th} street, an intersection very close to the Downtown area (<1km) but with a low passenger count.
    \item {Pflugerville}, a residential suburban with low economic activity, located 30km away from Downtown.
\end{enumerate}

We also show reconstructed densities at two selected times: Saturday midnight and Monday noon, characterized respectively by high and low demand across the city. Figure \ref{fig:densities-sample} shows that the method can discover complicated spatial and temporal interactions---for example, these two times exhibit a significant difference in upper-tail thickness for downtown, but much less so at the airport.  Further, it is still possible to interpolate a sensible probability distribution even with little or no data observed by borrowing information from adjacent regions in a flexible---for example, during Monday noon in Red River \&12th street, which is close to the Downtown area but has only one observation in the data. The posterior credibility intervals reflect that the uncertainty increases with less data, compensating for the fact that more interpolation took place. The fitted densities highlight some statistically significant differences in risk measures of interest, even with relatively low data count. For example, the estimate that the immediate productivity may fall below \$20 at Monday 12 AM at the airport is 76.1\% (90\% CI [73.2\%, 78.8\%) whereas the estimate for the University area at the same time is 51.7\% (90\% CI [46.1\%, 58.6\%].) An important remark illustrated by the figure is that the difference in expected productivity (dashed vertical lines) is actually rather small when comparing the same TAZ between Saturday 12 AM and Monday 12 AM. Yet, analyzing the full probability distribution of productivity can uncover important statistically significant differences in risk measures across space and time.

The rest of the paper is organized as follows.  In Section \ref{section:methodology}, we explain the method in detail. In Sections \ref{section:case-study} and \ref{sec:case-study:results}, we go beyond Figure \ref{fig:densities-sample} to present our complete analysis of the RideAustin dataset, using our method to glean meaningful insights about spatiotemporal variation in driver earnings.  Section \ref{sec:conc} then concludes with some final remarks.  All code and data for the paper are \href{https://github.com/mauriciogtec/gfen-reproduce}{publicly available}, as a \href{https://mauriciogtec.shinyapps.io/gfen/}{Shiny app for visualizing the results} and a \href{https://github.com/mauriciogtec/GraphFusedElasticNet.jl}{software package} in Julia for implementing the GFEN.

\section{Spatiotemporal Density Smoothing at Scale}
\label{section:methodology}

\subsection{Background \& Related Work}

Our methods build on two independent lines of work: first, spatial density smoothing over a graph; second, extensions of spatial models to spatiotemporal settings. We do not attempt to provide a detailed review of these mature fields. Instead, we provide a tailored outline of prior work closely related to our approach.

Kriging with the Gaussian process (GP) \citep{cressie-1993} is often considered the gold standard for spatial smoothing. Kriging jointly models every observed point as a multivariate distribution, where the covariance structure is determined by a kernel function, typically specified as a function of an underlying distance between two observations. Kriging can thus naturally interpolate to spatial locations with missing data.%
, which is relevant to the application considered in this paper.
 Relevant extensions of kriging and GPs are: dealing with non-Gaussian observations \citep{rasmussen2003gaussian}, spatio-temporal data \citep{gneiting2006geostatistical,cressie-wikle-2011}, graph data \citep{chua2006network}, and anisotropic variation \citep{ecker1999bayesian, friedland2017isotropic}. Anisotropic smoothing encourages signals with sharp contrasts, where standard isotropic smoothing promotes diffused estimates equivariantly in all directions. Sharp contrasts with geostatistical data can arise from rivers, mountains, and high-traffic roads.

The complexity of kriging with GPs is prohibitive for large scale problems. For network-based data \citep{chua2006network, okabe2012spatial}, the computational complexity is $\mathcal{O}(n^2m)$ where $n$ and $m$ are the number of nodes and edges. Memory is too an important limitation when working with GPs; for example, a covariance matrix with $n\approx 225,000$ nodes (like the one used in this paper) requires over 400GB of RAM memory. This capacity is not available today, even for advanced high-performance computing resources. The complexity of GPs has spawned a vast literature seeking scalable approximations and alternative models \citep{hensman2013gaussian}. Many of such approximations rely on sparsity or low-rank assumptions on the kernel matrix paired with special tensor factorization and matrix inversion techniques (e.g., \citet{zhang2019embarrassingly, cressie-2010, hensman2013gaussian}.) Note that many scalable alternatives are often tailored towards Gaussian observation likelihoods and rely on approximations for non-Gaussian likelihoods, except in special cases \citep{hensman2013gaussian}.

Markov Random Fields (MRFs) \citep{hartman2008fast, rue-held-2005} are one of the alternative models that can be used for approximate kriging. In particular, a Gaussian Markov random field (GMRF) is a type of GP that assumes sparsity in the inverse kernel (partial correlation or ``precision" matrix). As pointed out by \citet{xu-2015}, GMRFs are also capable of dealing with missing observations like full GPs, which makes them an attractive more scalable kriging alternative for spatiotemporal modeling of geostatistical data where many locations have missing data for at least some point in time, and which is also one of the challenges we face in working with the RideAustin dataset.

 We now discuss two main approaches in the statistics literature for extending spatial models to the spatio-temporal setting. The first approach typically involves estimating or specifying a covariance or kernel matrix that includes both space and time dimensions \citet{cressie-huang-1999, gneiting2006geostatistical}. The simplest approach is to specify the kernel matrix as the product of the temporal and spatial kernels \citep{gneiting2006geostatistical}. However, since the kernel specifies the covariance structure, this would not allow for interactions between time and space. As a result, non-separable kernel functions have been proposed to allow for such interactions \citep{cressie-huang-1999}.

Similarly, a separable spatio-temporal precision matrix can be specified in MRFs from the product of spatial and temporal precision matrices \citep{blangiardo2013spatial}. However, it is worth remarking that a separable precision matrix does not preclude spatio-temporal interactions nor does it imply a separable covariance kernel. Explicitly modeling spatio-temporal interactions in MRFs is primarily beneficial when seeking an explicit factorization of main (spatial and temporal) effects and interaction effects; for an example of this approach, see \citep{knorr2000bayesian}. Our examples will highlight that a separable penalty function on spatial and temporal differences is still fully capable of modelling spatio-temporal interactions.

A different approach to spatio-temporal modeling is to use dynamic probabilistic models, where the parameters of a spatial model are assumed to change smoothly over time. Some examples of this work include \citet{cressie-2010} and \citet{katzfuss-cressie-2011} for kriging with low-rank approximations and \citet{rushworth-2017} for GMRFs in the context of spatiotemporal disease modeling.

Note that kriging alone does not solve the problem of non-parametric density estimation at each spatial location. The kriging methods previously discussed in this section (GPs, MRFs, etc.) are primarily concerned with smoothing scalar quantities (regression). The scalable algorithm for smoothing non-parametric density estimates over a general graph that we rely upon heavily in this paper was proposed by \citet{tansey-scott-2015} and \citet{tansey-etal-2017}, who in turn build on work by \citet{tibshirani-taylor-2011}, \citet{ramdas-tibshirani-2016}, and \citet{wang-2016}. This line of research relies in turn on the Graph-fused Lasso (GFL) \citep{wang-2016}, also known as total-variation (TV) denoising in the image denoising literature \citep{getreuer-2012}. As an optimization method, the GFL uses the TV penalty based on the $\ell_1$-norm to denoise images by penalizing differences across smoothed estimates and shrinking towards a piecewise-constant spatial estimate. As a statistical method, it can be regarded as an anisotropic MRF. \citet{zuniga-etal-2019} previously conducted an analysis with the RideAustin data set using the GFL, and showed it to be effective in modeling purely mean spatial effects. However, their framework cannot directly be applied to neither the spatio-temporal setting nor full density estimation. The authors treat time in an ad-hoc manner by splitting the dataset into peak hours, mid-day, overnight, and weekend periods. The analysis presented here will extend the analysis of \citet{zuniga-etal-2019} to high temporal resolution and density estimation, revealing important risk scenarios for the drivers.

\subsection{Spatiotemporal Graphs and Densities}
\label{subsection:densities}

We now describe our analytical approach to the RideAustin data set. Our analysis uses an undirected graph $G$ to encode the spatial and temporal structure. The graph $G=(V, E)$ is a spatiotemporal graph in the sense that it has the following structure: First, there is exactly one node for every Austin TAZ and in each hour of the week (starting at midnight Sunday). Thus the set of vertices $V$ can be written as $V = S \times T$ where $S$ and $T$ are the sets of locations and times, respectively. Second, edges are either spatial or temporal. Thus the set of edges $E$ can be written as a disjoint union $E = E_S \cup E_T$, where $E_S$ connects adjacent nodes in the same spatial slices of the form $\{s\} \times T$ and $E_T$ connects nodes in the same temporal slice of the form $S \times \{t\}$.

At each vertex $v = (s,t)$, we observe data points of the form $\textbf{y}^{(s,t)}=\left\{y^{(s,t)}_1,\hdots,y^{(s,t)}_{N^{(s,t)}} \right\}$ where
$y^{(s,t)}_i \overset{iid}{\sim} f(s,t)$. In our case, each $y^{(s,t)}_i$ represents a measure of driver productivity (defined below) for a single ride originating at location $s$ during hour $t$.  Based on these data, we seek to estimate each distribution $f(s,t)$ across the whole graph.  However, $N^{(s,t)}$ is small or zero in many vertices, and thus independently estimating $f(s,t)$ at each vertex of the graphs leads to poor estimates.  This estimation necessitates some form of graph smoothing, in which we borrow information from neighboring regions encoded by $G$. 

To do so in a scalable way, we represent each density $f(s,t)$ using a recursive dyadic partition, as follows.  Suppose that all densities have a common sample space $B$.  At the first level of the partitioning scheme, we write $B$ as a union of disjoint non-empty ``child'' sets $B= B_0 \cup B_1$.  At the second level, we then recursively partition each child as $B_0 = B_{00} \cap B_{01}$ and $B_1 = B_{10} \cap B_{11}$ into two child sets each---and so on, partitioning each child set recursively until we reach a given max depth $K$.  We can arrange these sets in a tree and index the nodes of the tree by binary sequences: formally, for every $k \in\{1, \hdots, K - 1\}$ and for every $B_\gamma$ where $\gamma\in\{0,1\}^k$ is a length-$k$ binary sequence, we have that $B_\gamma = B_{\gamma0} \cup B_{\gamma1}$ is a union of disjoint non-empty sets. We refer to $B_{\gamma0}$ and $B_{\gamma1}$ as the left children and right children of $B_\gamma$. To convey the intuition of this approach, consider the specific example of $B=[0,1)$. We could then define $B_0 = [0,1/2)$ and $B_1=[1/2,1)$. Similarly, we could write $B_{00}=[0,1/4)$, $B_{01}=[1/4,1/2)$, $B_{10}=[1/2,3/4)$, and $B_{11}=[3/4,1)$---and so on.  Note that this is the same kind of recursive tree structure used in defining a Polya-tree prior in nonparametric Bayesian analysis.

 As \citet{tansey-etal-2017} point out, the key feature of such a tree structure is that it turns the problem of smoothing a functional object $f$ across a graph into a series of scalar smoothing problems, each of which may be solved independently in parallel. To see this, assume that $Y \sim f$ is some random variable with sample space $B$, and that we are using a tree $B^{(K)}$ to partition $B$ in the manner defined above. We then approximate $f$ in terms of quantities of the form
$\omega_\gamma = P(Y \in B_{\gamma0} \mid Y \in B_\gamma)$
which is the probability that $Y$ falls in the left child $B_{\gamma0}$ given that it falls in the parent $B_\gamma$. We refer to these $\omega_{\gamma}$'s as the splitting probabilities in the tree representation of $f$.  These give us a finite approximation of $f$, whose resolution is determined by the depth of the tree $K$. More precisely, for any $\gamma=\gamma_1\cdots\gamma_K\in\{0,1\}^K$ we have
\begin{equation}\label{eq:recover-probability}
\begin{aligned}
P(Y \in B_\gamma)
& = \prod_{j=0}^{K-1} \omega_{\gamma_1\cdots\gamma_j}^{\gamma_{j+1}}\left(1 - \omega_{\gamma_1\cdots\gamma_j}\right)^{1 - \gamma_{j+1}}.
\end{aligned}
\end{equation}

We now put the above formulation back into the setting of our RideAustin data.  In our model, we have a family of probability distributions $f(s,t)$ indexed by space (TAZs) and time (hours of the week), assumed to have common support $B$.  Therefore, given a tree representation of $B$, we have a family of splitting probabilities $\Omega^{(s,t)} = \bigcup_{k=1}^K \left\{\omega^{(s,t)}_\gamma : \gamma \in \left\{0,1\right\}^k\right\}
$ for each vertex on the graph $v = (s,t)$. If we define $m^{(s,t)}_\gamma$ as the count of total observations in $\mathbf{y}^{(s,t)}$ that fall within $B_\gamma$. Then for each non-terminal ($|\gamma| < K$) node $B_\gamma$ we have
\begin{equation}\label{eq:binomial-model}
m_{\gamma0}^{(s,t)} \sim \mathrm{Binomial}(\omega^{(s,t)}_\gamma, m_\gamma^{(s,t)}) \, ,
\end{equation}
independently of $m_{\gamma}$.  Thus \eqref{eq:binomial-model} enables us to estimate the $\omega_\gamma$ from the data: we simply count the number of occurrences in each bin and estimate a set of binomial probabilities for the tree structure, while smoothing these probabilities across the graph. Expression \eqref{eq:recover-probability} can then be used to recover the full density estimates. In practice, we parameterize \eqref{eq:binomial-model} in terms of a log-odds variable $\beta_\gamma^{(s,t)}$ such that $\omega^{(s,t)}_\gamma = (1 + \exp(-\beta_\gamma^{(s,t)}))^{-1}$. 

Section \ref{sec:tree-selection} in our case study with the RideAustin data discusses additional considerations for selecting an adequate tree splitting scheme.

\subsection{The Behavior of the GFL and GMRF with Missing Data}\label{sec:gfl}

We now turn to the question of smoothing the tree-splitting log-odds $\beta_\gamma^{(s,t)}$ across the graph. Our starting point for smoothing is the GFL.  We briefly recall the method, and we then discuss the complications that arise with the missing data scenario that we confront in analyzing the RideAustin data.  Our discussion in this section will motivate the introduction of the GFEN in the next section. Since the smoothing step will be identical in form for every node $B_\gamma$ in the tree representation of $f(s,t)$, we drop the subscript $\gamma$ to lighten the notation.

Recall that $E$ is the set of edges in our graph, and let $\lambda > 0$ be a penalization parameter.  The \textbf{GFL objective} is
\begin{equation}\label{eq:gfl}
\begin{aligned}
\operatornamewithlimits{minimize}_{\boldsymbol{\beta}} \quad \sum_{v \in V} l(\mathbf{y}^{(v)}, \beta^{(v)} ) + \lambda \sum_{vw\in E} \lvert \beta^{(v)} - \beta^{(w)} \rvert.
\end{aligned}    
\end{equation}
where $l$ is any strictly convex loss function used to estimate a model. In our case, it will be the negative loglikelihood implied by model \eqref{eq:binomial-model}. It is worth comparing this objective with the one arising under an $\ell_2$ penalty on edge differences, which is closely connected with the log posterior arising under a GMRF prior. The \textbf{GMRF objective} is
\begin{equation}\label{eq:gmrf}
\begin{aligned}
\operatornamewithlimits{minimize}_{\boldsymbol{\beta}} \quad \sum_{v \in V} l(\mathbf{y}^{(v)}, \beta^{(v)} ) +  \lambda \sum_{vw\in E}(\beta^{(v)} - \beta^{(w)})^2.
\end{aligned}    
\end{equation}

Both models have been used and studied extensively for smoothing and denoising problems, and we shall not attempt to provide a systematic review or comparison. Instead, we will use a simple example to illustrate the difference between these models, with particular attention to the missing data case, to motivate our approach. To develop intuition on the behavior of the smoothing penalties, will first analyze the closed-form solution for toy problem with Gaussian likelihood and Binomial likelihoods.

\begin{prob}[Denoising a minimal graph with missing data and Gaussian likelihood] \label{prob:missing-data} Consider a chain graph with three vertices $V=\{1,2,3\}$ and two edges $E=\{e_{12}, e_{23}\}$. At vertices $1$ and $3$ we observe the data points $y_1$ and $y_3$ respectively, assuming wlog that $y_1 < y_3$. But at vertex $2$ we observe no data. Given a fixed $\lambda > 0$, we will study the vector $\boldsymbol{\beta}=(\beta_1,\beta_2,\beta_3)$ minimizing the total loss function
$l(\boldsymbol{\beta})=(y_1 - \beta_1)^2 + (y_3 - \beta_3)^2 + \lambda\left(\lvert \beta_2 - \beta_1 \rvert^p + \lvert \beta_3 - \beta_2 \rvert^p\right)
$ where $p=1$ corresponds to the GFL and $p=2$ corresponds to the GMRF.
\end{prob}

The following facts exemplify that the GFL may fail to perform any interpolation, arising from the lack of a unique solution for the missing data point.
\begin{prop}[GFL solution - Gaussian likelihood]\label{prop:gfl} If $\lambda < \frac{1}{2}(y_3 - y_1)$ then there is no unique solution to the GFL objective \eqref{eq:gfl} and the solution set can be described as $\hat{\beta}_1 = y_1 + \lambda$, $\hat{\beta}_3 = y_3 - \lambda$ and $\hat{\beta}_2  \in (\hat{\beta}_1, \hat{\beta}_3)$. If $\lambda \geq \frac{1}{2}(y_3 - y_1)$ then there is a unique solution $\hat{\beta}_1=\hat{\beta}_2=\hat{\beta}_3=\frac{1}{2}(y_1 + y_3)$.
\end{prop}

\begin{prop}[GMRF solution - Gaussian likelihood]\label{prop:gmrf} For every $\lambda > 0$ the solution set of the GMRF objective \eqref{eq:gmrf} is $\hat{\beta}_1 = y_1 + \frac{\lambda}{1 + \lambda}\cdot\frac{y_3 - y_1}{2}$, $\hat{\beta}_3 = y_3 - \frac{\lambda}{1 + \lambda}\cdot\frac{y_3 - y_1}{2}$, and $\hat{\beta}_2 = \frac{\hat{\beta}_1 + \hat{\beta}_3}{2}$. Thus the vertex with missing point is assigned to the middle point, regardless of the value of $\lambda$. Moreover, all the points converge to the middle point as $\lambda \to \infty$.
\end{prop}

Fact \ref{prop:gmrf} shows two fundamental differences in comparison with the GFL. First, the GMRF performs interpolation: regardless of the value of $\lambda$, the missing data vertex is assigned to the middle point of its neighbors. Second, the difference between $y_1$ and $y_3$ determines the magnitude of the smoothing effect. As a consequence, outliers will have a stronger impact on  the GMRF solution than on the GFL. Also, here $\beta_1$ and $\beta_3$ are only asymptotically converging to the middle point, whereas the GFL would collapse them for high values of $\lambda$.

\begin{prob}[Denoising a minimal graph with missing data and Binomial likelihood] \label{prob:missing-data-binom}  Consider the same chain graph as in problem \ref{prob:missing-data} and consider the statistical model $y_i\sim \mathrm{Binomial}(\omega_i, N_i)$ with $\omega_i=(1 + e^{-\beta_i})^{-1}$. As in problem \ref{prob:missing-data} there is missing data for $i=2$, or equivalently, $N_2 = 0$ and $N_1, N_3 > 0$. Assume also for simplicity that $y_1, y_3 > 0$. The regularized binomial negative loglikelihood loss is
$$
l(\boldsymbol{\beta})=\sum_{i\in\{1,3\}}(-y_i\log(\omega_i) - (N_i - y_i) \log(1 - \omega_i)) + \lambda\left(\lvert \beta_2 - \beta_1 \rvert^p + \lvert \beta_3 - \beta_2 \rvert^p\right).
$$
\end{prob}

\begin{prop}[GFL and GFEN solution - binomial likelihood]\label{prop:gfl-binom} Let $b_i = \mathrm{logit}(y_i / N_i)$ for $i\in\{1,3\}$. Then the solutions to problem \ref{prob:missing-data-binom} for the GFL and GFEN are the same as in facts \ref{prop:gfl} and \ref{prop:gmrf} replacing the observed data $y_i$ in the Gaussian case by $b_i=\mathrm{logit}(y_i / N_i)$ where $\mathrm{logit(p)}:=\log(p/(1-p))$.
\end{prop}

We conclude this section with a remark that a commonly used regularization technique with fused lasso models is to add a shrinkage prior of the form $\lVert \beta \rVert$ (see, for example, \citet{mitov2013afused}.) However, this approach has undesirable side effects when fitting the density model based on binomial splitting probabilities (as in equation \eqref{eq:binomial-model}.) Doing so will contract the fitted probabilities towards 1/2 at every split, inducing an unwanted dependency on the splitting values of the binary tree. %
In consequence, these types of regularization priors are not advised for the density estimation technique considered in this paper.

\subsection{The Graph-fused Elastic Net (GFEN)}

The GFEN arises from the straightforward combination of the GFL and GMRF penalties.

\begin{mydef}[GFEN]
Given a graph $G=(V, E)$, and a collection of strictly convex loss functions $l(\mathbf{y}^{(v)}, \beta^{v})$ at each node $v\in V$. The \textbf{GFEN objective} is defined as 
\begin{equation}\label{eq:gfen}
    \operatornamewithlimits{minimize}_{\boldsymbol{\beta}} \, \sum_{v \in V} l(\mathbf{y}^{(v)}, \beta^{(v)}) + \sum_{p \in \{1,2\}} \sum_{vw\in E} \lambda_p\lvert\beta^{(v)}-\beta^{(w)}\rvert^p
\end{equation}
where  $\lambda_{1}, \lambda_{2} > 0$ are penalty hyperparameters for each norm. For spatiotemporal graphs, this definition is slightly modified to include different penalization parameters for spatial and temporal edges. Thus, if $G$ is a spatiotemporal graph with vertex set $S\times T$ and spatial and temporal edges $E_S$ and $E_T$ respectively. The spatiotemporal GFEN objective is
\begin{equation}\label{eq:gfen-spatiotemporal}
    \operatornamewithlimits{minimize}_{\boldsymbol{\beta}} \, \sum_{v \in V} l(\mathbf{y}^{(v)}, \beta^{(v)}) + \sum_{p \in \{1,2\}} \sum_{d\in \{S,T\}} \sum_{vw\in E_d} \lambda_{d,p}\lvert\beta^{(v)}-\beta^{(w)}\rvert^p
\end{equation}
where $\lambda_{S, 1}, \lambda_{S, 2}, \lambda_{T, 1}, \lambda_{T, 2} > 0$ are the penalty hyperparameters.
\end{mydef}

\begin{figure}[tb]
    \centering
    \begin{subfigure}[tb]{.48\linewidth}
        \centering
        \includegraphics[width=0.6\linewidth]{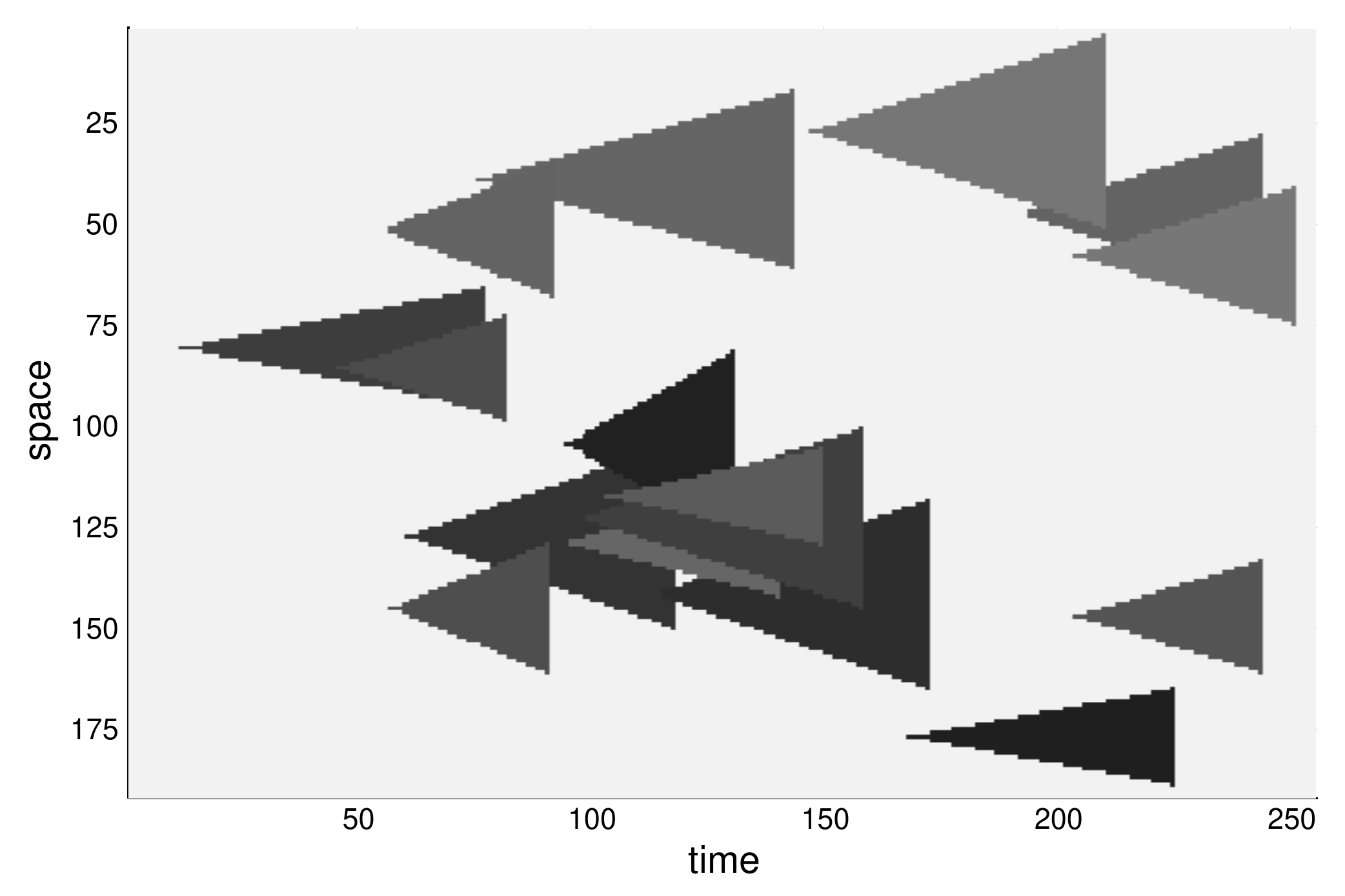}
        \caption{True probability (binomial model)}
        \label{fig:geom:truth}
    \end{subfigure}%
    ~
    \begin{subfigure}[tb]{.48\linewidth}
        \centering
        \includegraphics[width=0.6\linewidth]{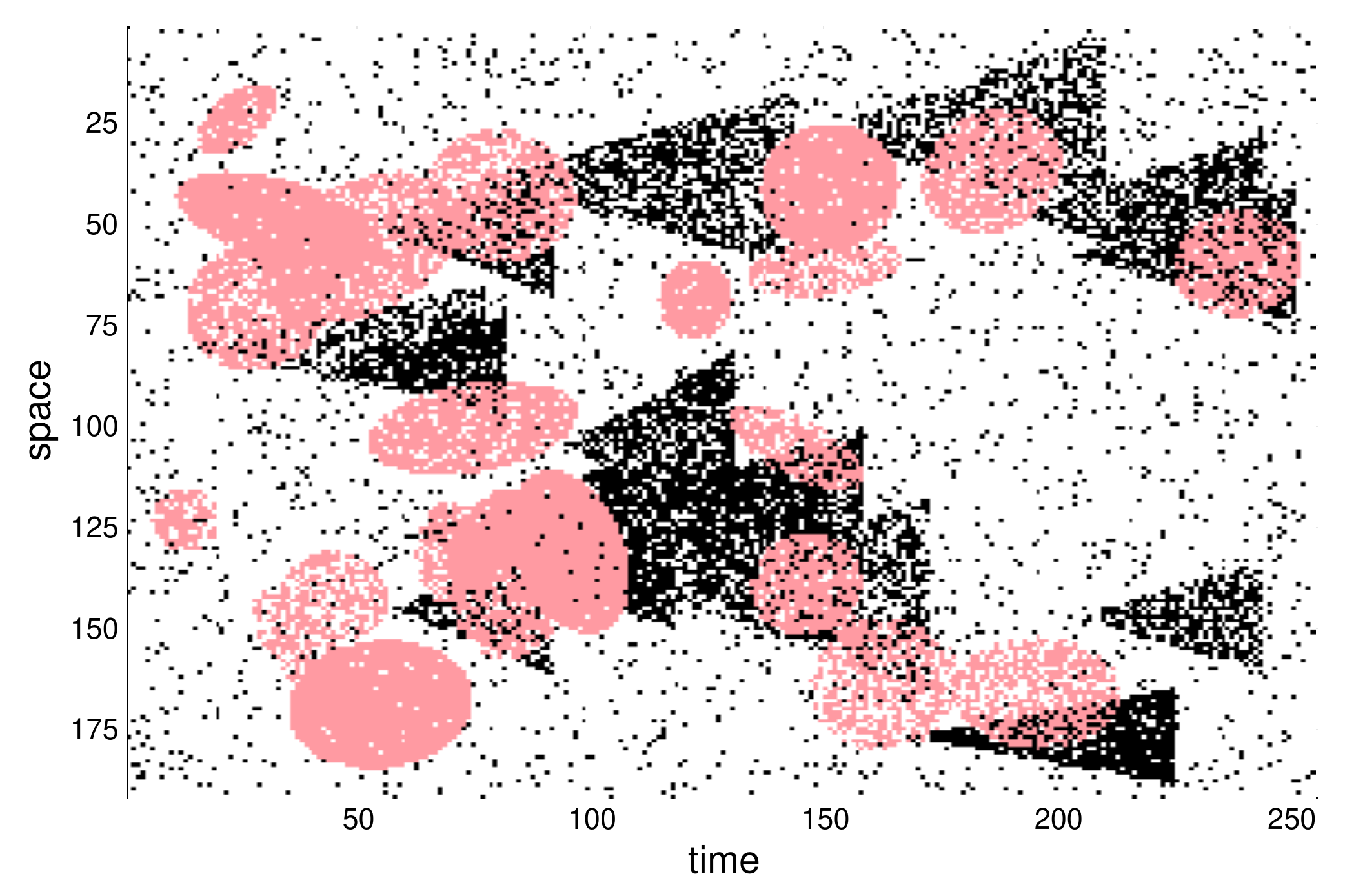}
        \caption{Observed (black/white),  missing (red)}
        \label{fig:geom:data}
    \end{subfigure}
    ~
    \begin{subfigure}[tb]{.48\linewidth}
        \centering
        \includegraphics[width=0.6\linewidth]{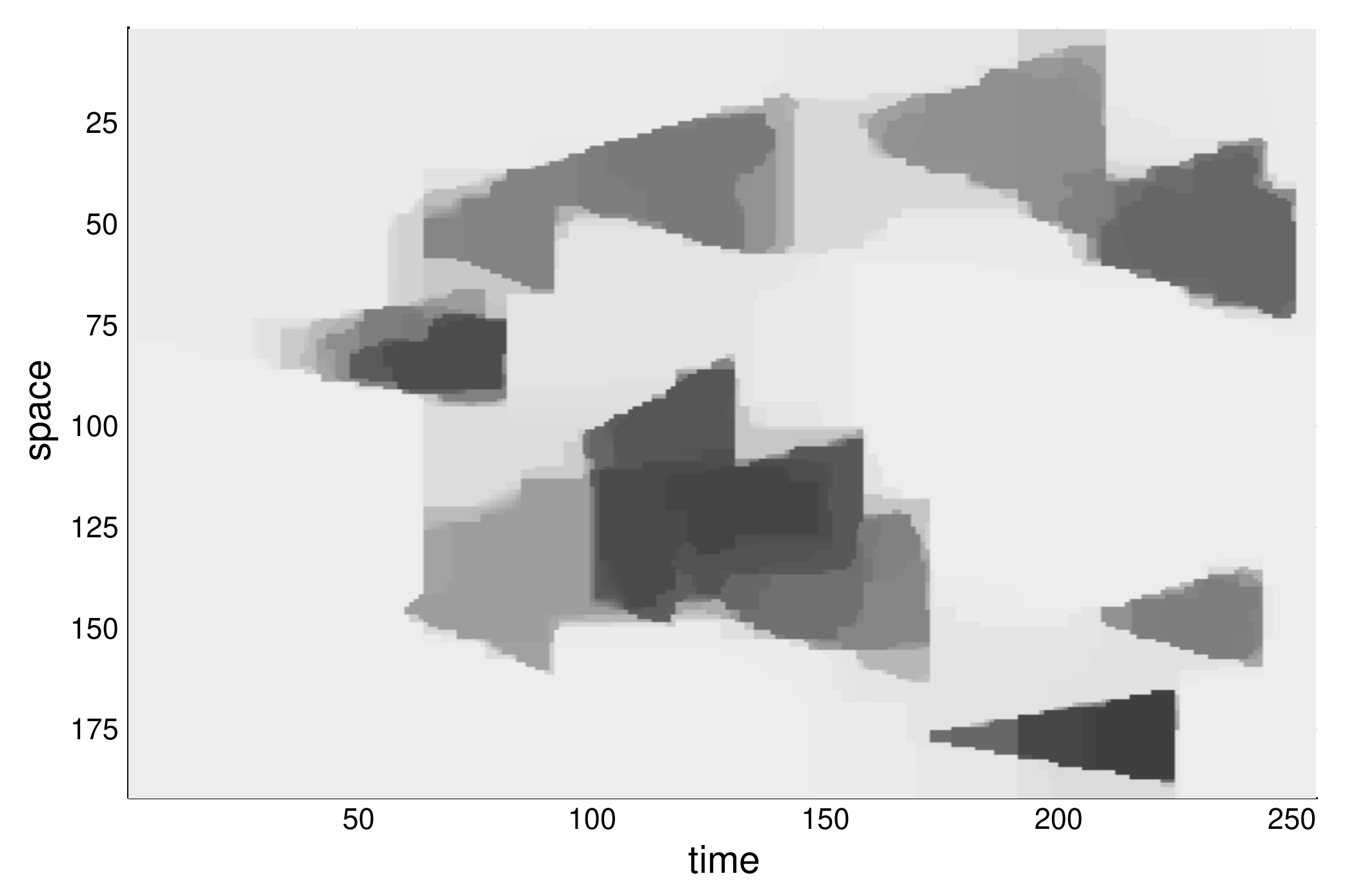}
        \caption{\footnotesize $\lambda_{1}, \lambda_{2}=1.0, 0.001$}
        \label{fig:geom:gfl}
    \end{subfigure}%
    ~
    \begin{subfigure}[tb]{.48\linewidth}
        \centering
        \includegraphics[width=0.6\linewidth]{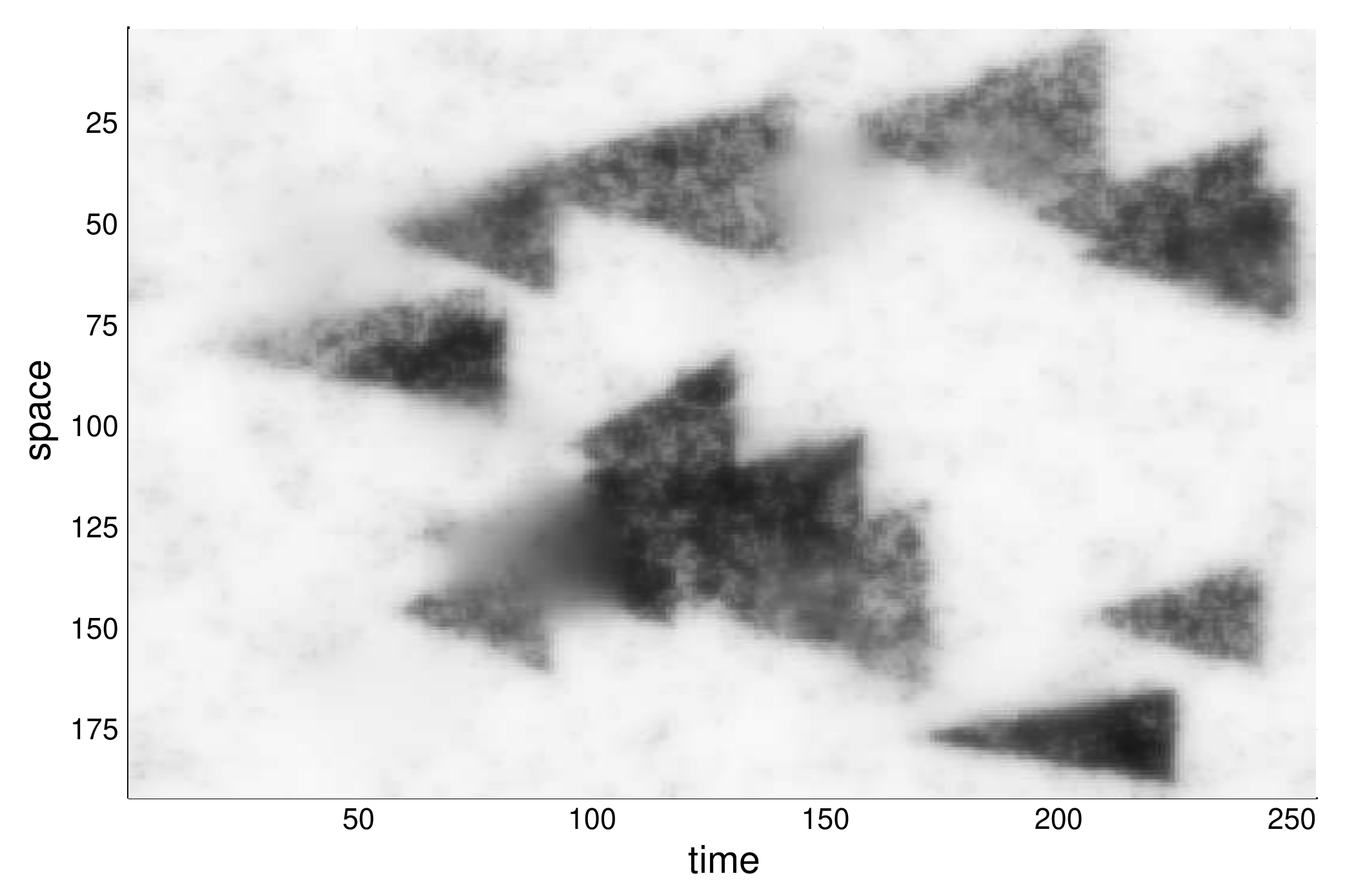}
        \caption{\footnotesize $\lambda_{1}, \lambda_{2}=0.001, 2.0$}
        \label{fig:geom:gmrf}
    \end{subfigure}
    ~
    \begin{subfigure}[tb]{.48\linewidth}
        \centering
        \includegraphics[width=0.6\linewidth]{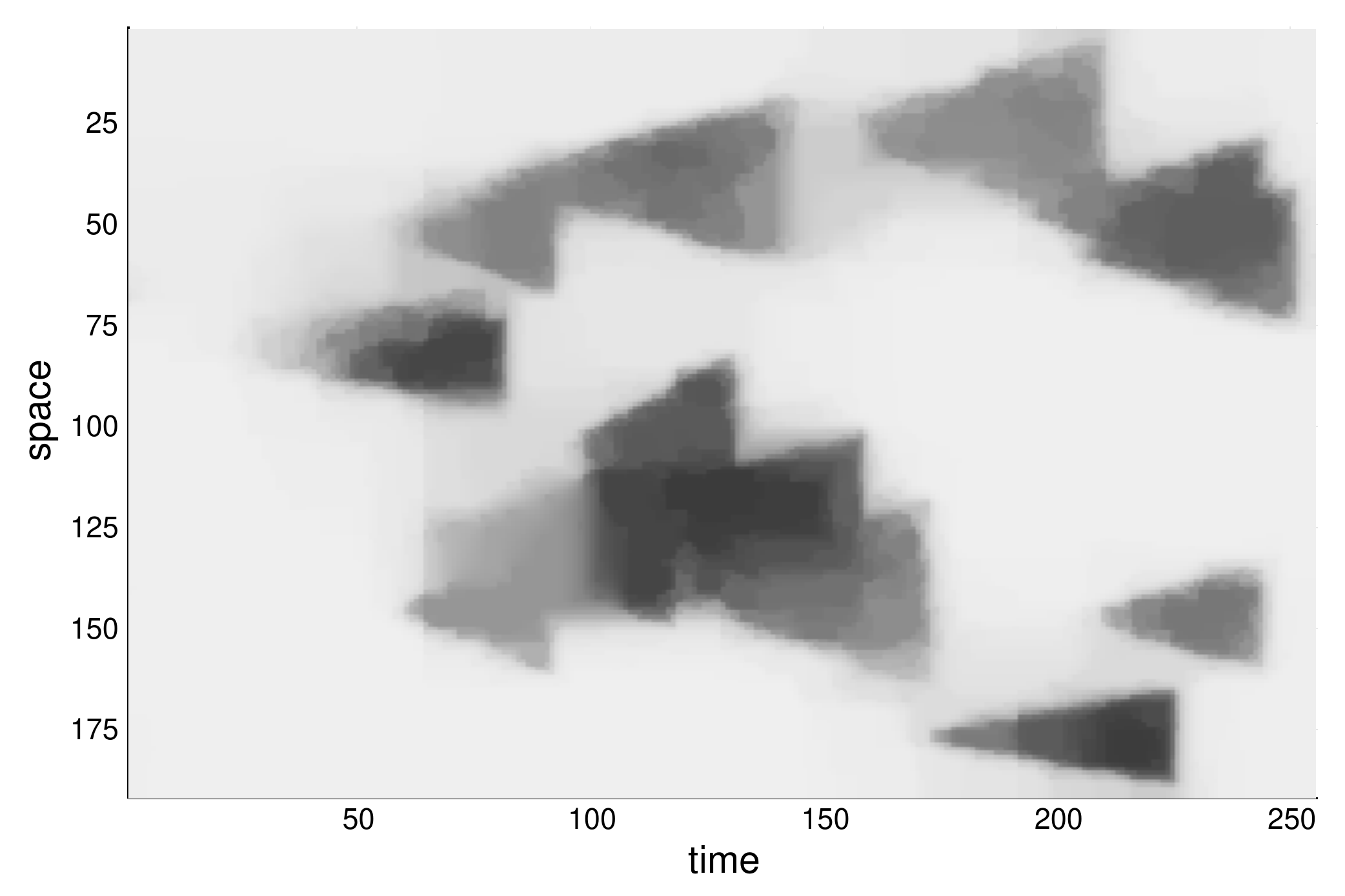}
        \caption{\footnotesize $\lambda_{1},  \lambda_{2}=0.75, 1.5$}
        \label{fig:geom:gfen}
    \end{subfigure}%
    ~
    \begin{subfigure}[tb]{.48\linewidth}
        \centering
        \includegraphics[width=0.6\linewidth]{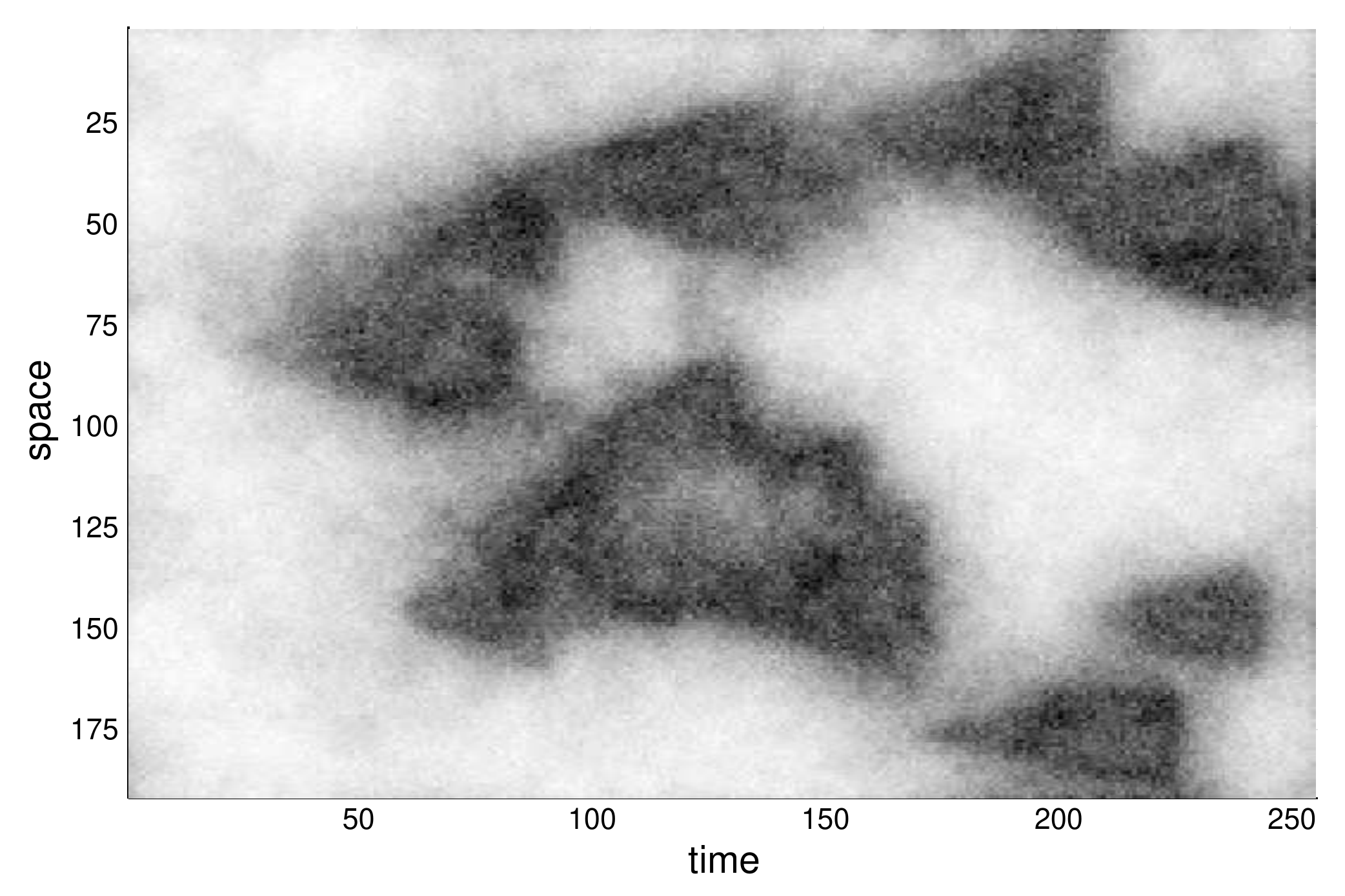}
        \caption{\footnotesize Posterior uncertainty ($\lambda_{1},  \lambda_{2}=0.75, 1.5$)}
        \label{fig:geom:gfen-uncomment}
    \end{subfigure}%
    \caption{Examples of inference with variants of the Binomial GFEN. The horizontal and vertical dimensions represent time and space. The data shown in (a) are simulated samples from $\mathrm{Bernoulli}(1, p)$ at each pixel. Black dots are successes and white dots failures. $p=5\%$ for the background and $p\sim \mathrm{Unif}(50\%, 95\%)$ for the triangles. Red pixels are missing data nodes.$\lambda_{p}$ is the parameter value assigned to both $\lambda_{Sp}$ and $\lambda_{Tp}$. (f) is the IQR of the posterior computed at each pixel. Black dots in (f) mean higher uncertainty.} 
    \label{fig:example-image-binomial}
\end{figure}

\begin{figure}[tb]
    \centering
    \includegraphics[width=0.93\linewidth]{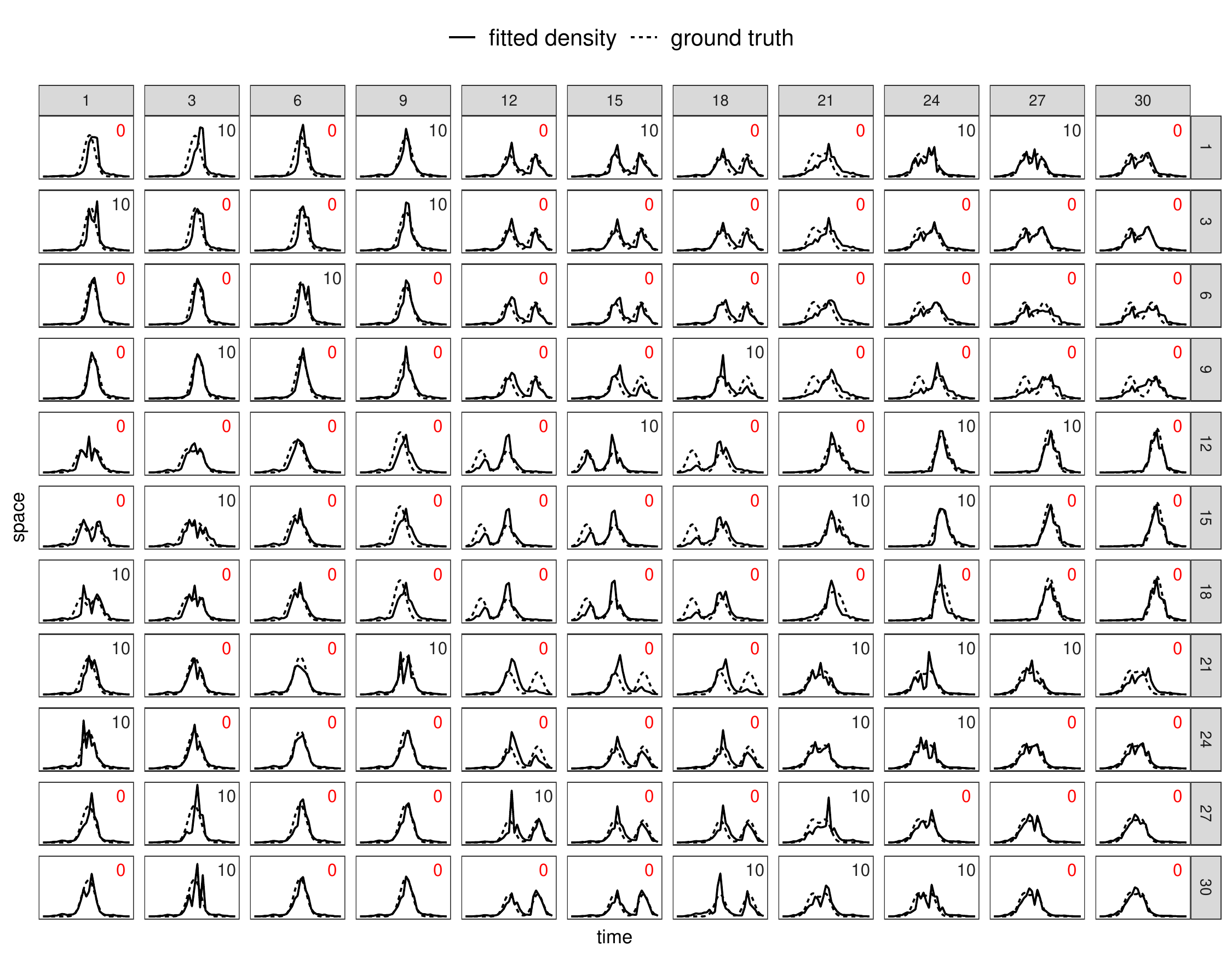}
    \caption{Example of simulated density smoothing task on a spatio-temporal grid of $30 \times 30$ with non-separable space-time effects. Each density plot corresponds to a vertex on the graph. The number of samples at each vertex is indicated in the top-right corner of each curve. 80\% randomly selected vertices have no data, and the rest have sample size 10. The solid line shows the GFEN's MAP estimates for $\lambda_{S1}=0.25, \lambda_{S2}=0.05, \lambda_{T1}=0.25, \lambda_{T2}=0.05$.}
    \label{fig:benchmarks-spt}
\end{figure}

The relative weights assigned to the $\ell_1$ and $\ell_2$-total variation penalties in expressions \eqref{eq:gfen} and  \eqref{eq:gfen-spatiotemporal} have a very intuitive interpretation: their ratio controls the degree of sharpness and smoothness in the solutions. The GFEN solutions will typically have both smooth transitions and collapse to piece-wise constant estimates regions with low variation relative to the signal and penalty parameter. The GFEN also interpolates through missing data like the GMRF. This intermediate behavior is illustrated in Figure \ref{fig:benchmarks-linear} in the Appendix, which shows a scenario similar to Problem \ref{prob:missing-data} with missing data vertices on a chain graph. Figure \ref{fig:example-image-binomial} illustrates inference with the GFEN for the Binomial likelihood as in Problem \ref{prob:missing-data-binom}, which is the model we use to estimate the conditional splitting probability for our density estimation technique. The figure shows how the $\ell_1$-norm promotes sharpness, and the $\ell_2$-norm promotes smoothness. In all cases, the smoothed signal reflects interactions in space and time and interpolates missing data. The figure also displays the interquartile range for the posterior estimates at each point in the grid, showing that the uncertainty concentrates along the edges between different success probabilities, as well as regions of missing data.  To offer further intuition, in proposition \ref{prop:gfen} we also state the appendix we state the analogous closed-form solution of the GFEN to the toy problems problems \ref{prob:missing-data} and \ref{prob:missing-data-binom} which captures this behavior.

We now show an illustration of the GFEN for a full density estimation task with a high rate of missing data vertices and complex space-time interactions, mimicking the challenges in the RideAustin dataset. Figure \ref{fig:benchmarks-spt} shows a subsample of a $30\times 30$ grid where at each vertex $(s,t)$ a density $f(s,t)$ is estimated using the dyadic decomposition technique explained in section \ref{subsection:densities}. The figure shows that: (1) the method does a good job at recovering the true densities even at the missing data vertices; (2) the smoothing adapts to complex space-time interactions. The details of the simulation are given in the following section and the Appendix.

From the Bayesian point of view, the GFEN can be seen as a MRF prior
$$
p(\beta) \propto \exp(-\textstyle{\sum_{p \in \{1,2\}} \sum_{vw\in E} \lambda_p\lvert\beta^{(v)}-\beta^{(w)}\rvert^p}).
$$
As other MRF priors, it carries the Markovian assumption that  a variable at a given location is independent of the rest of the graph conditional on its neighbors. Unconditionally, MRF priors allow correlation between any two arbitrary nodes in the graph.

\subsection{Fast MAP estimation and Bayesian inference}
\label{sec:fitting}
 The focus of this work is on large-scale inference. For this purpose, we develop a technique for obtaining maximum \textit{a posteriori} (MAP) at scale by extending the fast convex optimization technique for the GFL presented by \citet{tansey-scott-2015} to the objective in \eqref{eq:gfen}. We briefly outline the procedure. First, we start from a decomposition of the edges $E$ into a set of non-overlapping trails $E = \bigcup \{\tau \mid \tau\in\mathcal{T}\}$. The core of this strategy is to reduce the optimization objective to solving individual, non-overlapping smoothing problems along each trail $\tau$ for each $\ell_p$ penalty separately. The advantage of doing this is that the subproblem in each trail is 1-dimensional and can be solved in linear time \citep{barbero-sra-2018}. Second, a principled mathematical way is necessary to combine the solutions in each trail and guarantee that the original objective is minimized. The ADMM algorithm \citep{boyd2011distributed} provides a formal optimization framework for this purpose. Full details of our extension are in Appendix \ref{appendix:algorithm}.

We remark that although our focus in this work is on scalable estimation, Bayesian inference with the GFEN is still possible. While the MAP can be obtained in a few minutes for a graph with a million nodes, MCMC sampling typically takes many hours or days. The MAP estimate can be used to provide a smart initiation scheme for the Markov chains. Gibbs sampling is easy to implement, noting that the negative loglikelihood in \eqref{eq:gfen} is convex. Therefore it is easy to draw iid samples for the Gibbs conditional using Adaptive Rejection Sampling for sampling log-concave densities \citep{gilks1992adaptive}. See appendix \ref{appendix:algorithm} for further details. Parallelism can be implemented with asynchronous Gibbs steps \citep{terenin2020asynchronous}. This is the approach we take to provide uncertainty quantification for our case study after tuning the hyperparameters using our fast algorithm for point estimates. In section \ref{sec:hypersearch} we provide further details of the hyperparameter selection strategy.

\subsection{Simulation study}\label{section:simulation}

We conducted a brief series of experiments to understand better how the GFEN compares with respect to the GFL and a GMRF on simulated data sets with features broadly similar to the RideAustin data.  The question we seek to answer here is: in the presence of missing data and both smooth and sharp spatiotemporal transitions, how does the GFEN compare with both the GFL and $\ell^2$ smoothing of the kind arising in a GMRF? 

We designed a density smoothing task on a $30\times 30 $ grid representing the time and space dimensions. At each node $(s,t)$, the true density $f$ consists of a mixture of Gaussians of the form $f(s,t) = \frac{1}{2}N(\nu_{1}(s)\mu_{1}(t), \sigma) + \frac{1}{2}N(\nu_{2}(s)\mu_{2}(t), \sigma)$. The space and time effects $\mu_i$ and $\nu_i$ vary according to three possible behaviors: piece-wise constant, piece-wise linear, and a mix of the former two. The space and time effects interact with each other, so they cannot be additively separated. Each task is evaluated under two missing-data regimes with 10\% or 80\% of the vertices having missing data. The sample size is 10 in the nodes where data is observed. Figure \ref{fig:benchmarks-spt} exemplifies a simulated task using this recipe. In addition, to assess the robustness of the methods with respect to outliers, we designed a scheme where 50\% of the vertices have one outlier, sampled from an over-dispersed distribution with more than ten times the standard deviation.

\begin{table}[!tb]
  \centering
  \footnotesize
  \begin{tabular}{l|l|ccc|ccc}
    \multicolumn{2}{c}{} & \multicolumn{3}{c}{10\% missing} & \multicolumn{3}{c}{80\% missing} \\
    \itshape spatial effect & \itshape time effect & \itshape GFL & \itshape GFEN & \itshape GMRF & \itshape GFL & \itshape GFEN & \itshape GMRF \\ \midrule
    pw constant & pw constant & 0.903 & \textbf{0.767} &  0.843  & 0.949 & \textbf{0.886} &  0.906 \\ 
    pw constant & pw linear & 0.869 & \textbf{0.756} &  0.839 & 0.945 & \textbf{0.832} &  0.887 \\ 
        pw constant & mixed & 0.888 & \textbf{0.802} & 0.865 & 0.946 & \textbf{0.903} & 0.913 \\ 
    pw linear & pw linear & 0.858 & \textbf{0.721} & 0.782  & 0.812 & \textbf{0.758} & 0.795 \\ 
    pw linear & mixed & 0.813 & \textbf{0.780} & 0.814 & 0.926 & \textbf{0.826} & 0.859 \\ 
    mixed  & mixed & 0.840 & \textbf{0.748} & 0.831 & 0.937 & \textbf{0.858} & 0.916 \\ 
    mixed+outliers  & mixed+outliers & 0.886 & \textbf{0.789} &  0.840 & 0.965 & \textbf{0.858} & 0.885  \\
 \end{tabular}
 \normalsize
\caption{Simulated density estimations task with Dyadic decomposition. Spatiotemporal grids with missing data and non-additively separable space-time effects. The table shows the average out-of-sample negative loglikelihood (less is better) for each task combination.}
\label{tbl:benchmarks}
\end{table}

Table \ref{tbl:benchmarks} shows the results of the experiment with a comparison of the performance of each method on different possible combinations of spatial and temporal dynamics. The performance value reported in the table is the best out-of-sample negative loglikelihood estimated using a validation dataset sampled independently from the true distribution, averaged over 48 simulated datasets for each task. For each dataset, the hyperparameters are selected with 5-fold cross-validation from a set of random candidates. Overall, the GFEN is the best performing method in all tasks. A perhaps surprising result is that the GFL performs overall worst than the GMRF even for the piecewise constant effects. A plausible hypothesis is that the tiny shrinkage regularization added to the GFL for identifiability  is causing biased estimates
(see note at the end of section \ref{sec:gfl} and Appendix \ref{appendix:simulation_task}.)

\newpage
\section{Case Study: Driver Productivity}
\label{section:case-study}

\subsection{Ride-Sourcing Data}

The non-profit TNC RideAustin, based in Austin, Texas, published data about their ride-sourcing service in early 2017 \citep{dataworld-2017}. The dataset records rides that happened between June 2nd, 2016, and April 13th, 2017. Each trip corresponds to a row in the database and includes information about the origin and destination coordinates, starting and ending time, driver number, cost, and request time. During this period, RideAustin had no major competition since rival companies such as Uber and Lyft were temporarily restricted from operating in Austin.  Since the demand during the first month was limited, we restricted our analysis to data from September 1st, 2016, to April 13th, 2017. We selected rides having the origin and destination coordinates within the traffic analysis zones (TAZs) of Austin.
Since a trip can have different vehicle categories and rates (standard, premium, luxury, and sport utility vehicle [SUV]), to make every trip comparable, we standardized all of them to the regular car category using RideAustin's public pricing formula. We also used this formula to normalize trips with surge price. Our motivation for removing the surge price is that whereas the pricing scheme is known and dependent on standard features such as mileage and time rates, the surge price onset and offset are less predictable.

\subsection{Measuring Productivity}\label{subsection:productivity}

To measure driver productivity, we follow the strategy suggested by \citet{zuniga-etal-2019}. Our \textit{productivity} measurement is taken \textit{prospectively} from the driver's notionally future rides. %
This metric is an indirect measure of a driver's expected continuation payoff. It intends to capture the market conditions of the trip destination and further spatial dynamics for ending in this specific region. It assesses the value or productivity of a region, inherent to characteristics of an area, such as expected trip revenue and unproductive time. 

\begin{figure}[tb]
    \centering
    \includegraphics[width=.5\linewidth]{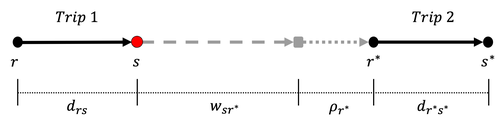}
    \caption{Description of the driver time used for the analysis}
    \label{fig:driver}
\end{figure}

Consider a fixed driver that at time $t$ has just finished a trip (Trip 1) from pick-up and drop-off locations $s$ and $r$ respectively, and whose subsequent trip (Trip 2) has pick-up and drop-off locations $r^*$ and $s^*$ respectively. First, we define the driver's \textit{idle time} $w_{sr^*}$ as the time in hours that the driver will wait until a subsequent trip is assigned. Second, we define the \textit{reach time} $\rho_{r^*}$ as the time between the assignment of Trip 2 and the pick-up moment for Trip 2; note that the driver is not earning income during this period. Third, let $d_{r^*s^*}$ be the duration in hours of Trip 2. Finally, define $F_{r^*s^*}$ as the total final fare that a customer pays for Trip 2 as determined by the tariff system. Our variable of interest is then defined as $\pi_{s,t} := \frac{F_{r^*s^*}}{w_{sr^*} + \rho_{r^*} + d_{r^*s^*}}$.  Figure \ref{fig:driver} shows a graphic representation of the driver time used to estimate this measure. 

This definition of $\pi_{s,t}$ captures an interesting aspect of productivity since it contrasts the time that the driver will stay unproductive with the quality/value of the subsequent trip. Moreover, its values are given naturally in dollars/hour. This metric is assigned to the destination of the first trip ($s$ in Figure \ref{fig:driver}), even if the second trip started in another location because we are interested in understanding the attractiveness of a trip's destination from a driver perspective.
The idea behind this approach is that when a trip ends, the driver starts searching for new riders. This prospective measurement gives the expected earnings given that a driver is at the specific location and time the last trip ended. If a trip ends in a low-demand location, the idle time will be significant, but subsequent trips could be longer. 
We acknowledge that this definition deliberately ignores the characteristics of the trip that led to that position. It does not account for the first trip (Trip 1) distance and duration because we want to isolate the effects of the destination location on further driver earnings.
In Figures \ref{fig:prod:idlehist} and \ref{fig:prod:idlemap} in the Supplement show the distribution of the idle time variable, both aggregating locations and times together, as well as grouped by TAZ. 

For the final analysis, we only consider trips in which the waiting time for a subsequent trip was less than one hour. This assumption was necessary in order to exclude the cases where the driver took a break or stopped working for the day. As previously remarked, during the time of data collection, RideAustin did not have a major competing company. This reduces the cases of extraneous long inter-trip times due to app switching.

\subsection{Construction of the Spatio-Temporal Graph}
\label{section:graph}
The TAZs of Austin provide the advantage of using a size that varies according to the traffic intensity, which is also correlated to the number of trips present in the dataset. We thus have a high resolution near downtown and a low resolution in more rural areas. A total of $1,342$ TAZs were considered, requiring that at least one trip originated or ended in that location. Time was discretized hourly, with a periodicity of one week, for a total of $168=24\times7$ periods. In total, we considered $225,456=168\times1,342$ units of observation, which we used as the vertices of an undirected spatiotemporal graph. Figures \ref{fig:prod:timely} and \ref{fig:prod:spatial}, in Appendix \ref{appendix:global-distribution}, show the total counts of trips aggregating marginally for each time unit and each space unit. When considered marginally, all space units and all the time units have some data. However, once we split by both space and time, only 101,928 vertices (45.2\%) have some data. 

We construct a set of edges $E$ as the union of a disjoint set of spatial and temporal edges $E=E_S\cup E_T$. The set $E_S$ of edges in the spatial direction was constructed geographically, drawing an edge for all geographically adjacent TAZs for a spatial slice at every time. We excluded a few TAZs that were disconnected to the largest connected component of the graph. Edges in the temporal direction $E_T$ were built for every time slice, i.e., joining the vertices for the same TAZ in subsequent hours. An extra edge joining the last and first hour of the week for every location is used to account for the weekly periodicity.  

\subsection{Choice of binary splits}\label{sec:tree-selection} To select the tree structure, we adopt an empirical approach based on the data. The aim is to select a splitting scheme with the following three properties: (1) it generates splits without extreme data unbalances, protecting from instability in the binomial subproblems; (2) it assigns additional resolution to values of productivity where there is more density, so it can more likely detect small spatio-temporal variations; (3) it covers the full support of the densities at a minimally acceptable resolution. If these three properties hold, the density estimates should be robust to the choice of splits. 

As a first approach, we consider using a recursive split based on quantiles of dyadic numbers, using as reference for computing the quantiles the ``global distribution'' where all data points in the RideAustin dataset are pooled together. This approach to splitting the output space using a balanced binary tree is commonly applied in the literature of $kd$-trees, for example, see \citep{bentley1975multidimensional}. This strategy achieves the desired characteristics (1) and (2), but it results in poor detail in the tails of the distribution. Therefore, we adopt a hybrid approach starting from a balanced quantile tree of depth five and add five additional splits to improve resolution in the tails. Appendix \ref{appendix:tree-splitting} further explains the addition of splits in the tails. See figure \ref{fig:quantile-splitting} for a visual comparison of the distinct splitting schemes. In the supplement we also discuss an alternative principled approach for deciding the tree structure based on the Hierarchical MAP Optional Polya Tree (HMAP-OPT) proposed by \citet{wong2010optional}, also shown in the figure. Although we ultimately adopted the quantile approach for our analysis, the results did not change substantially using the HMAP-OPT splits, showing a reassuring robustness to the choice of splits.

\begin{figure}[tb]
    \centering
    \includegraphics[width=0.98\linewidth]{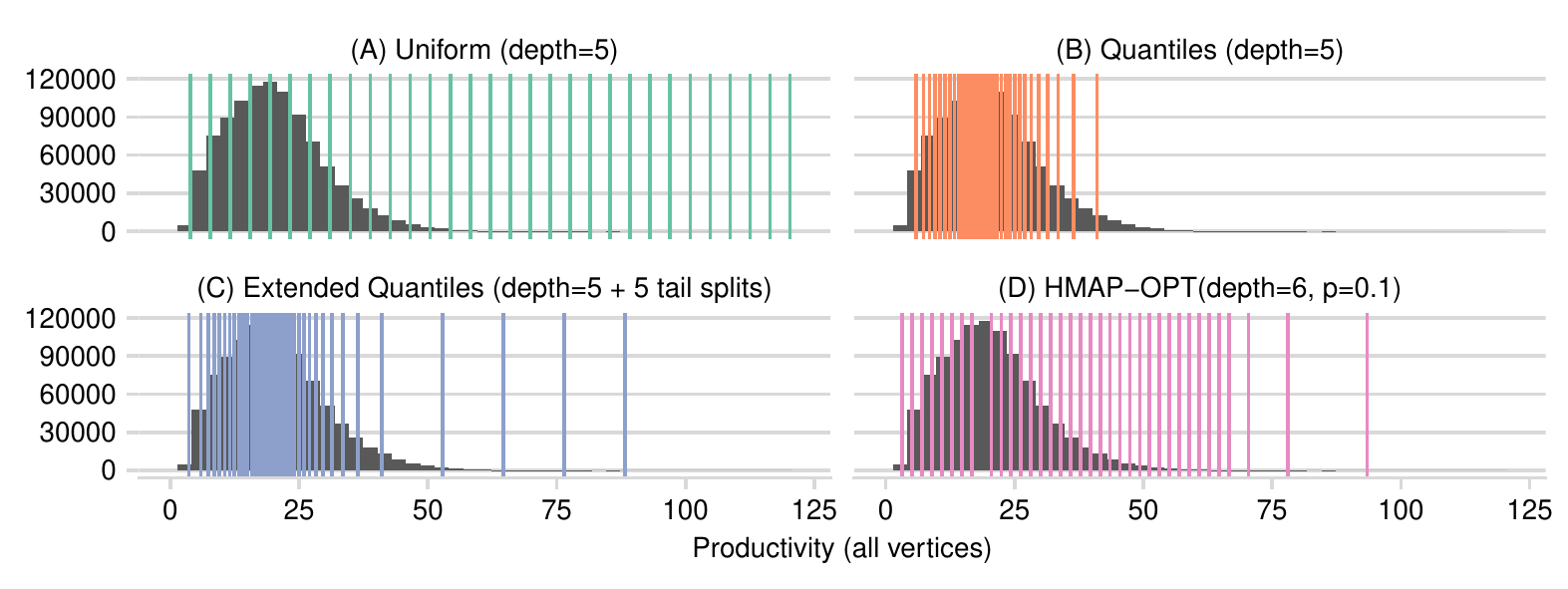}
    \caption{Different choices of tree decompositions. The histogram shows the distribution of all the observations of the productivity variable from the RideAustin dataset.}
    \label{fig:quantile-splitting}
\end{figure}

\subsection{Choice of regularization values with the MAP estimates and cross-validation}\label{sec:hypersearch}

We briefly overview the penalty hyperparameter selection strategy, with full details given in Appendix \ref{appendix:bayes-opt}. The fact that the GFEN can handle smooth and non-smooth transitions and dealing with differentiated spatial and temporal dynamics comes at the expense of four hyperparameters to tune. First, the solution path approaches typically used to tune the GFL penalty hyperparameter are no longer effective. These approaches are based on gradual increments in the value of a single hyperparameter, see \citet{tibshirani-2015}. Further, the smoothness induced by the additional $\ell_2$ penalty means that there is no simple expression for the effective degrees of freedom corresponding to a given combination of hyperparameters. Therefore, model selection techniques based on information criteria that depend on degrees of freedom (e.g., AIC) are more difficult to use. Here we propose to optimize the hyperparameters using a node-wise cross-validation framework, taking advantage of our fast MAP estimation framework, and show that it can be effective in selecting good regularization penalties.

Regular cross-validation leaves out a portion of the number of data points. However, since a significant fraction of the data is concentrated in a few graph vertices, the proposed cross-validation strategy leaves out entire vertices instead. The cross-validation error is the average loglikelihood of the data from the nodes left out during training. We use $k$-fold cross-validation with $k=5$ for each combination of hyperparameters to evaluate the average out-of-sample loglikelihood in the vertices left out using their respective predicted densities. To select promising candidates of hyperparameters to evaluate, we use an active search strategy using the Bayesian optimization framework \citep{shahriari-etal-2016}. This strategy is based on modeling the cross-validated loss using a Gaussian Process as a function of the hyperparameters.  It gives a full predictive surface for the out-of-sample loss at unseen combinations of hyperparameters. This surface and the estimated uncertainty are used to select new candidates from the predictive posterior of the maximum of the surface. To exploit parallelism, we proceed in generations of size $6$ for each of the $36$ splits, for a total of $48$ generations. In the end, the model with the best-predicted loss is selected for each split. Taking into account the number of folds in the cross-validation, a total of $51,840$ models are evaluated and compared in only a few hours using the super-computing environment Stampede 2 at the Texas Advanced Computing Center (TACC).

\begin{figure}[tb]
    \centering
    \begin{subfigure}[tb]{0.48\linewidth}
        \centering
        \includegraphics[width=0.99\linewidth]{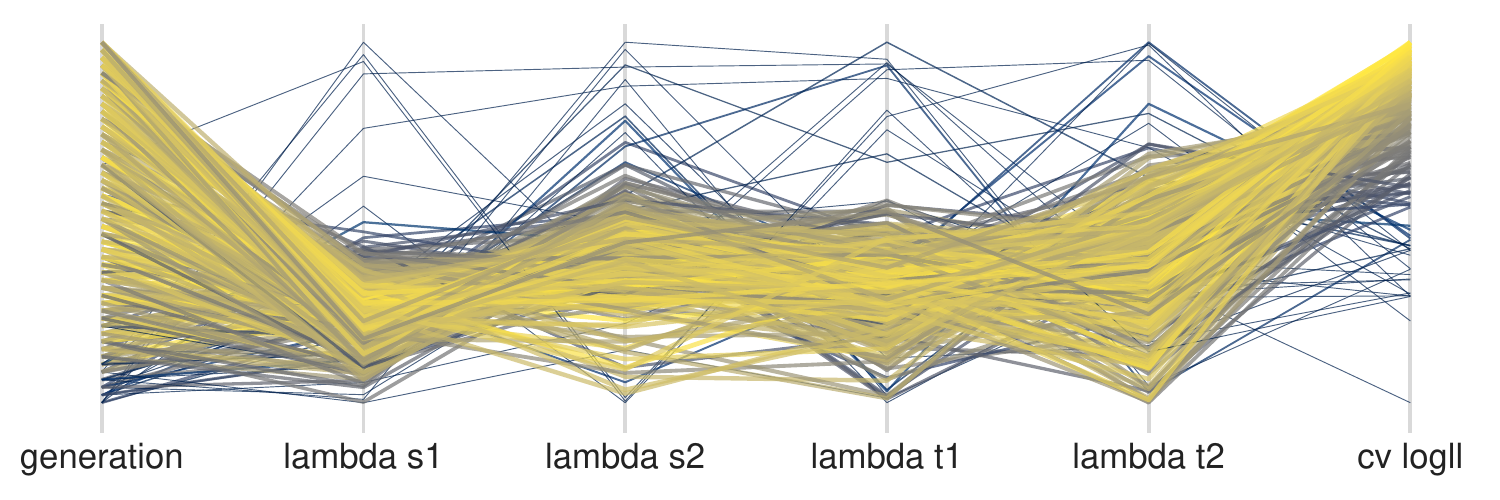}
        \caption{Split 9: (9.6, 11.6, 13.3).}
        \label{fig:bayes-opt:a}
    \end{subfigure}
    ~
    \begin{subfigure}[tb]{0.48\linewidth}
        \centering
        \includegraphics[width=0.99\linewidth]{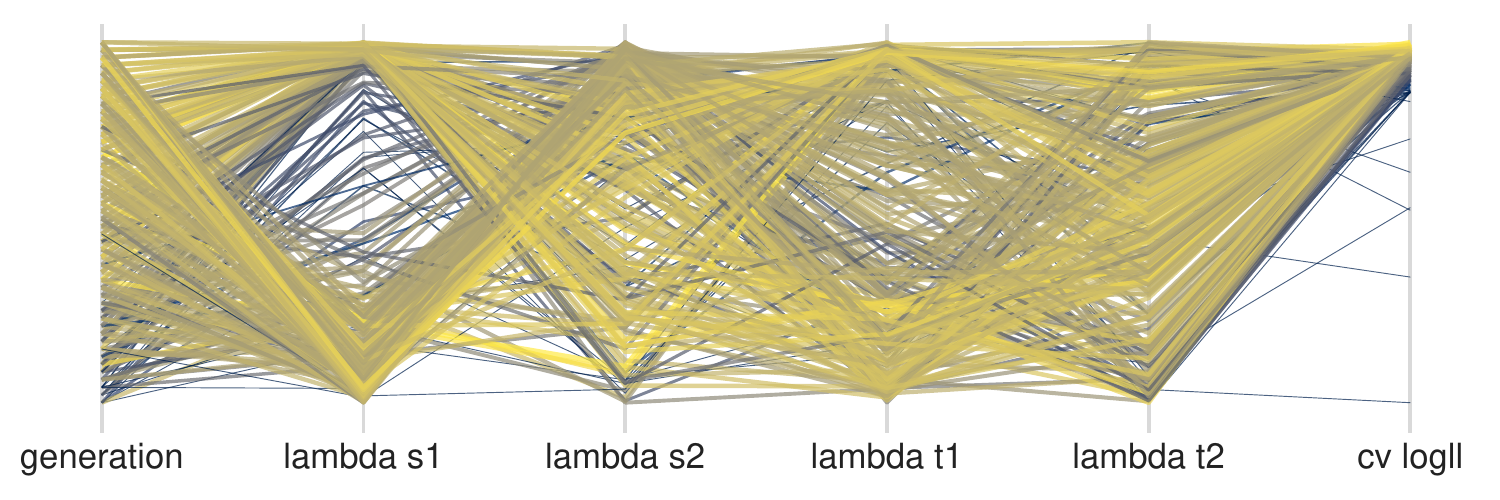}
        \caption{Split 13: (22.6, 24.1, 26.0).}
        \label{fig:bayes-opt:b}
    \end{subfigure}
    ~
    \begin{subfigure}[tb]{0.48\linewidth}
        \centering
        \includegraphics[width=0.99\linewidth]{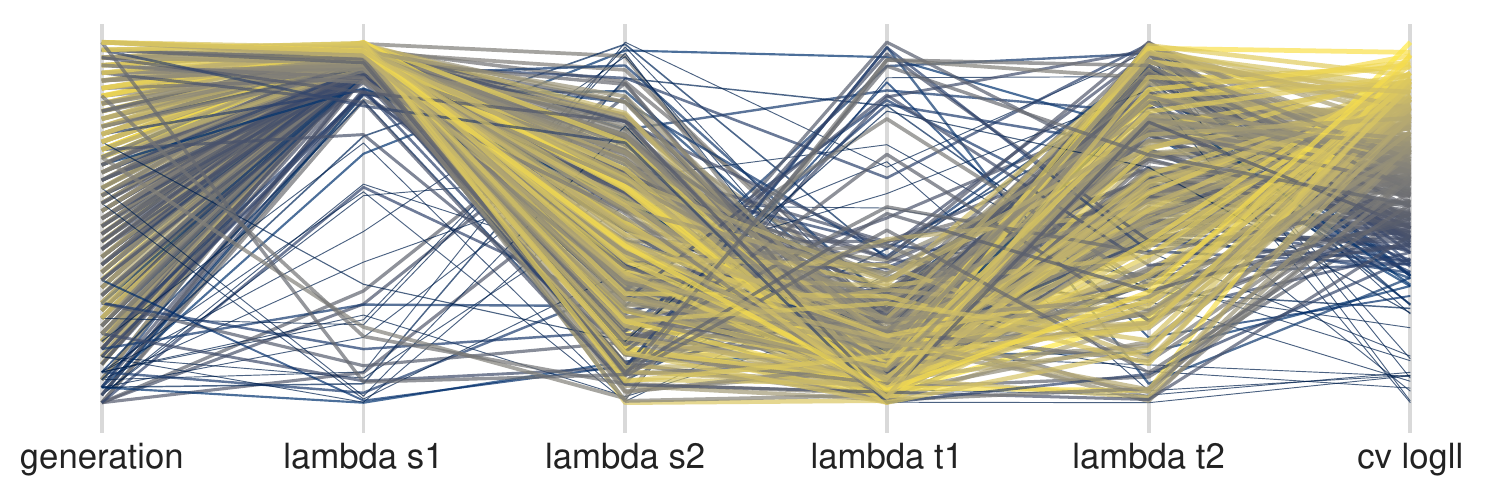}
        \caption{Split 21: (9.57, 11.55, 13.34).}
        \label{fig:bayes-opt:c}
    \end{subfigure}
    ~
    \begin{subfigure}[tb]{0.48\linewidth}
        \centering
        \includegraphics[width=0.99\linewidth]{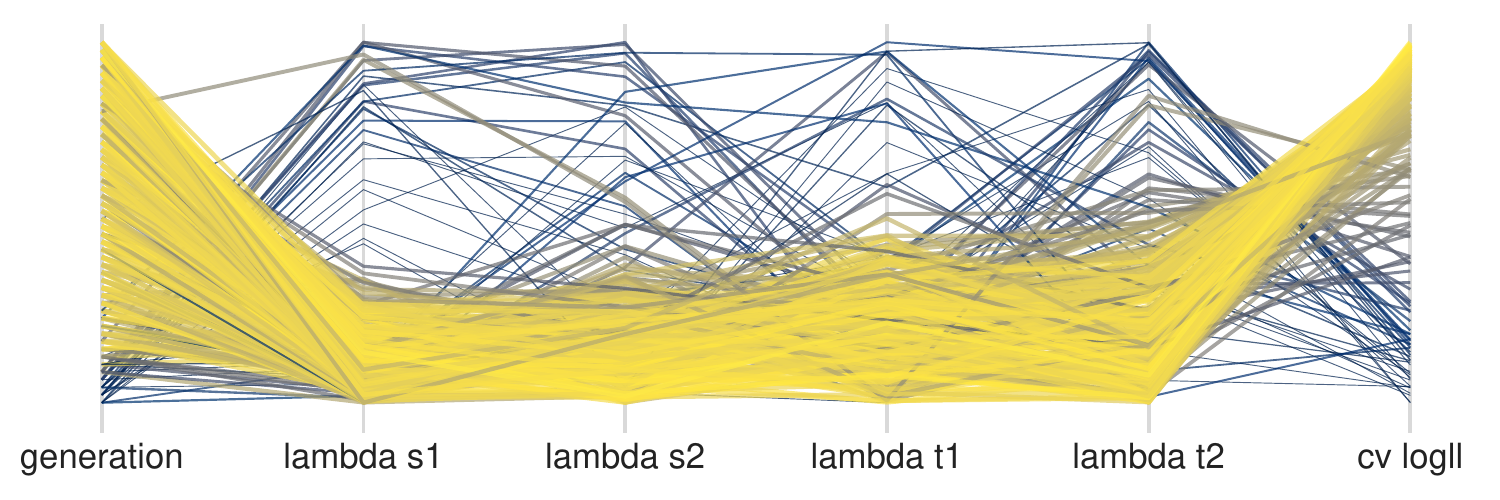} 
        \caption{Split 33: (41.0, 52.8, 100.0)}
        \label{fig:bayes-opt:d}
    \end{subfigure}
    \caption{Parallel coordinates plot of results of the Bayesian Optimization for four examples splits. Each line represents a combination of hyperparameters and the results of the cross-validation loglikelihood evaluation. The leftmost vertical axis is the generation of the Bayesian optimization procedure. The rightmost vertical axis is the evaluated cross-validation loglikelihood, which is also represented by the color of each line for ease of visualization. A brighter color means better outcomes (higher loglikelihood.) The range of values of each hyperparameter and the loglikelihood are normalized to the same range for comparability.}
    \label{fig:bayes-opt}
\end{figure}

Figure \ref{fig:bayes-opt} shows an illustration of the results of the Bayesian optimization process for 4 out of the 36 splits. Each line in the plot represents a combination of parameters. The image illustrates different types of behavior for the optimization strategy. Panel \ref{fig:bayes-opt:a} shows a split where the the best solutions found use a balanced combination of penalization parameters, \ref{fig:bayes-opt:b} shows a substitution effect where good solutions cluster in two groups; \ref{fig:bayes-opt:c} shows a preference for one regularization type; finally, \ref{fig:bayes-opt:d} shows best results with low regularization. Overall, all splits show that the cross-validated loglikelihood improves for later generations.

\section{Case Study: Results}\label{sec:case-study:results}

\subsection{Overview of the inference results}
\label{section:inference-results} 
\begin{figure}[!tb]
\centering
\includegraphics[width=10.5cm, height=11.5cm]{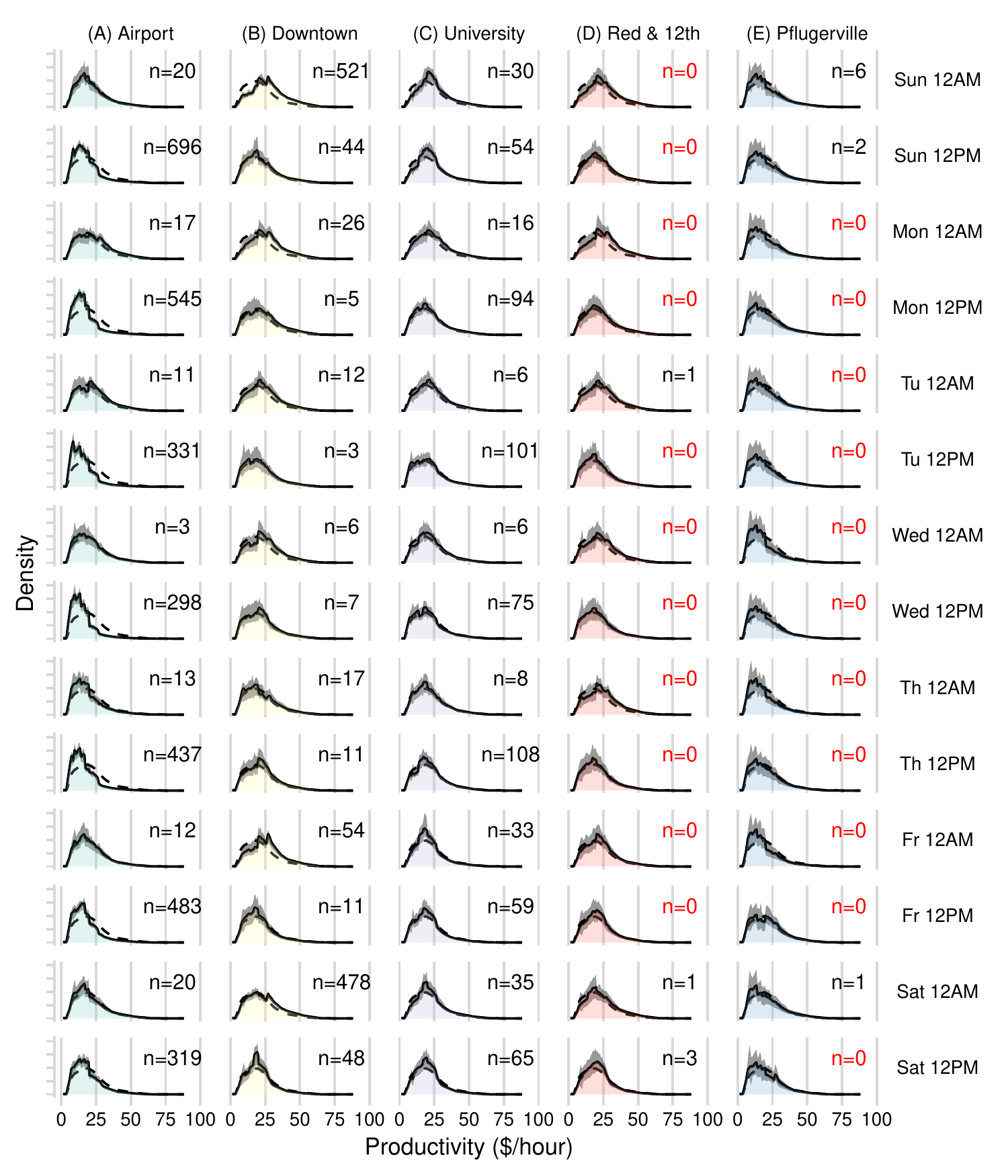}
\caption{Driver Productivity by Time and Location. The global distribution is shown in dashes; the number of observed data points in the corresponding node of the graph is shown in the upright corner of each density plot (n). Time is shown every 12 hours.}
\label{fig:spacetime-densities}
\end{figure}

We show a sample of the results for the representative set of locations described in section \ref{sec:overview} which consider central areas of high and low demand (University, Downtown, Red River \& 12th), the airport, and a suburb distant from downtown (Pflugerville.) Figure \ref{fig:spacetime-densities} shows the estimated densities in twelve-hour intervals. Given the importance of the airport in the ride-sourcing trip demand.  From these figure, several interesting qualitative remarks are readily available.

First, \textit{not all days are the same}: We see that on a typical Wednesday at noon, most locations have a density close to the global mean, whereas, in a typical Saturday night, the locations in the central area (University, downtown, Red River) have higher productivity. Nor are all weekdays exchangeable; Mondays and have higher productivity than, say, Wednesdays. 
 
Second, \textit{smoothed results are sensible even with very sparse data}: We observe similar estimates of the three locations chosen in central Austin (columns B, C, D), despite the fact that Red River had almost no observations (column D).  We also see that in Saturday night in this location is shrunk to its spatial neighbors downtown rather than towards the mean.

\begin{figure}[!tb]
    \centering
    \includegraphics[width=12cm, height=11cm]{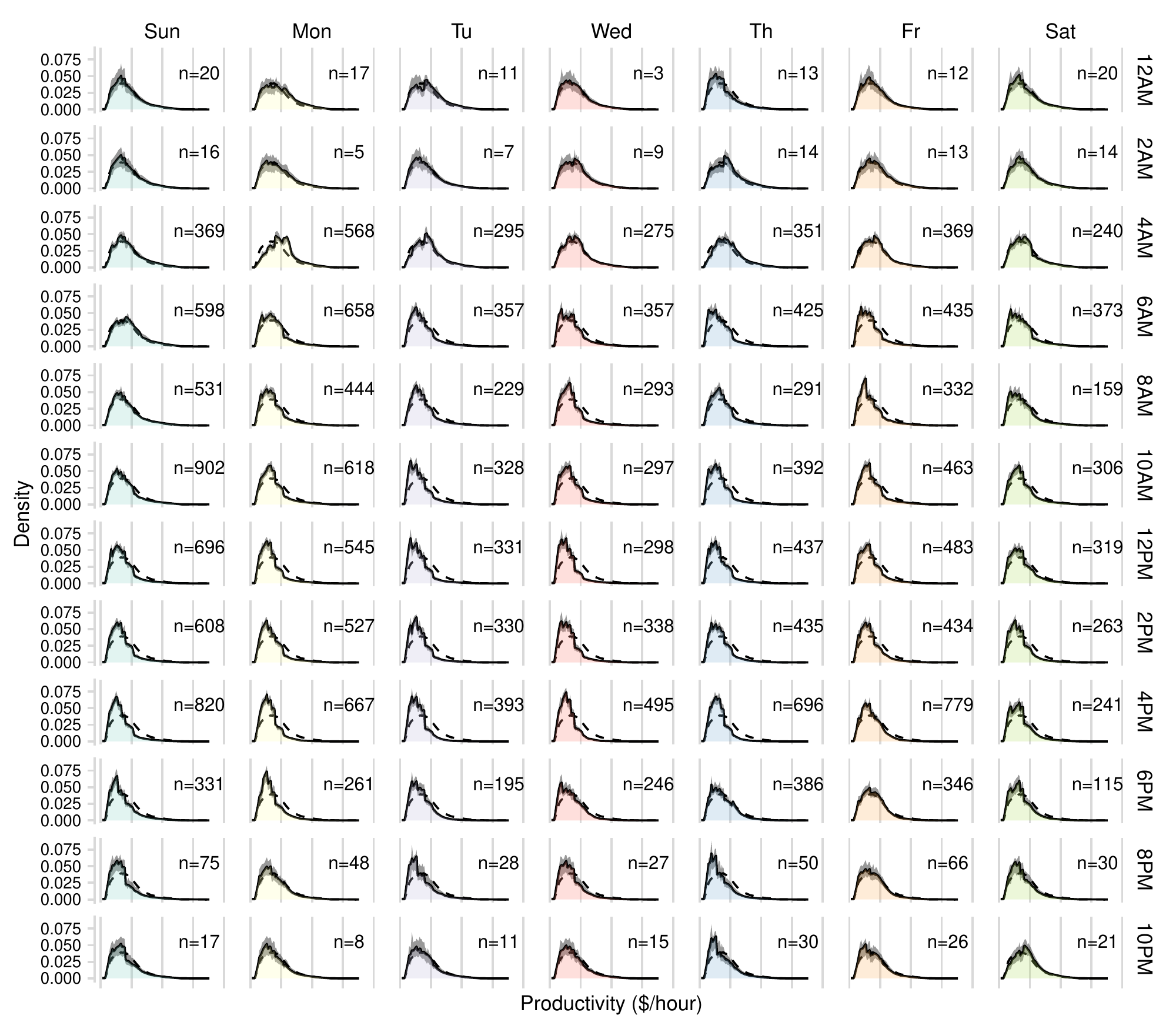}
    \caption{Driver Productivity of the Airport TAZ. The global distribution (dashed); the number of observations in the dataset (n). Time is shown every two hours for a full week.}
    \label{fig:airport-densities}
\end{figure}

Third, \textit{we recover periodicities in the data without assuming them}: We chose not to create edges between different days of the week at the same time of the day (e.g., there is no direct edge between a Monday 12 AM and Tuesday 12 AM). Nevertheless, we do observe periodic patterns for some locations, notably at the airport. Every morning around 3 AM, its distribution is close to the global mean. However, every afternoon it is shifted downwards. Figure \ref{fig:airport-densities} also shows the results in two-hour intervals at the airport. In the figure, one observes a smooth transition between mornings, when the distribution of productivity is closer to the global mean, to afternoons, when it is generally smaller. Interestingly, while other locations showed a clearly different distribution during weekends and business days, the airport is more similar every day of the week.

\subsection{Investigating Driver Productivity}

 We present a list of interesting scientific inquiries that can be answered using the full distributions of productivity at each location and time. First, \textit{what is the probability of not exceeding a specific salary?} To answer this question, we use \textit{tail probabilities}, and we compare to standardized living wages in the Austin area. Second, we seek to know \textit{how many dollars per hour do the least productive fraction $\alpha$ of drivers earn at the most?} For this task, we study \textit{quantiles} from the fitted distributions.

{\bfseries Tail probabilities: the risk of not attaining a living wage}. Having access to the full distribution of earnings allows us to estimate the probability that a driver will obtain a minimum living salary as a function of space and time. Table \ref{tbl:livingwages} shows estimated living wages for families living in Austin in 2017 \citep{nadeau-2017}. 
To these wages, we must add the activity-specific additional costs, such as the fixed fee of \$0.99 charged per trip charged by RideAustin as well as car maintenance, which on average is \$4.78 hourly after-tax deductions \citep{mishel-2018, hall-etal-2016}. Since a driver completes a bit more than one trip per hour on average, we rounded up the total hourly costs to \$6.00. The final reference values, including costs, are also presented in Table \ref{tbl:livingwages}.

\begin{table}[!tb]
    \centering
    \footnotesize
    \begin{tabular}{l|l|l|l|l}
        \textbf{} & \textit{1 adult} &  \textit{2 adults} & \textit{2 adults (1 working)} & \textit{1 adult}  \\ %
        \textbf{} & \textit{0 children}  &   \textit{2 children} & \textit{2 children} & \textit{2 children } \\
       \hline
        \textit{living wage} (\$/hour) & 12.56  & 15.64 & 26.73 & 28.74  \\ 
        \textit{living wage+costs} (\$/hour) & 18.56  & 21.64 & 32.73 & 34.74  \\ 
    \end{tabular}
    \normalsize
    \caption{Estimated living hourly wages in Austin TX for 2017 \citep{nadeau-2017} with estimated maintenance costs and RideAustin's fees.}
    \label{tbl:livingwages}
\end{table}

\begin{figure}[!tb]
    \centering
    \begin{center}
    \begin{subfigure}[tb]{0.48\linewidth}
        \centering
        \includegraphics[width=0.99\linewidth]{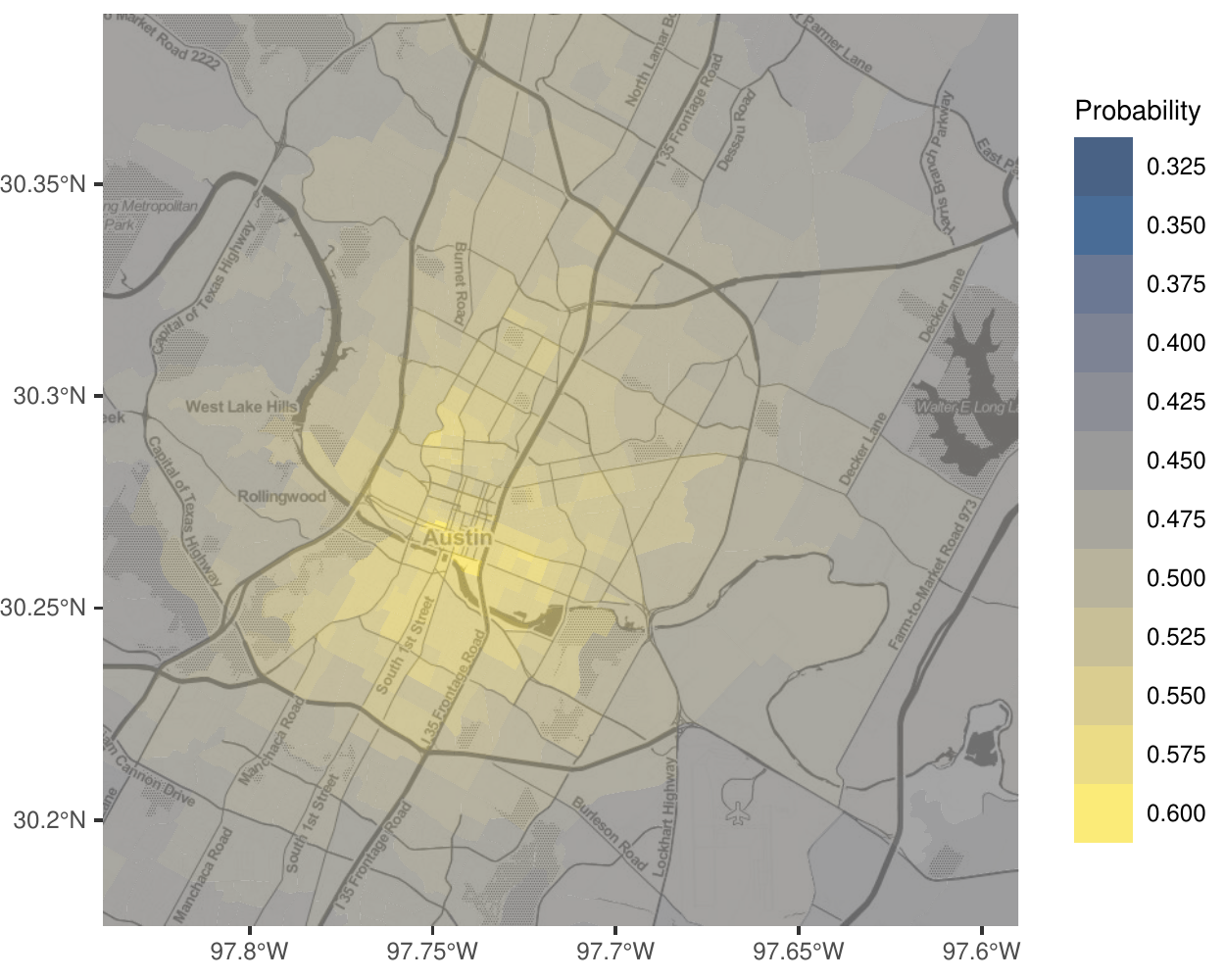}
        \caption{Sunday 12 AM}
        \label{fig:wages:a}
    \end{subfigure}
    \begin{subfigure}[tb]{0.48\linewidth}
        \centering
        \includegraphics[width=0.99\linewidth]{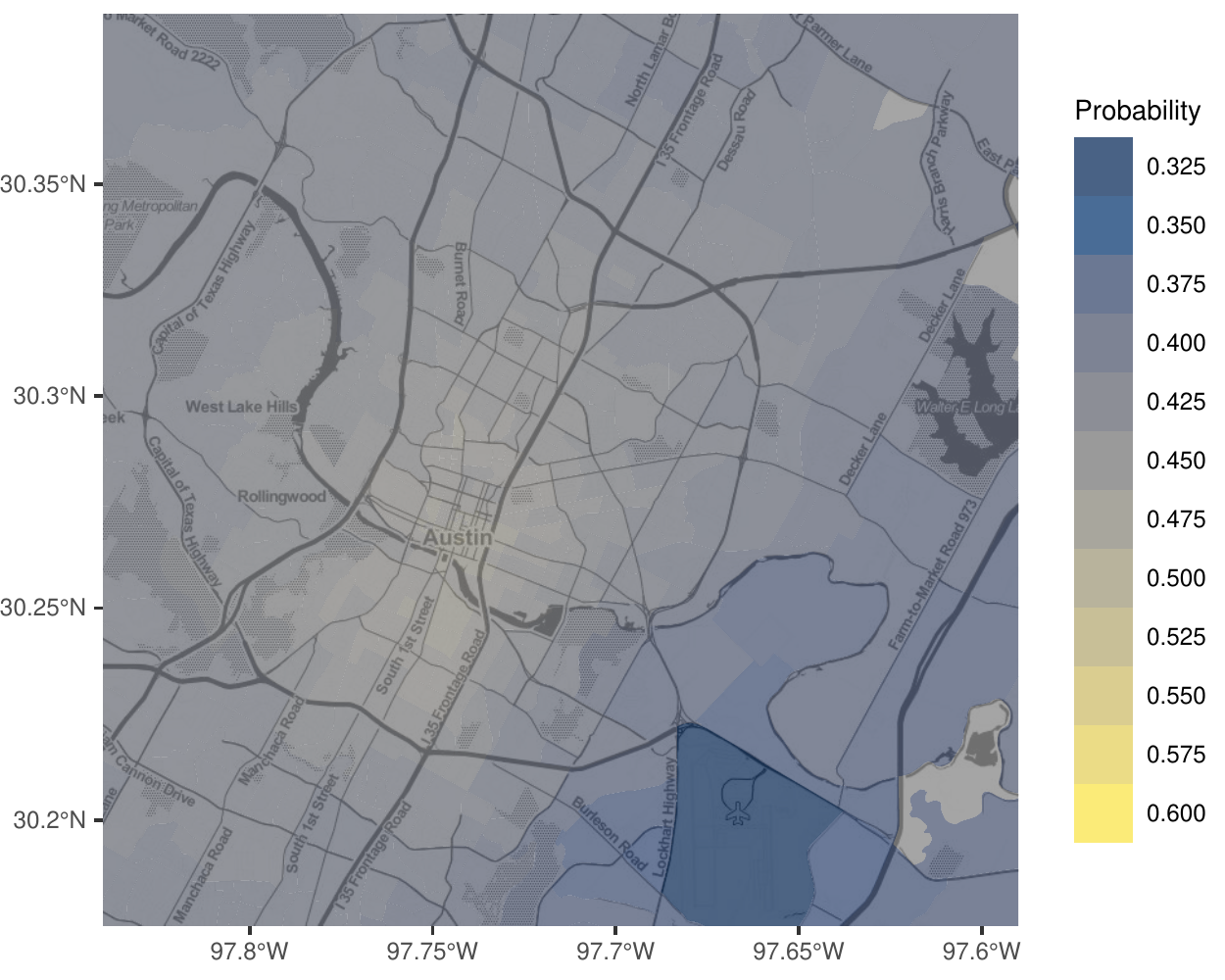}
        \caption{Monday 12 PM}
        \label{fig:wages:b}
    \end{subfigure}\hfill
    \caption{Probability of exceeding \$21.64 in the next hour given a current location (living wage with costs for two working adults with two children).}
    \label{fig:wages}
    \end{center}
\end{figure}

Figure \ref{fig:wages} shows examples of the results for the case of two working adults with two children (\$21.64). Other cases are included in Appendix \ref{appendix:maps_livingwages}. We observe that during a Saturday night when there is a high demand (\textit{c.f.} Figure \ref{fig:prod:timely}) the probability of exceeding the living wage is close to 60\% near downtown, and it decreases to around 45\% as a driver lies farther away from central Austin. In contrast, during Monday 12 PM, with low demand, the probability ranges from 30\% to 45\%, being worst at the airport. These results suggest that drivers are at a high risk of not earning a living wage.

{\bfseries Quantiles: How bad are the least productive doing?} RideAustin is a non-profit company with the explicit goal of seeking to ensure the well-being of its workers.  It makes sense for RideAustin to target the population at specific levels of economic risk. Along these lines, one may ask: what is the expected income of the lowest fraction $\alpha$ of drivers? That is, we seek to estimate the quantiles $q_\alpha(s,t) := \min_q \mathbb{P}(\pi_{s,t} > q) > \alpha$ for each $(s, t)$. Examples of interesting values for $\alpha$ might be $\{0.1, 0.25, 0.5, 0.75, 0.9\}$. The case $\alpha = 0.1$ is shown in Figure \ref{fig:quantiles:0.1}; the rest are included in Appendix \ref{appendix:quantiles}. We can see that on a typical Monday at noon, which has low traffic and demand intensity, the lowest $10\%$ quantile approximately ranges from $\$7$ to $\$11$, with a flatter distribution across the city. In contrast when the demand it reaches $\$13$ in several areas of central Austin, likely because of the high concentration of weekend entertainment venues in central Austin (\textit{c.f.} Figure \ref{fig:prod:timely}). This result should be interpreted in light of Table \ref{tbl:livingwages}, which shows that a living wage in Austin for a single working adult with no children is over $\$18$ in the same year where the data was recorded. Thus, these results suggests again high risk for the drivers to earnin below the living wage. 

\begin{figure}[!tb]
    \centering
    \begin{minipage}[tb]{0.48\linewidth}
        \includegraphics[width=0.99\linewidth]{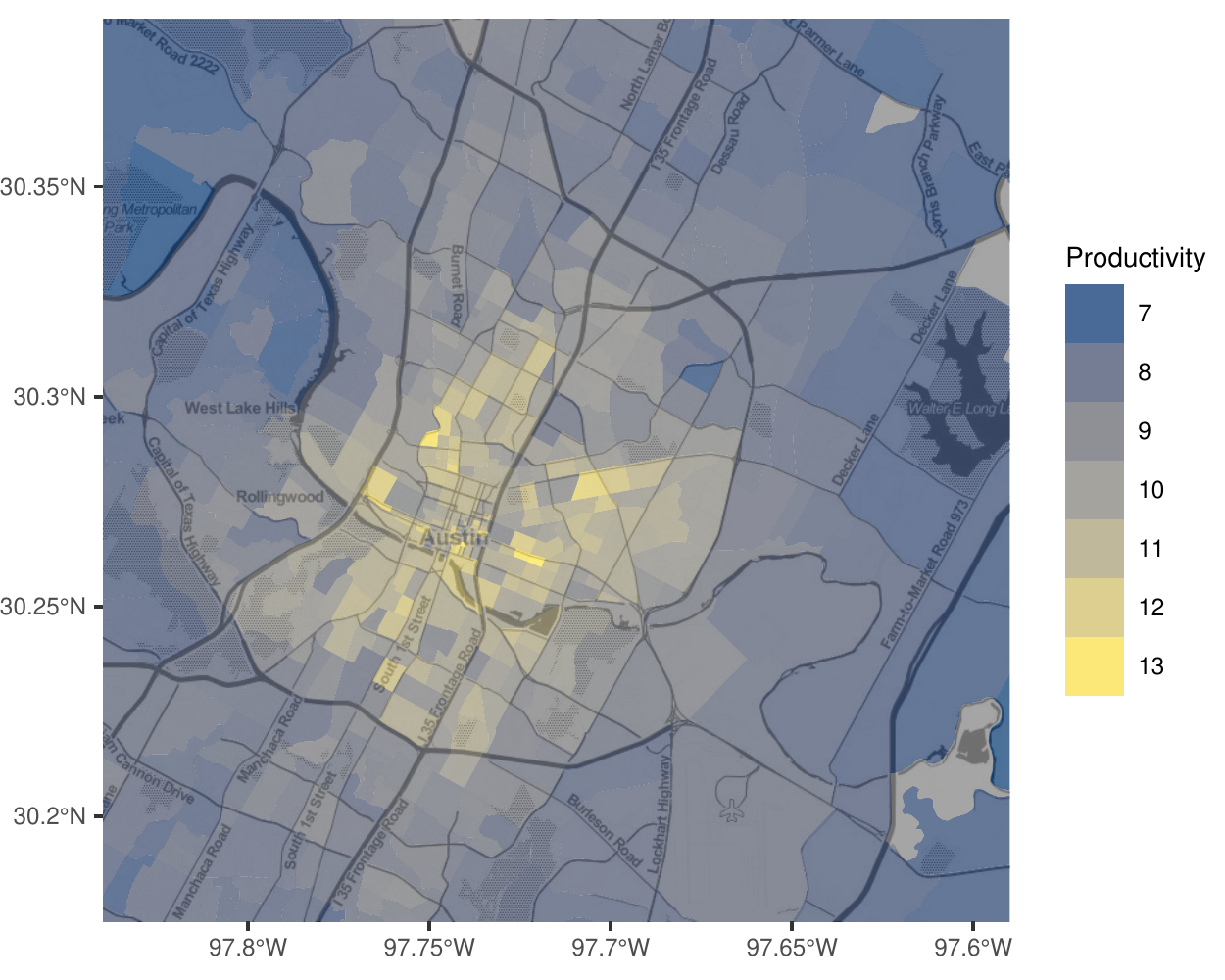}
        \subcaption{Sunday 12 AM}
        \label{fig:quantiles:0.1:a}
    \end{minipage}
    \begin{minipage}[tb]{0.48\linewidth}
        \centering
        \includegraphics[width=0.99\linewidth]{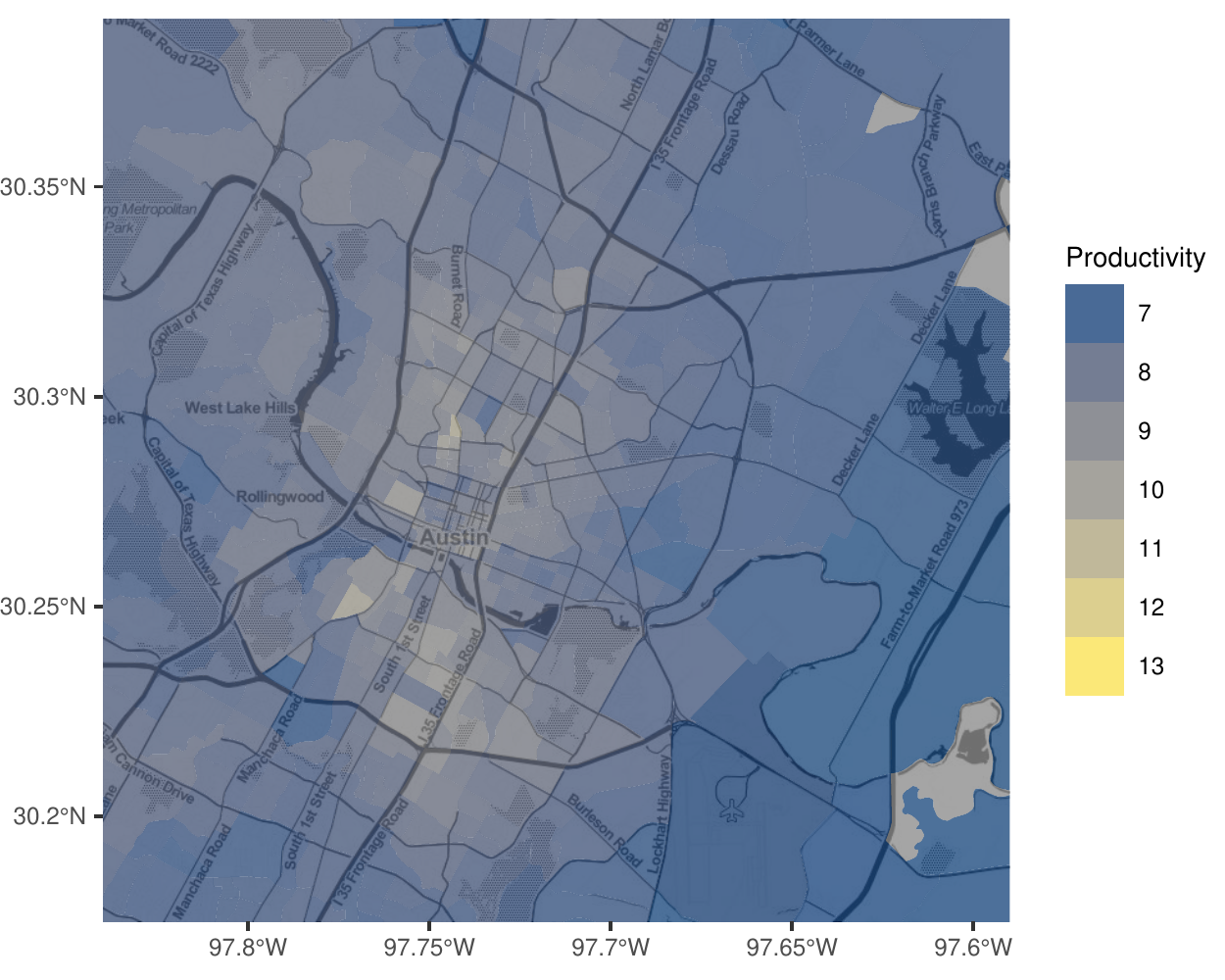}
        \subcaption{Monday 12 PM}
        \label{fig:quantiles:0.1:b}
    \end{minipage}\hfill
    \caption{Lower 10\% quantile of productivity for different times and locations.}
    \label{fig:quantiles:0.1}
\end{figure}

\begin{figure}[!tb]
    \centering
    \begin{minipage}[tb]{0.48\linewidth}
        \includegraphics[width=0.99\linewidth]{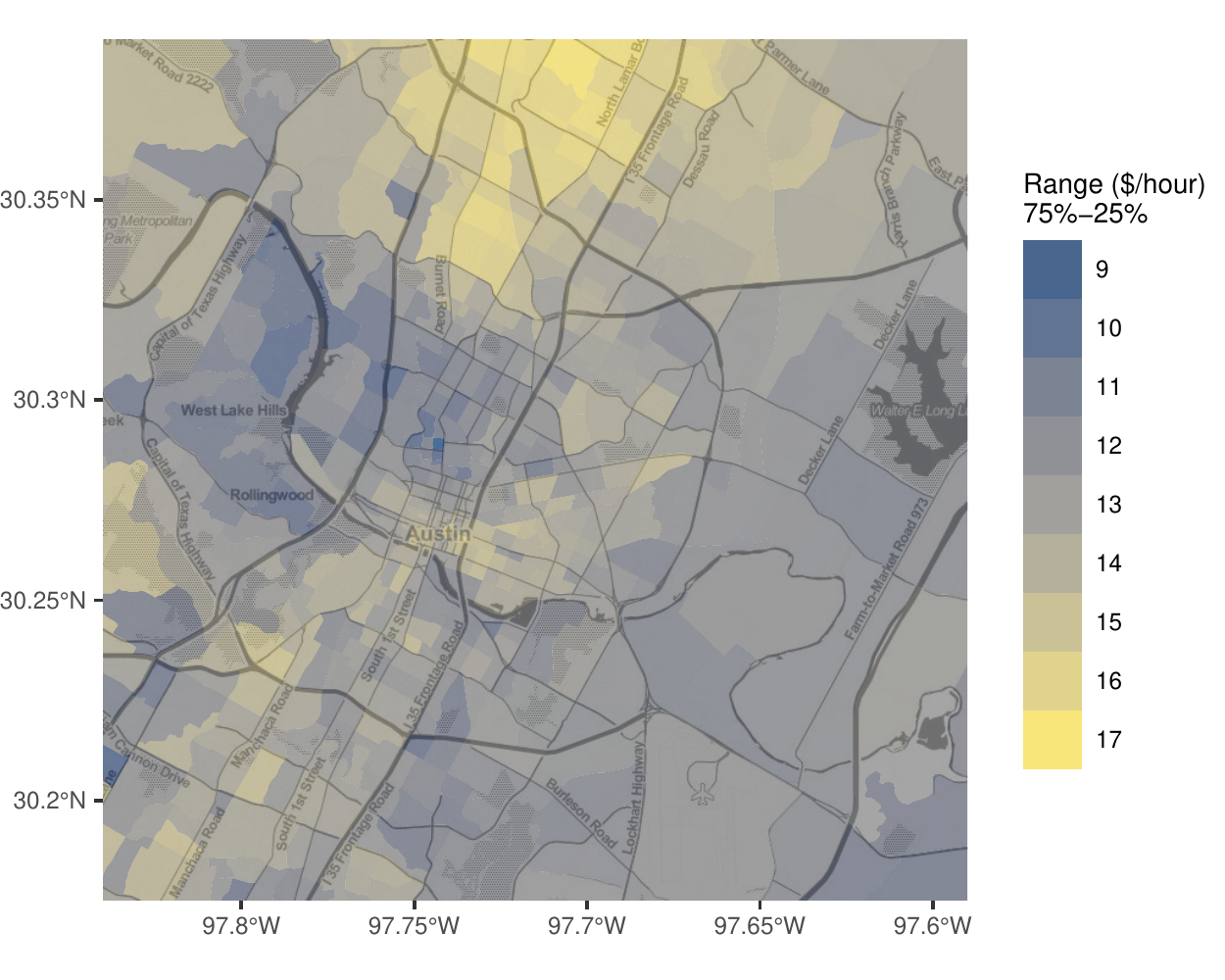}
       \subcaption{Sunday 12 AM}
    \end{minipage}
    \begin{minipage}[tb]{0.48\linewidth}
        \includegraphics[width=0.99\linewidth]{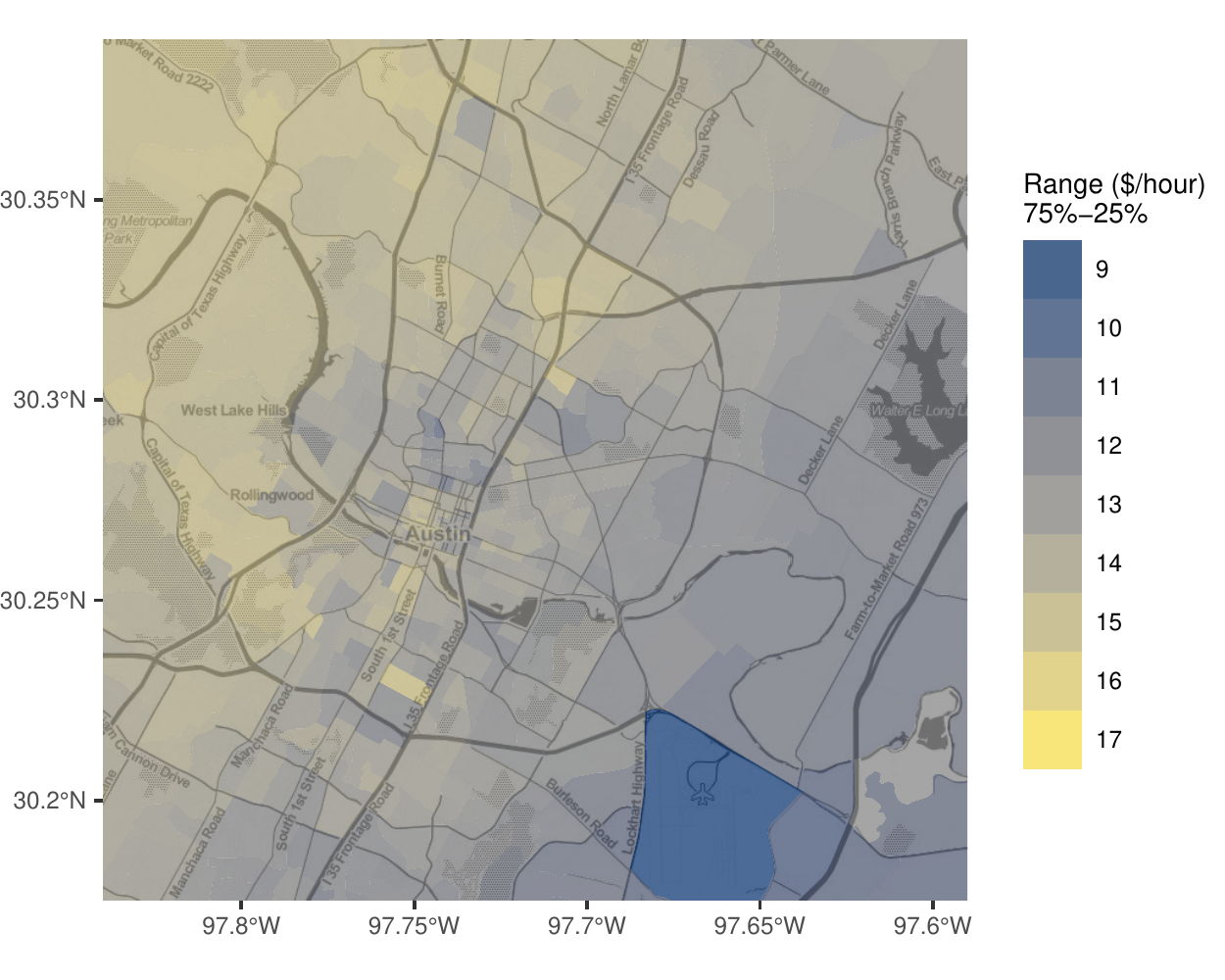}
        \subcaption{Monday 12 PM}
        \label{fig:iqr:b}
    \end{minipage}\hfill
    \caption{Variability of the distribution of productivity for different times and locations (measured by the inter-quartile range of the MAP density estimates).}
    \label{fig:iqr}
\end{figure}

A frequently used measure of spread is the inter-quartile range $\text{IQR} = q_{0.75} - q_{0.25}$. This quantity is typically preferred over standard deviation for skewed distributions, such as our measure of productivity. Figure \ref{fig:iqr} shows the IQR for Saturday night and Monday at noon. This figure complements our previous inquiry using tail probabilities and quantiles. For example, during Saturday night, it shows that variability differs significantly in central areas even if they have similar productivity values. Interestingly it also shows high variability in a suburban neighborhood in the north.

\section{Conclusions}
\label{sec:conc}

In this study, we presented a methodology for estimating the probability distribution of the productivity of a ride-sourcing driver as a function of space and time. We used information from more than 1.2 million trips in Austin, Texas, to provide a case study for the application of the proposed methodology. To the best of our knowledge, this article is the first academic study presenting a large-scale empirical analysis of spatiotemporal effects on the productivity of drivers of TNCs. Previous studies were either of theoretical nature, focused on mean spatial effects only and not spatiotemporal densities, or were applied to taxi data and therefore not specifically tailored to TNC data.

Our proposed method of analysis, based on the Graph-fused Elastic Net (GFEN), offers several advantages for spatiotemporal modeling. For example, its ability to interpolate enabled us to provide hourly estimates, even in locations and times with no observations. A high temporal resolution model could help detect periodic or near-periodic events, such as arrivals' peaks at the airport. Also, it can be used to detect the locations and periods with the lowest (or highest) productivity values. In addition, having full density estimates enabled valuable insights that could not be possible using mean effects only. For example, we have estimated the probability of not exceeding a specific salary (e.g., a living wage) and provided a value-at-risk-style calculation that answers the question of what salary threshold defines the worst-performing drivers specified by a given quantile. This methodology can help transportation engineers, policy-makers, and other ride-sourcing stakeholders to address multiple modeling challenges using trip-level data. The method can also be extended to other ride-sourcing metrics such as idle or ``deadheading'' time, driver reach time, or any other trip-level metric of interest. Furthermore, the methodology can be extended to analyze metrics from different transportation modes, such as public transit, rental scooters, or taxi services.

A direction for future research is to include covariates in the analysis to obtain richer conditional density estimates. For example, it would be helpful to include as covariates the presence of a major event (such as university football games, Austin's Formula 1 race, and South by Southwest) or weather data. This extension can be easily implemented by replacing the binomial model in \eqref{eq:binomial-model} with a logistic regression model or any binary prediction model. The drawback of this approach is that, with the obvious strategy of smoothing each parameter separately, the complexity of the smoothing problem would increase proportionally to the complexity of the conditional estimation model. Therefore, an interesting area of research is to find an efficient smoothing framework for this task.

\bibliographystyle{imsart-nameyear} %
\bibliography{ref}       %

\begin{thebibliography}{xx}

\harvarditem{Ando}{2011}{ando2011predictive}
Ando, T.  \harvardyearleft 2011\harvardyearright , `Predictive bayesian model
  selection', {\em American Journal of Mathematical and Management Sciences}
  {\bf 31}(1-2),~13--38.

\harvarditem[Arlot et~al.]{Arlot, Celisse et~al.}{2010}{arlot2010survey}
Arlot, S., Celisse, A. et~al.  \harvardyearleft 2010\harvardyearright , `A
  survey of cross-validation procedures for model selection', {\em Statistics
  surveys} {\bf 4},~40--79.

\harvarditem{Barbero \harvardand\ Sra}{2018}{barbero-sra-2018}
Barbero, A. \harvardand\ Sra, S.  \harvardyearleft 2018\harvardyearright ,
  `Modular proximal optimization for multidimensional total-variation
  regularization', {\em The Journal of Machine Learning Research} {\bf
  19}(1),~2232--2313.

\harvarditem[Boyd et~al.]{Boyd, Parikh, Chu, Peleato, Eckstein
  et~al.}{2011}{boyd2011distributed}
Boyd, S., Parikh, N., Chu, E., Peleato, B., Eckstein, J. et~al.
  \harvardyearleft 2011\harvardyearright , `Distributed optimization and
  statistical learning via the alternating direction method of multipliers',
  {\em Foundations and Trends in Machine learning} {\bf 3}(1),~1--122.

\harvarditem{Gilks \harvardand\ Wild}{1992}{gilks1992adaptive}
Gilks, W.~R. \harvardand\ Wild, P.  \harvardyearleft 1992\harvardyearright ,
  `Adaptive rejection sampling for gibbs sampling', {\em Journal of the Royal
  Statistical Society: Series C (Applied Statistics)} {\bf 41}(2),~337--348.

\harvarditem{Johnson}{2013}{johnson-2013}
Johnson, N.~A.  \harvardyearleft 2013\harvardyearright , `A dynamic programming
  algorithm for the fused lasso and l 0-segmentation', {\em Journal of
  Computational and Graphical Statistics} {\bf 22}(2),~246--260.

\harvarditem[{Shahriari} et~al.]{{Shahriari}, {Swersky}, {Wang}, {Adams}
  \harvardand\ {de Freitas}}{2016}{shahriari-etal-2016}
{Shahriari}, B., {Swersky}, K., {Wang}, Z., {Adams}, R.~P. \harvardand\ {de
  Freitas}, N.  \harvardyearleft 2016\harvardyearright , `Taking the human out
  of the loop: A review of bayesian optimization', {\em IEEE} {\bf
  104}(1),~148--175.

\harvarditem[Snoek et~al.]{Snoek, Larochelle \harvardand\
  Adams}{2012}{snoek-etal-2012}
Snoek, J., Larochelle, H. \harvardand\ Adams, R.~P.  \harvardyearleft
  2012\harvardyearright , Practical bayesian optimization of machine learning
  algorithms, {\em in} `Advances in Neural Information Processing Systems',
  pp.~2951--2959.

\harvarditem[Tansey et~al.]{Tansey, Athey, Reinhart \harvardand\
  Scott}{2017}{tansey-etal-2017}
Tansey, W., Athey, A., Reinhart, A. \harvardand\ Scott, J.~G.  \harvardyearleft
  2017\harvardyearright , `Multiscale spatial density smoothing: An application
  to large-scale radiological survey and anomaly detection', {\em Journal of
  the American Statistical Association} {\bf 112}(519),~1047--1063.

\harvarditem{Tansey \harvardand\ Scott}{2015}{tansey-scott-2015}
Tansey, W. \harvardand\ Scott, J.~G.  \harvardyearleft 2015\harvardyearright ,
  `A fast and flexible algorithm for the graph-fused lasso', {\em
  arXiv:1505.06475} .

\harvarditem[Tibshirani et~al.]{Tibshirani, Saunders, Rosset, Zhu \harvardand\
  Knight}{2005}{tibshirani-2015}
Tibshirani, R., Saunders, M., Rosset, S., Zhu, J. \harvardand\ Knight, K.
  \harvardyearleft 2005\harvardyearright , `Sparsity and smoothness via the
  fused lasso', {\em Journal of the Royal Statistical Society: Series B
  (Statistical Methodology)} {\bf 67}(1),~91--108.

\harvarditem{Welch \harvardand\ Bishop}{1995}{welch-1995}
Welch, G. \harvardand\ Bishop, G.  \harvardyearleft 1995\harvardyearright , An
  introduction to the kalman filter, Technical report, University of North
  Carolina at Chapel Hill.

\harvarditem{Wohlberg}{2017}{wohlberg2017admm}
Wohlberg, B.  \harvardyearleft 2017\harvardyearright , `Admm penalty parameter
  selection by residual balancing', {\em arXiv:1704.06209} .

\end{thebibliography}


\begin{thebibliography}{48}

\bibitem[\protect\citeauthoryear{Barbero and Sra}{2018}]{barbero-sra-2018}
\begin{barticle}[author]
\bauthor{\bsnm{Barbero},~\bfnm{Alvaro}\binits{A.}} \AND
  \bauthor{\bsnm{Sra},~\bfnm{Suvrit}\binits{S.}}
(\byear{2018}).
\btitle{Modular proximal optimization for multidimensional total-variation
  regularization}.
\bjournal{The Journal of Machine Learning Research}
\bvolume{19}
\bpages{2232--2313}.
\end{barticle}
\endbibitem

\bibitem[\protect\citeauthoryear{Bentley}{1975}]{bentley1975multidimensional}
\begin{barticle}[author]
\bauthor{\bsnm{Bentley},~\bfnm{Jon~Louis}\binits{J.~L.}}
(\byear{1975}).
\btitle{Multidimensional binary search trees used for associative searching}.
\bjournal{Communications of the ACM}
\bvolume{18}
\bpages{509--517}.
\end{barticle}
\endbibitem

\bibitem[\protect\citeauthoryear{Bezanson et~al.}{2017}]{bezanson2017julia}
\begin{barticle}[author]
\bauthor{\bsnm{Bezanson},~\bfnm{Jeff}\binits{J.}},
  \bauthor{\bsnm{Edelman},~\bfnm{Alan}\binits{A.}},
  \bauthor{\bsnm{Karpinski},~\bfnm{Stefan}\binits{S.}} \AND
  \bauthor{\bsnm{Shah},~\bfnm{Viral~B}\binits{V.~B.}}
(\byear{2017}).
\btitle{Julia: A fresh approach to numerical computing}.
\bjournal{SIAM review}
\bvolume{59}
\bpages{65--98}.
\end{barticle}
\endbibitem

\bibitem[\protect\citeauthoryear{Blangiardo
  et~al.}{2013}]{blangiardo2013spatial}
\begin{barticle}[author]
\bauthor{\bsnm{Blangiardo},~\bfnm{Marta}\binits{M.}},
  \bauthor{\bsnm{Cameletti},~\bfnm{Michela}\binits{M.}},
  \bauthor{\bsnm{Baio},~\bfnm{Gianluca}\binits{G.}} \AND
  \bauthor{\bsnm{Rue},~\bfnm{H{\aa}vard}\binits{H.}}
(\byear{2013}).
\btitle{Spatial and spatio-temporal models with R-INLA}.
\bjournal{Spatial and spatio-temporal epidemiology}
\bvolume{4}
\bpages{33--49}.
\end{barticle}
\endbibitem

\bibitem[\protect\citeauthoryear{Boyd et~al.}{2011}]{boyd2011distributed}
\begin{barticle}[author]
\bauthor{\bsnm{Boyd},~\bfnm{Stephen}\binits{S.}},
  \bauthor{\bsnm{Parikh},~\bfnm{Neal}\binits{N.}},
  \bauthor{\bsnm{Chu},~\bfnm{Eric}\binits{E.}},
  \bauthor{\bsnm{Peleato},~\bfnm{Borja}\binits{B.}},
  \bauthor{\bsnm{Eckstein},~\bfnm{Jonathan}\binits{J.}} \betal{et~al.}
(\byear{2011}).
\btitle{Distributed Optimization and Statistical Learning via the Alternating
  Direction Method of Multipliers}.
\bjournal{Foundations and Trends in Machine learning}
\bvolume{3}
\bpages{1--122}.
\end{barticle}
\endbibitem

\bibitem[\protect\citeauthoryear{Chua, Kolaczyk and
  Crovella}{2006}]{chua2006network}
\begin{barticle}[author]
\bauthor{\bsnm{Chua},~\bfnm{David~B}\binits{D.~B.}},
  \bauthor{\bsnm{Kolaczyk},~\bfnm{Eric~D}\binits{E.~D.}} \AND
  \bauthor{\bsnm{Crovella},~\bfnm{Mark}\binits{M.}}
(\byear{2006}).
\btitle{Network kriging}.
\bjournal{IEEE Journal on Selected Areas in Communications}
\bvolume{24}
\bpages{2263--2272}.
\end{barticle}
\endbibitem

\bibitem[\protect\citeauthoryear{Cradeur}{2019}]{filter1}
\begin{bmisc}[author]
\bauthor{\bsnm{Cradeur},~\bfnm{Jay}\binits{J.}}
(\byear{2019}).
\btitle{{Big changes could be coming to Uber’s destination filter}}.
\bnote{Available online. Accessed Dec 1 2019}.
\end{bmisc}
\endbibitem

\bibitem[\protect\citeauthoryear{{Cressie}}{1993}]{cressie-1993}
\begin{bbook}[author]
\bauthor{\bsnm{{Cressie}},~\bfnm{Noel}\binits{N.}}
(\byear{1993}).
\btitle{Statistics for Spatial Data}.
\bpublisher{J. Wiley}.
\end{bbook}
\endbibitem

\bibitem[\protect\citeauthoryear{Cressie and Huang}{1999}]{cressie-huang-1999}
\begin{barticle}[author]
\bauthor{\bsnm{Cressie},~\bfnm{Noel}\binits{N.}} \AND
  \bauthor{\bsnm{Huang},~\bfnm{Hsin-Cheng}\binits{H.-C.}}
(\byear{1999}).
\btitle{Classes of nonseparable, spatio-temporal stationary covariance
  functions}.
\bjournal{Journal of the American Statistical Association}
\bvolume{94}
\bpages{1330--1339}.
\end{barticle}
\endbibitem

\bibitem[\protect\citeauthoryear{Cressie, Shi and Kang}{2010}]{cressie-2010}
\begin{barticle}[author]
\bauthor{\bsnm{Cressie},~\bfnm{Noel}\binits{N.}},
  \bauthor{\bsnm{Shi},~\bfnm{Tao}\binits{T.}} \AND
  \bauthor{\bsnm{Kang},~\bfnm{Emily~L}\binits{E.~L.}}
(\byear{2010}).
\btitle{Fixed rank filtering for spatio-temporal data}.
\bjournal{Journal of Computational and Graphical Statistics}
\bvolume{19}
\bpages{724--745}.
\end{barticle}
\endbibitem

\bibitem[\protect\citeauthoryear{Cressie and Wikle}{2011}]{cressie-wikle-2011}
\begin{bbook}[author]
\bauthor{\bsnm{Cressie},~\bfnm{N}\binits{N.}} \AND
  \bauthor{\bsnm{Wikle},~\bfnm{CK}\binits{C.}}
(\byear{2011}).
\btitle{Statistics for Spatio-Temporal Data}.
\bpublisher{John Wiley \& Sons}.
\end{bbook}
\endbibitem

\bibitem[\protect\citeauthoryear{Ecker and Gelfand}{1999}]{ecker1999bayesian}
\begin{barticle}[author]
\bauthor{\bsnm{Ecker},~\bfnm{Mark~D}\binits{M.~D.}} \AND
  \bauthor{\bsnm{Gelfand},~\bfnm{Alan~E}\binits{A.~E.}}
(\byear{1999}).
\btitle{Bayesian modeling and inference for geometrically anisotropic spatial
  data}.
\bjournal{Mathematical Geology}
\bvolume{31}
\bpages{67--83}.
\end{barticle}
\endbibitem

\bibitem[\protect\citeauthoryear{Friedland
  et~al.}{2017}]{friedland2017isotropic}
\begin{barticle}[author]
\bauthor{\bsnm{Friedland},~\bfnm{Carol~J}\binits{C.~J.}},
  \bauthor{\bsnm{Joyner},~\bfnm{T~Andrew}\binits{T.~A.}},
  \bauthor{\bsnm{Massarra},~\bfnm{Carol}\binits{C.}},
  \bauthor{\bsnm{Rohli},~\bfnm{Robert~V}\binits{R.~V.}},
  \bauthor{\bsnm{Trevi{\~n}o},~\bfnm{Anna~M}\binits{A.~M.}},
  \bauthor{\bsnm{Ghosh},~\bfnm{Shubharoop}\binits{S.}},
  \bauthor{\bsnm{Huyck},~\bfnm{Charles}\binits{C.}} \AND
  \bauthor{\bsnm{Weatherhead},~\bfnm{Mark}\binits{M.}}
(\byear{2017}).
\btitle{Isotropic and anisotropic kriging approaches for interpolating
  surface-level wind speeds across large, geographically diverse regions}.
\bjournal{Geomatics, Natural Hazards and Risk}
\bvolume{8}
\bpages{207--224}.
\end{barticle}
\endbibitem

\bibitem[\protect\citeauthoryear{Getreuer}{2012}]{getreuer-2012}
\begin{barticle}[author]
\bauthor{\bsnm{Getreuer},~\bfnm{Pascal}\binits{P.}}
(\byear{2012}).
\btitle{Rudin-Osher-Fatemi Total Variation Denoising using Split Bregman}.
\bjournal{Image Processing On Line}
\bvolume{2}
\bpages{74--95}.
\end{barticle}
\endbibitem

\bibitem[\protect\citeauthoryear{Gilks and Wild}{1992}]{gilks1992adaptive}
\begin{barticle}[author]
\bauthor{\bsnm{Gilks},~\bfnm{Walter~R}\binits{W.~R.}} \AND
  \bauthor{\bsnm{Wild},~\bfnm{Pascal}\binits{P.}}
(\byear{1992}).
\btitle{Adaptive rejection sampling for Gibbs sampling}.
\bjournal{Journal of the Royal Statistical Society: Series C (Applied
  Statistics)}
\bvolume{41}
\bpages{337--348}.
\end{barticle}
\endbibitem

\bibitem[\protect\citeauthoryear{{Gneiting}, {Genton} and
  {Guttorp}}{2006}]{gneiting2006geostatistical}
\begin{barticle}[author]
\bauthor{\bsnm{{Gneiting}},~\bfnm{Tilmann}\binits{T.}},
  \bauthor{\bsnm{{Genton}},~\bfnm{Marc~G}\binits{M.~G.}} \AND
  \bauthor{\bsnm{{Guttorp}},~\bfnm{Peter}\binits{P.}}
(\byear{2006}).
\btitle{Geostatistical space-time models, stationarity, separability, and full
  symmetry}.
\bjournal{Monographs On Statistics and Applied Probability}
\bvolume{107}
\bpages{151}.
\end{barticle}
\endbibitem

\bibitem[\protect\citeauthoryear{Hall and Krueger}{2016}]{hall-etal-2016}
\begin{btechreport}[author]
\bauthor{\bsnm{Hall},~\bfnm{Jonathan~V}\binits{J.~V.}} \AND
  \bauthor{\bsnm{Krueger},~\bfnm{Alan~B}\binits{A.~B.}}
(\byear{2016}).
\btitle{An Analysis of the Labor Market for Uber’s Driver-Partners in the
  United States}
\btype{Technical Report} No. \bnumber{22843},
\bpublisher{National Bureau of Economic Research}.
\end{btechreport}
\endbibitem

\bibitem[\protect\citeauthoryear{Hartman and
  H{\"o}ssjer}{2008}]{hartman2008fast}
\begin{barticle}[author]
\bauthor{\bsnm{Hartman},~\bfnm{Linda}\binits{L.}} \AND
  \bauthor{\bsnm{H{\"o}ssjer},~\bfnm{Ola}\binits{O.}}
(\byear{2008}).
\btitle{Fast kriging of large data sets with Gaussian Markov random fields}.
\bjournal{Computational Statistics \& Data Analysis}
\bvolume{52}
\bpages{2331--2349}.
\end{barticle}
\endbibitem

\bibitem[\protect\citeauthoryear{Hensman, Fusi and
  Lawrence}{2013}]{hensman2013gaussian}
\begin{binproceedings}[author]
\bauthor{\bsnm{Hensman},~\bfnm{James}\binits{J.}},
  \bauthor{\bsnm{Fusi},~\bfnm{Nicol{\`o}}\binits{N.}} \AND
  \bauthor{\bsnm{Lawrence},~\bfnm{Neil~D}\binits{N.~D.}}
(\byear{2013}).
\btitle{Gaussian processes for Big data}.
In \bbooktitle{Proceedings of the Twenty-Ninth Conference on Uncertainty in
  Artificial Intelligence}
\bpages{282--290}.
\end{binproceedings}
\endbibitem

\bibitem[\protect\citeauthoryear{Katzfuss and
  Cressie}{2011}]{katzfuss-cressie-2011}
\begin{barticle}[author]
\bauthor{\bsnm{Katzfuss},~\bfnm{Matthias}\binits{M.}} \AND
  \bauthor{\bsnm{Cressie},~\bfnm{Noel}\binits{N.}}
(\byear{2011}).
\btitle{Spatio-temporal Smoothing and EM Estimation for Massive Remote-sensing
  Data Sets}.
\bjournal{Journal of Time Series Analysis}
\bvolume{32}
\bpages{430--446}.
\end{barticle}
\endbibitem

\bibitem[\protect\citeauthoryear{Knorr-Held}{2000}]{knorr2000bayesian}
\begin{barticle}[author]
\bauthor{\bsnm{Knorr-Held},~\bfnm{Leonhard}\binits{L.}}
(\byear{2000}).
\btitle{Bayesian modelling of inseparable space-time variation in disease
  risk}.
\bjournal{Statistics in medicine}
\bvolume{19}
\bpages{2555--2567}.
\end{barticle}
\endbibitem

\bibitem[\protect\citeauthoryear{{Li} et~al.}{2019}]{li-2019}
\begin{barticle}[author]
\bauthor{\bsnm{{Li}},~\bfnm{Sen}\binits{S.}},
  \bauthor{\bsnm{{Tavafoghi}},~\bfnm{Hamidreza}\binits{H.}},
  \bauthor{\bsnm{{Poolla}},~\bfnm{Kameshwar}\binits{K.}} \AND
  \bauthor{\bsnm{{Varaiya}},~\bfnm{Pravin}\binits{P.}}
(\byear{2019}).
\btitle{{Regulating TNCs: Should Uber and Lyft Set Their Own Rules?}}
\bjournal{arXiv:1312.7750}.
\end{barticle}
\endbibitem

\bibitem[\protect\citeauthoryear{{Mishel}}{2018}]{mishel-2018}
\begin{btechreport}[author]
\bauthor{\bsnm{{Mishel}},~\bfnm{Lawrence}\binits{L.}}
(\byear{2018}).
\btitle{Uber Drivers' Compensation, Wages, and the Scale of Uber and the Gig
  Economy}
\btype{Technical Report} No. \bnumber{145552},
\bpublisher{Economic Policy Institute Working Paper Series}.
\end{btechreport}
\endbibitem

\bibitem[\protect\citeauthoryear{{Mitov} and
  {Claassen}}{2013}]{mitov2013afused}
\begin{barticle}[author]
\bauthor{\bsnm{{Mitov}},~\bfnm{Venelin}\binits{V.}} \AND
  \bauthor{\bsnm{{Claassen}},~\bfnm{Manfred}\binits{M.}}
(\byear{2013}).
\btitle{{A Fused Elastic Net Logistic Regression Model for Multi-Task Binary
  Classification}}.
\bjournal{arXiv:1312.7750}.
\end{barticle}
\endbibitem

\bibitem[\protect\citeauthoryear{{Nadeau}}{2017}]{nadeau-2017}
\begin{btechreport}[author]
\bauthor{\bsnm{{Nadeau}},~\bfnm{Carey~Anne}\binits{C.~A.}}
(\byear{2017}).
\btitle{Living Wage Calculator: User's Guide, Update 2017}
\btype{Technical Report},
\bpublisher{Open Data Nation}.
\bnote{Online; accessed March 1 2019}.
\end{btechreport}
\endbibitem

\bibitem[\protect\citeauthoryear{Okabe and Sugihara}{2012}]{okabe2012spatial}
\begin{bbook}[author]
\bauthor{\bsnm{Okabe},~\bfnm{Atsuyuki}\binits{A.}} \AND
  \bauthor{\bsnm{Sugihara},~\bfnm{Kokichi}\binits{K.}}
(\byear{2012}).
\btitle{Spatial analysis along networks: statistical and computational
  methods}.
\bpublisher{John Wiley \& Sons}.
\end{bbook}
\endbibitem

\bibitem[\protect\citeauthoryear{Perea}{2017}]{filter2}
\begin{bmisc}[author]
\bauthor{\bsnm{Perea},~\bfnm{Christian}\binits{C.}}
(\byear{2017}).
\btitle{Uber Drops Destination Filters Back to two Trips Per Day}.
\bnote{Available online. Accessed Dec 1 2019}.
\end{bmisc}
\endbibitem

\bibitem[\protect\citeauthoryear{Polson and Scott}{2010}]{polson2010shrink}
\begin{barticle}[author]
\bauthor{\bsnm{Polson},~\bfnm{Nicholas~G}\binits{N.~G.}} \AND
  \bauthor{\bsnm{Scott},~\bfnm{James~G}\binits{J.~G.}}
(\byear{2010}).
\btitle{Shrink globally, act locally: Sparse Bayesian regularization and
  prediction}.
\bjournal{Bayesian statistics}
\bvolume{9}
\bpages{105}.
\end{barticle}
\endbibitem

\bibitem[\protect\citeauthoryear{Ramdas and
  Tibshirani}{2016}]{ramdas-tibshirani-2016}
\begin{barticle}[author]
\bauthor{\bsnm{Ramdas},~\bfnm{Aaditya}\binits{A.}} \AND
  \bauthor{\bsnm{Tibshirani},~\bfnm{Ryan~J.}\binits{R.~J.}}
(\byear{2016}).
\btitle{Fast and Flexible ADMM Algorithms for Trend Filtering}.
\bjournal{Journal of Computational and Graphical Statistics}
\bvolume{25}
\bpages{839--858}.
\end{barticle}
\endbibitem

\bibitem[\protect\citeauthoryear{Rasmussen}{2003}]{rasmussen2003gaussian}
\begin{binproceedings}[author]
\bauthor{\bsnm{Rasmussen},~\bfnm{Carl~Edward}\binits{C.~E.}}
(\byear{2003}).
\btitle{Gaussian processes in machine learning}.
In \bbooktitle{Summer school on machine learning}
\bpages{63--71}.
\bpublisher{Springer}.
\end{binproceedings}
\endbibitem

\bibitem[\protect\citeauthoryear{Rue}{2001}]{rue2001fast}
\begin{barticle}[author]
\bauthor{\bsnm{Rue},~\bfnm{H{\aa}vard}\binits{H.}}
(\byear{2001}).
\btitle{Fast sampling of Gaussian Markov random fields}.
\bjournal{Journal of the Royal Statistical Society: Series B (Statistical
  Methodology)}
\bvolume{63}
\bpages{325--338}.
\end{barticle}
\endbibitem

\bibitem[\protect\citeauthoryear{Rue and Held}{2005}]{rue-held-2005}
\begin{bbook}[author]
\bauthor{\bsnm{Rue},~\bfnm{Havard}\binits{H.}} \AND
  \bauthor{\bsnm{Held},~\bfnm{Leonhard}\binits{L.}}
(\byear{2005}).
\btitle{Gaussian Markov random fields}.
\bpublisher{Chapman and Hall/CRC}.
\end{bbook}
\endbibitem

\bibitem[\protect\citeauthoryear{Rushworth, Lee and
  Sarran}{2017}]{rushworth-2017}
\begin{barticle}[author]
\bauthor{\bsnm{Rushworth},~\bfnm{Alastair}\binits{A.}},
  \bauthor{\bsnm{Lee},~\bfnm{Duncan}\binits{D.}} \AND
  \bauthor{\bsnm{Sarran},~\bfnm{Christophe}\binits{C.}}
(\byear{2017}).
\btitle{An Adaptive Spatiotemporal Smoothing Model for Estimating Trends and
  Step Changes in Disease Risk}.
\bjournal{Journal of the Royal Statistical Society: Series C}
\bvolume{66}
\bpages{141--157}.
\end{barticle}
\endbibitem

\bibitem[\protect\citeauthoryear{{Shahriari}
  et~al.}{2016}]{shahriari-etal-2016}
\begin{barticle}[author]
\bauthor{\bsnm{{Shahriari}},~\bfnm{B.}\binits{B.}},
  \bauthor{\bsnm{{Swersky}},~\bfnm{K.}\binits{K.}},
  \bauthor{\bsnm{{Wang}},~\bfnm{Z.}\binits{Z.}},
  \bauthor{\bsnm{{Adams}},~\bfnm{R.~P.}\binits{R.~P.}} \AND \bauthor{\bsnm{{de
  Freitas}},~\bfnm{N.}\binits{N.}}
(\byear{2016}).
\btitle{Taking the Human Out of the Loop: A Review of Bayesian Optimization}.
\bjournal{IEEE}
\bvolume{104}
\bpages{148--175}.
\end{barticle}
\endbibitem

\bibitem[\protect\citeauthoryear{Smith}{2019}]{smithuber}
\begin{bmisc}[author]
\bauthor{\bsnm{Smith},~\bfnm{C}\binits{C.}}
(\byear{2019}).
\btitle{Amazing Uber Stats and Facts (2019)}.
\end{bmisc}
\endbibitem

\bibitem[\protect\citeauthoryear{Tansey and Scott}{2015}]{tansey-scott-2015}
\begin{barticle}[author]
\bauthor{\bsnm{Tansey},~\bfnm{Wesley}\binits{W.}} \AND
  \bauthor{\bsnm{Scott},~\bfnm{James~G}\binits{J.~G.}}
(\byear{2015}).
\btitle{A Fast and Flexible Algorithm for the Graph-fused Lasso}.
\bjournal{arXiv:1505.06475}.
\end{barticle}
\endbibitem

\bibitem[\protect\citeauthoryear{Tansey et~al.}{2017}]{tansey-etal-2017}
\begin{barticle}[author]
\bauthor{\bsnm{Tansey},~\bfnm{Wesley}\binits{W.}},
  \bauthor{\bsnm{Athey},~\bfnm{Alex}\binits{A.}},
  \bauthor{\bsnm{Reinhart},~\bfnm{Alex}\binits{A.}} \AND
  \bauthor{\bsnm{Scott},~\bfnm{James~G.}\binits{J.~G.}}
(\byear{2017}).
\btitle{Multiscale Spatial Density Smoothing: An Application to Large-Scale
  Radiological Survey and Anomaly Detection}.
\bjournal{Journal of the American Statistical Association}
\bvolume{112}
\bpages{1047--1063}.
\end{barticle}
\endbibitem

\bibitem[\protect\citeauthoryear{Tec}{2018}]{tec2018ars}
\begin{bmisc}[author]
\bauthor{\bsnm{Tec},~\bfnm{Mauricio}\binits{M.}}
(\byear{2018}).
\btitle{AdaptiveRejectionSampling.jl}.
\bnote{Julia Package v0.11}.
\end{bmisc}
\endbibitem

\bibitem[\protect\citeauthoryear{Terenin, Simpson and
  Draper}{2020}]{terenin2020asynchronous}
\begin{binproceedings}[author]
\bauthor{\bsnm{Terenin},~\bfnm{Alexander}\binits{A.}},
  \bauthor{\bsnm{Simpson},~\bfnm{Daniel}\binits{D.}} \AND
  \bauthor{\bsnm{Draper},~\bfnm{David}\binits{D.}}
(\byear{2020}).
\btitle{Asynchronous gibbs sampling}.
In \bbooktitle{International Conference on Artificial Intelligence and
  Statistics}
\bpages{144--154}.
\bpublisher{PMLR}.
\end{binproceedings}
\endbibitem

\bibitem[\protect\citeauthoryear{Tibshirani and
  Taylor}{2011}]{tibshirani-taylor-2011}
\begin{barticle}[author]
\bauthor{\bsnm{Tibshirani},~\bfnm{Ryan~J.}\binits{R.~J.}} \AND
  \bauthor{\bsnm{Taylor},~\bfnm{Jonathan}\binits{J.}}
(\byear{2011}).
\btitle{The Solution Path of the Generalized Lasso}.
\bjournal{The Annals of Statistics}
\bvolume{39}
\bpages{1335--1371}.
\end{barticle}
\endbibitem

\bibitem[\protect\citeauthoryear{Tibshirani et~al.}{2005}]{tibshirani-2015}
\begin{barticle}[author]
\bauthor{\bsnm{Tibshirani},~\bfnm{Robert}\binits{R.}},
  \bauthor{\bsnm{Saunders},~\bfnm{Michael}\binits{M.}},
  \bauthor{\bsnm{Rosset},~\bfnm{Saharon}\binits{S.}},
  \bauthor{\bsnm{Zhu},~\bfnm{Ji}\binits{J.}} \AND
  \bauthor{\bsnm{Knight},~\bfnm{Keith}\binits{K.}}
(\byear{2005}).
\btitle{Sparsity and Smoothness via the Fused Lasso}.
\bjournal{Journal of the Royal Statistical Society: Series B (Statistical
  Methodology)}
\bvolume{67}
\bpages{91--108}.
\end{barticle}
\endbibitem

\bibitem[\protect\citeauthoryear{Wang et~al.}{2016}]{wang-2016}
\begin{barticle}[author]
\bauthor{\bsnm{Wang},~\bfnm{Yu-Xiang}\binits{Y.-X.}},
  \bauthor{\bsnm{Sharpnack},~\bfnm{James}\binits{J.}},
  \bauthor{\bsnm{Smola},~\bfnm{Alexander~J.}\binits{A.~J.}} \AND
  \bauthor{\bsnm{Tibshirani},~\bfnm{Ryan~J.}\binits{R.~J.}}
(\byear{2016}).
\btitle{Trend Filtering on Graphs}.
\bjournal{Journal of Machine Learning Research}
\bvolume{17}
\bpages{1--41}.
\end{barticle}
\endbibitem

\bibitem[\protect\citeauthoryear{Wong et~al.}{2010}]{wong2010optional}
\begin{barticle}[author]
\bauthor{\bsnm{Wong},~\bfnm{Wing~H}\binits{W.~H.}},
  \bauthor{\bsnm{Ma},~\bfnm{Li}\binits{L.}} \betal{et~al.}
(\byear{2010}).
\btitle{Optional P{\'o}lya tree and Bayesian inference}.
\bjournal{The Annals of Statistics}
\bvolume{38}
\bpages{1433--1459}.
\end{barticle}
\endbibitem

\bibitem[\protect\citeauthoryear{{Data World}}{2017}]{dataworld-2017}
\begin{bmisc}[author]
\bauthor{\bsnm{{Data World}}}
(\byear{2017}).
\btitle{{RideAustin Dataset}}.
\bnote{[dataset]}.
\end{bmisc}
\endbibitem

\bibitem[\protect\citeauthoryear{Xu, Liang and Genton}{2015}]{xu-2015}
\begin{barticle}[author]
\bauthor{\bsnm{Xu},~\bfnm{Ganggang}\binits{G.}},
  \bauthor{\bsnm{Liang},~\bfnm{Faming}\binits{F.}} \AND
  \bauthor{\bsnm{Genton},~\bfnm{Marc~G}\binits{M.~G.}}
(\byear{2015}).
\btitle{A Bayesian Spatio-temporal Geostatistical Model with an Auxiliary
  Lattice for Large Datasets}.
\bjournal{Statistica Sinica}
\bpages{61--79}.
\end{barticle}
\endbibitem

\bibitem[\protect\citeauthoryear{Zhang and
  Williamson}{2019}]{zhang2019embarrassingly}
\begin{barticle}[author]
\bauthor{\bsnm{Zhang},~\bfnm{Michael~Minyi}\binits{M.~M.}} \AND
  \bauthor{\bsnm{Williamson},~\bfnm{Sinead~A}\binits{S.~A.}}
(\byear{2019}).
\btitle{Embarrassingly Parallel Inference for Gaussian Processes}.
\bjournal{Journal of Machine Learning Research}
\bvolume{20}
\bpages{1--26}.
\end{barticle}
\endbibitem

\bibitem[\protect\citeauthoryear{Zou and Hastie}{2005}]{zou-2005}
\begin{barticle}[author]
\bauthor{\bsnm{Zou},~\bfnm{Hui}\binits{H.}} \AND
  \bauthor{\bsnm{Hastie},~\bfnm{Trevor}\binits{T.}}
(\byear{2005}).
\btitle{Regularization and Variable Selection via the Elastic Net}.
\bjournal{Journal of the Royal Statistical Society: Series B}
\bvolume{67}
\bpages{301--320}.
\end{barticle}
\endbibitem

\bibitem[\protect\citeauthoryear{{Zuniga-Garcia}
  et~al.}{2020}]{zuniga-etal-2019}
\begin{barticle}[author]
\bauthor{\bsnm{{Zuniga-Garcia}},~\bfnm{Natalia}\binits{N.}},
  \bauthor{\bsnm{{Tec}},~\bfnm{Mauricio}\binits{M.}},
  \bauthor{\bsnm{{Scott}},~\bfnm{James~G.}\binits{J.~G.}},
  \bauthor{\bsnm{{Ruiz-Juri}},~\bfnm{Natalia}\binits{N.}} \AND
  \bauthor{\bsnm{{Machemehl}},~\bfnm{Randy~B.}\binits{R.~B.}}
(\byear{2020}).
\btitle{Evaluation of Ride-Sourcing Search Frictions and Driver Productivity: A
  Spatial Denoising Approach}.
\bvolume{110}
\bpages{346--367}.
\end{barticle}
\endbibitem

\end{thebibliography}

\renewcommand\thefigure{\thesection.\arabic{figure}} 

\newpage

\begin{appendix}
    \section{Closed-from solution of the GFEN to Problems  \ref{prob:missing-data} and \ref{prob:missing-data-binom}}
    \label{appendix:gfen-solution}
    \setcounter{figure}{0}
    \setcounter{table}{0}

The following proposition states the solution of the GFEN to the toy denoising scenario described in Problems  \ref{prob:missing-data} and \ref{prob:missing-data-binom} for a graph with three nodes for which the middle node has missing data. Despite the more complicated solution, the reader can appreciate by comparing with the solution of the GFL and GMRF how the solution combines both behaviors.

\begin{prop}[GFEN solution - Gaussian and binomial likelihoods]\label{prop:gfen} The solution of the GFEN objective \eqref{eq:gfen} to the Gaussian case in toy problem \ref{prob:missing-data} is given by $\hat{\beta}_1 =  y_1 + \lambda_1 + \frac{\lambda_2}{1 + \lambda_2}\cdot\frac{y_3 - y_1}{2}$, $\hat{\beta}_3 = y_3 - \lambda_1 - \frac{\lambda_2}{1 + \lambda_2}\cdot\frac{y_3 - y_1}{2}$ and $\hat{\beta}_2 = \frac{\hat{\beta}_1 + \hat{\beta}_3}{2}$ for all $\lambda_1, \lambda_2 \geq 0$ such that $\lambda_1 + \frac{\lambda_2}{1 + \lambda_2}\cdot\frac{y_3 - y_1}{2} < \frac{y_1 + y_3}{2}$. If $(\lambda_1, \lambda_2)$ do not satisfy this condition, then there is a unique solution $\hat{\beta}_1=\hat{\beta}_2=\hat{\beta}_3=\frac{1}{2}(y_1 + y_3)$ at the midpoint. For the binomial case (problem \ref{prob:missing-data-binom},) let $b_i = \mathrm{logit}(y_i / N_i)$ for $i\in\{1,3\}$. Then the solution is the same in the Gaussian case with $b_i$ replacing the role of $y_i$.
\end{prop}
\end{appendix}

\begin{appendix}
    \section{Algorithmic details of the GFEN}
    \label{appendix:algorithm}
    \setcounter{figure}{0}
    \setcounter{table}{0}
    
    \subsection*{Fast MAP estimates with the ADMM} For ease of presentation, in this section we will focus on the case of a general graph, which corresponds to the GFEN objective \eqref{eq:gfen}. The extension to the spatio-temporal case will be straightforward. In this case, it will be convenient to assume that $\mathcal{T}$ itself can be written as a disjoint union $\mathcal{T} = \mathcal{T}_S \cup \mathcal{T}_T$ where the trails of $\mathcal{T}_S$ and $\mathcal{T}_T$ consists of spatial and temporal edges respectively. This assumption is not strictly necessary. However, it simplifies computation since the total variation penalization hyperparameters will always be constant for a given trail.

    We begin with an overview of the main ideas of the three steps of the algorithm. First, we start from a decomposition of the edges $E$ into a set of non-overlapping trails $E = \bigcup \{\tau \mid \tau\in\mathcal{T}\}$. The core of this strategy is to reduce the optimization objective to solving individual, nonoverlapping smoothing problems along each trail for each $\ell_p$ penalty separately. This approach is sometimes referred to as ``proximal stacking'' (see \citetA{barbero-sra-2018}). The advantage of doing this is that the optimization problem in each trail is 1-dimensional and can be solved in linear time.
    
    Second, a principled mathematical way is necessary to combine the solutions in each trail and guarantee that the original objective is minimized. First, we introduce a slack variable $\mathbf{z}_{\tau, p}$ for each trail $\tau$ and for each $p\in\{1,2\}$. For each of these, we require linear constraints $\mathbf{z}_{\tau,p} = \boldsymbol{\beta}[\tau] :=(\beta^{(v_1)},\hdots,\beta^{(v_k)})$ where the trail $\tau=(v_1, \hdots, v_k)$ is represented by its sequence of visited vertices. Together, these restrictions define a consensus problem for which the ADMM is an attractive solution method because of its flexibility and convergence properties \citepA{boyd2011distributed}. The ADMM is an iterative technique for solving convex optimization problems of the form $\min_{\boldsymbol{\beta}, \mathbf{z}} f(\boldsymbol{\beta}) + g(\mathbf{z})$ subject to a linear constraint of the form $\mathbf{A}\boldsymbol{\beta} = \mathbf{B} \mathbf{z}$. Here, $f$ will take the role of the likelihood and $g$ of the total variation penalty. The variable $\mathbf{z}$ will be a concatenation of each variable $\mathbf{z}_{\tau, p}$ and the linear constraints will come from the restrictions $\mathbf{z}_{\beta,\tau}=\boldsymbol{\beta}[\tau]$. Each iteration of the ADMM will consist of three simple unconstrained optimization subproblems that only involve one variable at a time.
    
    Finally, the optimization subproblem corresponding to the update for $\boldsymbol{\beta}$ can be replaced by an iteration of Newton's method. For the updates corresponding to $\mathbf{z}$, which itself consists of parallel updates for each $\mathbf{z}_{\tau, p}$, we leverage exact linear-time solvers for 1-dimensional smoothing problems. For $p=1$ we use the method of \citetA{barbero-sra-2018} for $\ell_1$-TV denoising and for $p=2$ we use the Kalman smoother (see \citetA{welch-1995}).
    
    We now present the mathematical details of the algorithm. The first step is to rewrite the GFEN objective as a constrained optimization problem using a set of slack variables $\mathbf{z}_{\tau, p} = (z^{(v)}_{\tau, p})_{v\in \tau}$ exactly one for each trail\footnote{With a slight abuse of notation, we say that $v\in \tau$ if $v$ appears in some edge of $\tau$.} $\tau \in \mathcal{T}$ and for each norm $p\in\{1,2\}$, obtaining
    \begin{equation}\label{eq:gfen-constrained}
        \begin{aligned}
            \operatornamewithlimits{minimize}_{\beta} \quad\quad & \sum_{v \in V} l(\mathbf{y}^{(v)}, \beta^{(v)}) + \sum_{\tau \in \mathcal{T}}  \sum_{p \in \{1,2\}} \lambda_p \mathrm{TV}_p(\mathbf{z}_{\tau, p}, \tau) \\ 
            \text{subject to} \quad\;\; & \mathbf{z}_{\tau, p} = \beta[\tau] \quad\text{for all}\quad \tau \in \mathcal{T}, p \in \{1,2\}
        \end{aligned}
    \end{equation}
    where $\beta[\tau] =  (\beta^{(v)})_{v\in \tau}$. A direct application of the ADMM algorithm yields the following iterative updates:
    \begin{equation}\label{eq:algo}
        \begin{aligned}
            \beta^{(v)}_{[k + 1]} &= \operatornamewithlimits{argmin}_{\beta} \; l(\mathbf{y}^{(v)}, \beta) + \alpha \sum_{\{\tau \colon v \in \tau\}} \sum_{p\in\{1,2\}} (\beta - z_{{\tau, p, [k]}}^{(v)} + u_{{\tau, p, [k]}}^{(v)})^2  & \forall v\in V \\
            \mathbf{z}_{{\tau, 1, [k + 1]}} &= \operatornamewithlimits{argmin}_{\mathbf{z}} \; \lVert \mathbf{z} - \beta_{[k + 1]}[\tau] - \mathbf{u}_{{\tau, 1, [k]}}\rVert^2  + \lambda_1\mathrm{TV}_1(\mathbf{z}, \tau) \, & \forall \tau \in \mathcal{T} \\
            \mathbf{z}_{{\tau, 1, [k + 1]}} &= \operatornamewithlimits{argmin}_{\mathbf{z}} \; \lVert \mathbf{z} - \beta_{[k + 1]}[\tau] - \mathbf{u}_{{\tau, 2, [k]}}\rVert^2  + \lambda_2\mathrm{TV}_2(\mathbf{z}, \tau) \, & \forall \tau \in \mathcal{T} \\
            \mathbf{u}_{{t,p, [k+1]}} &= \mathbf{u}_{{t,p, [k]}} + \beta_{[k+1]}[\tau] - \mathbf{z}_{t,p, [k+1]}  \hspace{8em} \forall  p\in\{1,2\}  & \forall \tau \in \mathcal{T}
        \end{aligned}
    \end{equation}
    where $\alpha$ is the scalar of the ADMM step-size parameter, and $\mathbf{u}_{t,p}$ are the ADMM dual variables. The algorithm depends on a random initialization of the parameters. Step 1 corresponds to a binomial negative log-likelihood model with a quadratic regularization. Following \citetA{tansey-etal-2017}, we can substitute the full minimization in Step 1 with a single iteration of Newton's method. Step 2 corresponds to the fused lasso problem for chain graphs. We leverage available linear-time solvers such as \citepA{johnson-2013} and \citepA{barbero-sra-2018}, which have comparative performance. We prefer the latter since it can handle different values of $\lambda_1$ for each edge, although in the current formulation it is assumed to be constant. Step 3 can be solved in linear time using the Kalman smoothing algorithm (see \citetA{welch-1995}). Finally, step 4 is the dual variable update of the ADMM algorithm in scaled form and does not require any sophisticated computation. There are different strategies to dynamically change the value $\alpha$ to accelerate convergence; empirically, we found that the method of \citetA{wohlberg2017admm} worked best for our problem.

    The smoothing step can be performed in an embarrassingly parallel  way for every node of the tree. There are additional possibilities for parallelism. In a high-performance computing environment, a useful strategy would be to distribute the smoothing problem for each tree node into different computation nodes using distributed memory parallelism. Meanwhile, steps 2-4 can be parallelized for each trail using shared-memory parallelism. 

To choose the trails we tried two different methods, but did not notice any difference in performance. First, we used the ``Median trail" heuristic described in \citepA{tansey-scott-2015}. Since for most graphs it is impossible to find a perfectly balanced set of non-overlapping trails, this strategy finds a sequence of non-overlapping trails traversing every edge with a heuristic that tries to balance the length of each trail. To do so it proceeds recursively with a series of trail decompositions, selection of the median sized trail, and pruning of the visited edges. We also tried a new trail decomposition technique based on an approximate solution to the ``Route inspection problem'', which finds a unique trail that traverses every edge in the graph with the minimal number of repetitions, adjusting the smoothing weights for repeated edges. This latter approach yielded perfectly balanced trails convenient for parallelization. However, in practice we did not observe any substantial increase or decrease in time to convergence.
    
\subsection*{Adaptive Rejection Sampling} Bayesian inference is easy to implement using an Adaptive Rejection Sampling (ARS) \citepA{gilks1992adaptive} for log-concave likelihoods. As before, we drop the notation for the split of the density tree since the same procedure can be applied at every split. Denote $\beta^{(-v)}=\{\beta^{(w)} \mid w\in V, w\neq v\}$. Then the conditional Gibbs distribution for $\beta^{(v)}$ is
\begin{align*}
    & \log p(\beta^{(v)} \mid \beta^{(-v)}, \boldsymbol{y}) = C +\\
     & \hspace{.5cm} s^{(v)} \log (w^{(v)}) + (n^{(v)} - s^{(v)}) \log(1 - w^{(v)}) + \sum_{p \in \{1,2\}} \sum_{d\in \{S,T\}} \sum_{w\in \mathcal{N}_d(v)} \lambda_{d,p}\lvert\beta^{(v)}-\beta^{(w)}\rvert^p
\end{align*}
where $C$ is a constant, $n^{(v)}$ and $s^{(v)}$ are respectively the observed ``attempts'' and ``successes'' for the binomial conditional split, $w^{(v)}=(1 + \exp(-\beta^{(v)}))^{-1}$ and $\mathcal{N}_d$ for $d\in\{S,T\}$ is the set of neighbors of $v$ in either the spatial or temporal axes. The conditional distribution is strictly log-concave even with missing data ($n^{(v)}=s^{(v)}=0$) since 
$$
0 > \frac{d^2}{d(\beta^{(v)})^2}\log p(\beta^{(v)} \mid \beta^{(-v)}, \boldsymbol{y}) = -  w^{(v)} (1 - w^{(v)}) - 2 \sum_{d\in \{S,T\}}\left|\mathcal{N}_d(v)\right| \lambda_{d,2}.
$$
We implement the Gibbs sampler in Julia \citep{bezanson2017julia} using the \texttt{ARS.jl} package \citep{tec2018ars} in Julia. ARS for densities with unbounded supports needs to be initialized with two points that have respectively positive and negative slope. For the left point we choose a small negative perturbation from the minimum of the neighbors $\{\beta^{(w)}\mid w\in \mathcal{N}_d(v), d\in\{S,T\}\}$ and the local MLE solution (when defined) $\log(n^{(v)}/s^{(v)} - 1)$. We do the converse using the maximum and a positive perturbation for the right bound. To improve speed, we use parallel asynchronous updates for each $v$ \citep{terenin2020asynchronous}. For each split, we run 5000 iterations starting from the MAP solution with a burn-in of 4000. The full sampling procedure takes approximately 8-12 hours per split.

Finally, we also note that it is straightforward to extend the MCMC sampling technique described by \citet{tansey-etal-2017} for the GFL. The essence of this strategy is to use global/local shrinkage to transform the $\ell_1$-penalty into an $\ell_2$ penalty with an exponential mixing distribution (see \citet{polson2010shrink}), and then use the fast sampling strategy for GMRFs of \citet{rue2001fast}. In our case, the new $\ell_2$ penalty would combine with the existing $\ell_2$ portion of the GFEN penalty, still preserving the GMRF structure. A comparison of this alternative approach and the Adaptive Rejection Sampling strategy used here is left for future investigation.

\section{Choice often Binary Tree}
\label{appendix:tree-splitting}
\setcounter{figure}{0}
\setcounter{table}{0}

Figure \ref{fig:quantile-splitting} in the main texts shows that a uniform quantile decomposition using quantiles of dyadic numbers can lead to low resolution in the tails. For example, a quantile decomposition of depth 2 has three splits. The first one uses the $q_{.5}$ quantile as middle point, where as the second level uses the $q_{.25}$ and $q_{.75}$ quantiles as midpoints splits. A quantile decomposition of depth $d$ has $2^d-1$ splits. For our analysis we first the 31 splits corresponding to $d=5$. The motivation to add splits is to combine the coverage property of the uniform splits with the multi-scale benefits of the quantile splits. For this purpose, we add a level-($d$ + 1) split in the left tail with midpoint $0.5(m + q_{1/2^d})$ where $m=\min(\boldsymbol{y})$. We then add 5 splits to the right tail by taking 7 uniformly spaced points in the range $[q_{1-1/2^d}, M]$ with $M=\$100$. Then for each point $x_i$ ($i=1,\hdots, 5)$ in the set of uniform points with the endpoints removed we add a level-$(d + i)$ split with left children $[x_{i-1}, x_i)$ and right children $[x_i, \max(\boldsymbol{y}))$.

We also considered the Hierarchical MAP Optional Polya Tree (HMAP-OPT) framework proposed by \citet{wong2010optional}. This framework yields adaptive trees with varying depth by applying the Optional Stopping Theorem recursively to decide whether to continue splitting over a branch, assuming a Polya tree prior. Intuitively, it can assign more resolution where the scale of the density requires so. Figure \ref{fig:quantile-splitting} also shows an example of the HMAP-OPT with maximum depth six and stopping probability of 0.1. We chose to use the extended quantile approach over the HMAP-OPT for two reasons. First, quantiles allow even more resolution in regions of higher density for a similar number of splits. Second, regions of density that the HMAP-OPT recognizes as a coarser scale could still have high variation between spatial locations. The HMAP-OPT is designed for a single density, and we cannot apply it at every location since we need the splitting values to be comparable among all sites. For future work, it would be interesting to explore extensions of the HMAP-OPT for the multi-location case.

    \section{Additional Hyperparameter Tuning Details}
    \label{appendix:bayes-opt}
    \setcounter{figure}{0}
    \setcounter{table}{0}
    
    Hyperparameter tuning casn be performed using in-sample and out-of-sample criteria. In-sample tuning for the graph-fused lasso is typically done using information criteria such as the Akaike information criterion (AIC) or the Bayesian information criterion (BIC) \citepA{tibshirani-2015}. These methods rely on the fact the degrees of freedom are easy to compute in the graph-fused lasso since it reduces to counting the number of plateaus \citepA{tansey-scott-2015}. However, in the case of the GFEN, the $\ell_2$-norm penalty adds smoothness and makes the approach of counting plateaus unfeasible. One solution that appears in the GMRF literature is to use an approximate information criterion such as the Deviance Information Criterion (DIC). However, that solution is known to favor over-fitted models and assumes an approximately normal distribution predictive distribution \citepA{ando2011predictive}. Below we describe an alternative out-of-sample tuning approach based on cross-validation; for a survey, see \citetA{arlot2010survey}.
    
    The overall idea is that we can use $k$-cross-validation to maximize the out-of-sample average probability of the data with respect to the estimated distributions. We search a set of hyperparameters $\boldsymbol{\lambda}=\{\lambda_\gamma\in \mathbb{R}^4_{\geq 0} \mid \gamma \in \{0,1\}^D\}$ for a model with depth $D$. Our objective in cross-validation is to minimize the negative loglikelihood arising from the model described in Section \ref{subsection:densities}. We divide our data $\mathbf{y}$ into $k$ equally sized testing sets or folds $\{\mathbf{y}^{\{j\}}_\text{cv}\}_{j=1}^k$. For our analysis, we split the data based on randomly partitioning the vertices of the graph in $k$ groups, and treating the vertices in the missing fold as missing data vertices. For each fold $j$, we use the training data $\mathbf{y}^{\{j\}}_\text{train} = \bigcup_{j' \neq j} \mathbf{y}^{\{j'\}}_
    \text{cv}$ to fit the parameters of a statistical model. We then use these parameters to compute the out-of-sample loss for the fold, using the evaluation set $\mathbf{y}^{\{j\}}_\text{test} := \mathbf{y}^{\{j\}}_\text{cv}$. Finally, we average the losses over all test sets. Considering all splits, the out-of-sample loss we seek to minimize with cross-validation is given by
    \begin{equation}\label{eq:negll}
        l(\boldsymbol{\lambda}) := \frac{1}{N} \sum_{j=1}^k \sum_{y_i \in \mathbf{y}^{\{j\}}_\text{test}} -\log \hat{f}_i(y_i \mid \hat{\boldsymbol{\beta}}^{\{j\}})
    \end{equation}
    where $N$ is the total number of data points, $\{\hat{\boldsymbol{\beta}}^{\{j\}}\}$ are the estimates on the $k$-fold obtained using the hyperparameters $\boldsymbol{\lambda}$, and $\hat{f}_i$ is the estimated density function at the vertex where the test data $y_i$ is observed. Because of the conditional independence properties of the tree split decomposition, the out-of sample loss can be optimized by optimizing each tree split independently.
    
For our analysis of the RideAustin data, we tune independently the hyperparameters for each tree split. For the simulation study in section \ref{section:simulation}, the same regularization level is used in all splits due to the high volume of experiments.

As discussed in the main text, we use Bayesian Optimization \citepA{snoek-etal-2012, shahriari-etal-2016} to guide the search by modeling $l(\boldsymbol{\lambda})$ using a Gaussian Process (GP). The main idea of this approach is the following:

\begin{enumerate}[itemsep=0pt, partopsep=0pt]
    \item Suppose we have observed out-of-sample losses $\hat{l}_1,\hdots, \hat{l}_n$ corresponding to hyperparameters $\lambda_1,\hdots, \lambda_n$. The GP assumption is that $\hat{l}_1,\hdots, \hat{l}_n$ follow a multivariate Gaussian distribution.
    \item Moreover, the multivariate distribution is assumed to have the form $    (\hat{l}_1,...,\hat{l}_n)  \sim \mathrm{Normal}(0, K + \sigma^2 I)$ where $K:=(K_{ij})_{i,j=1}^n$ is some Kernel matrix depending on $\lambda_1,\hdots,\lambda_n$ and $\sigma^2$ models the uncertainty in the observations $\hat{l}_j$.
    \item Given a new point $l_*$ corresponding to an untested hyperparameter $\lambda_*$, the fact that $(l_*, \hat{l}_1,\hdots, \hat{l}_n)$ is multivariate Gaussian can be used to easily compute the predictive distribution of $l_*$ given the observed $\hat{l}_1,\hdots, \hat{l}_n$. Therefore, to propose new candidates we generate samples and select candidates with minimum sampled predicted loss. 
    \item At the end of the optimization rounds, we select the $\lambda_*$ that has the lowest predictive expected loss $E[l_* \mid \hat{l}_1,\hdots, \hat{l}_n]$. 
\end{enumerate}
The above steps give a high-level description of the idea of Bayesian optimization. For a detailed explanation, we refer the reader to the review by \citetA{shahriari-etal-2016}. 

We use a radial kernel $K_{ij}:=\exp(-a \lVert \lambda_i - \lambda_j\rVert^2_2)$ with $a=0.15$ and $\sigma=0.1$. At each step the data is standardized before evaluating the GP's posterior. we proceed in generations of size $6$ for each of the $36$ splits, for a total of $48$ generations, as explained in section \ref{sec:hypersearch}.

\section{Simulation Task}
\label{appendix:simulation_task}
\setcounter{figure}{0}
\setcounter{table}{0}

\subsection{Effect Dynamics}

\begin{figure}[!tb]
    \centering
    \begin{subfigure}[tb]{.3\linewidth}
        \centering
        \includegraphics[width=0.98\linewidth, height=100px]{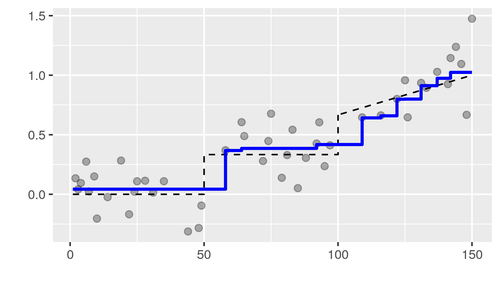}
        \caption{\footnotesize GFL}
    \end{subfigure}%
    ~
    \begin{subfigure}[tb]{.3\linewidth}
        \centering
        \includegraphics[width=0.98\linewidth, height=100px]{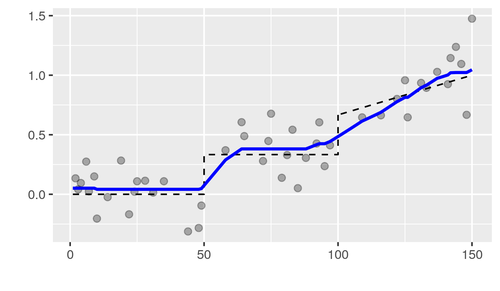}
        \caption{GFEN}
    \end{subfigure}
    ~
    \begin{subfigure}[tb]{.3\linewidth}
        \centering
        \includegraphics[width=0.98\linewidth, height=100px]{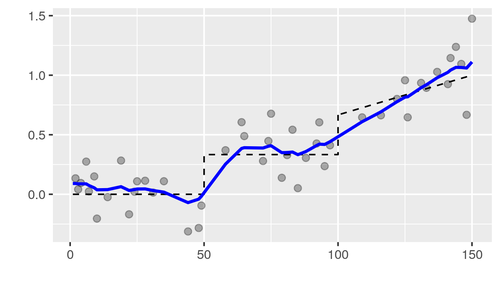}
        \caption{\footnotesize GMRF}
    \end{subfigure}%
    \caption{Comparison of the estimation of a signal on a chain graph. The true signal (dashed) is a mix of piece-wise constant and smooth regions. Data is observed with noise, with 66\% of the vertices having missing data. The hyperparameters for all models were chosen independently using cross-validation to minimize the quadratic prediction error.} 
    \label{fig:benchmarks-linear}
\end{figure}

The spatial effects $\mu_i$ and $\nu_i$ (see Section \ref{section:simulation}) were generated according to the following rule: Given a size $N$ (in the experiment $N=30$) we divided the range $\{1,\hdots,30\}$ into three equally sized segments. The signal values assigned to each segment will be crafted with the aim that the full observed vector follows one of three possible dynamics:

\begin{enumerate}[itemsep=0pt, partopsep=0pt]
    \item \textbf{Piece-wise constant}. The effect consists of \textit{plateaus} of constant values chosen at random with discontinuous jumps between plateaus. 
    \item \textbf{Piece-wise linear}. The effect consists of linearly interpolated shifts between random values. There are not discontinuous jumps.
    \item \textbf{Mixed}. The effect is a combination of the above. Some segments consist of constant-value plateaus with discontinuous jumps and others of linearly interpolated shifts.
\end{enumerate}
To achieve this, we start by assigning a value in the range $[-1, 1]$ to each of the points $\{0, N/3, 2N/3, N\}$, which connect the segments. Each segment can be either linear or constant. If a segment is chosen to be linear, the segment is filled with its corresponding linear interpolation for the value of the beginning and end of the segment. For constant segments, only the value of the beginning of the segment is considered and repeated in all the segments. Piece-wise linear signals have all segments linear, piece-wise constant signals have all segments constant, and mixed-type signals contain a random combination having at least one segment of each type. Figure \ref{fig:benchmarks-linear} shows an example of a mixed dynamic with $N=150$ where the first two segments are piece-wise constant, and the rightmost segment is linearly interpolated (smooth).

\subsection{Evaluation}

For each simulated dataset and for each model (GFL, GMRF and the GFEN), we sample 24 combinations of hyperparameters where taken from $\log_{10}(\lambda_{d,p}) \sim \mathrm{Uniform}(-2, 7)$ with the restriction $\lambda_{d,1}=0$ for the GFL and $\lambda_{d,2}=0$ for the GMRF. Since the GFL is ill-defined for problems with missing data (see Section \ref{sec:gfl}), a tiny regularization prior of the form $10^{-8}\lVert \beta \rVert^2$ was added to the optimization objective.

We chose the best hyperparameters for each method (GFL, GFEN, and GMRF) using 5-fold cross-validation (see equation \eqref{eq:negll}). The final evaluation uses 100 samples from the ground truth density at each location and evaluates the log-likelihood of the fitted model. We generated 48 simulated datasets for each of $6=\binom{3}{2}$ combinations of spatial and temporal effect. In addition,  we added another task to evaluate the resilience with respect to outliers, injecting one outlier in 50\% of the vertices in the \textit{mixed vs. mixed} task. An outlier was generated by sampling from an over-dispersed version of the true distribution with ten times the standard deviation. The experiments are conducted under two missing data regimes with 10\% and 80\% randomly selected missing data nodes respectively. In total, we conducted $32,256=3\times 24 \times 48 \times 7 \times 2$ experiments for comparison, each one resulting in a density estimate for each one of the vertices of the $30\times 30$ grid.

\section{Data distribution}
\label{appendix:global-distribution}

\begin{figure}[H]
    \centering
    \begin{subfigure}[h]{0.495\linewidth}
        \centering
        \includegraphics[width=.8\linewidth]{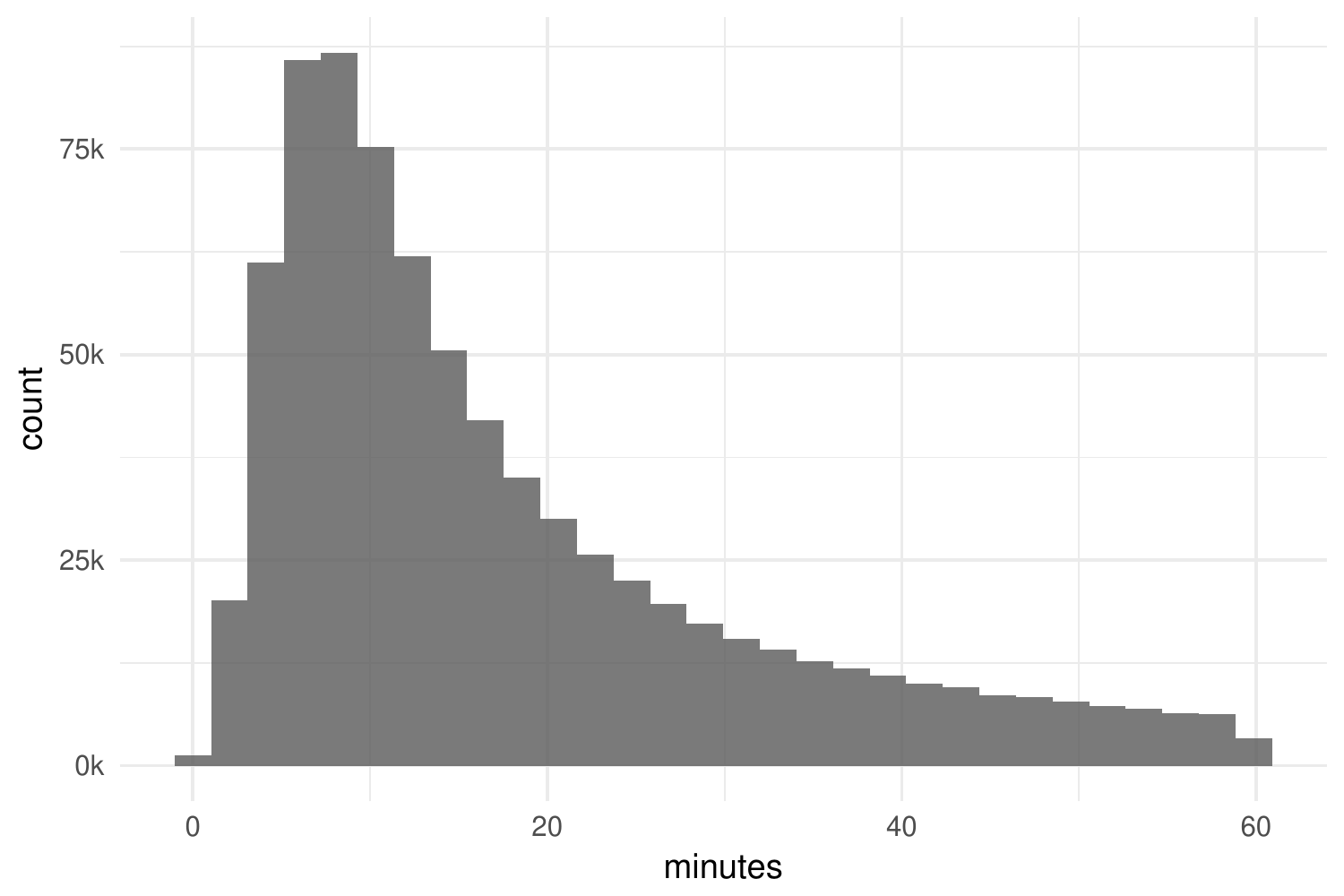}
        \subcaption{Distribution of idle time after trip end}
        \label{fig:prod:idlehist}
    \end{subfigure}
    \begin{subfigure}[h]{0.495\linewidth}
        \centering
        \includegraphics[width=.8\linewidth]{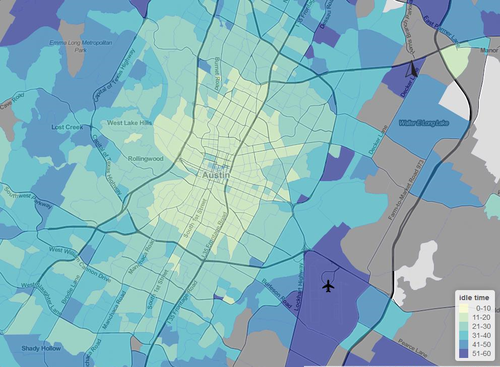}
        \subcaption{Median idle minutes in between trips by TAZ}
        \label{fig:prod:idlemap}
    \end{subfigure}
    \begin{subfigure}[h]{0.495\linewidth}
        \centering
        \includegraphics[width=.8\linewidth]{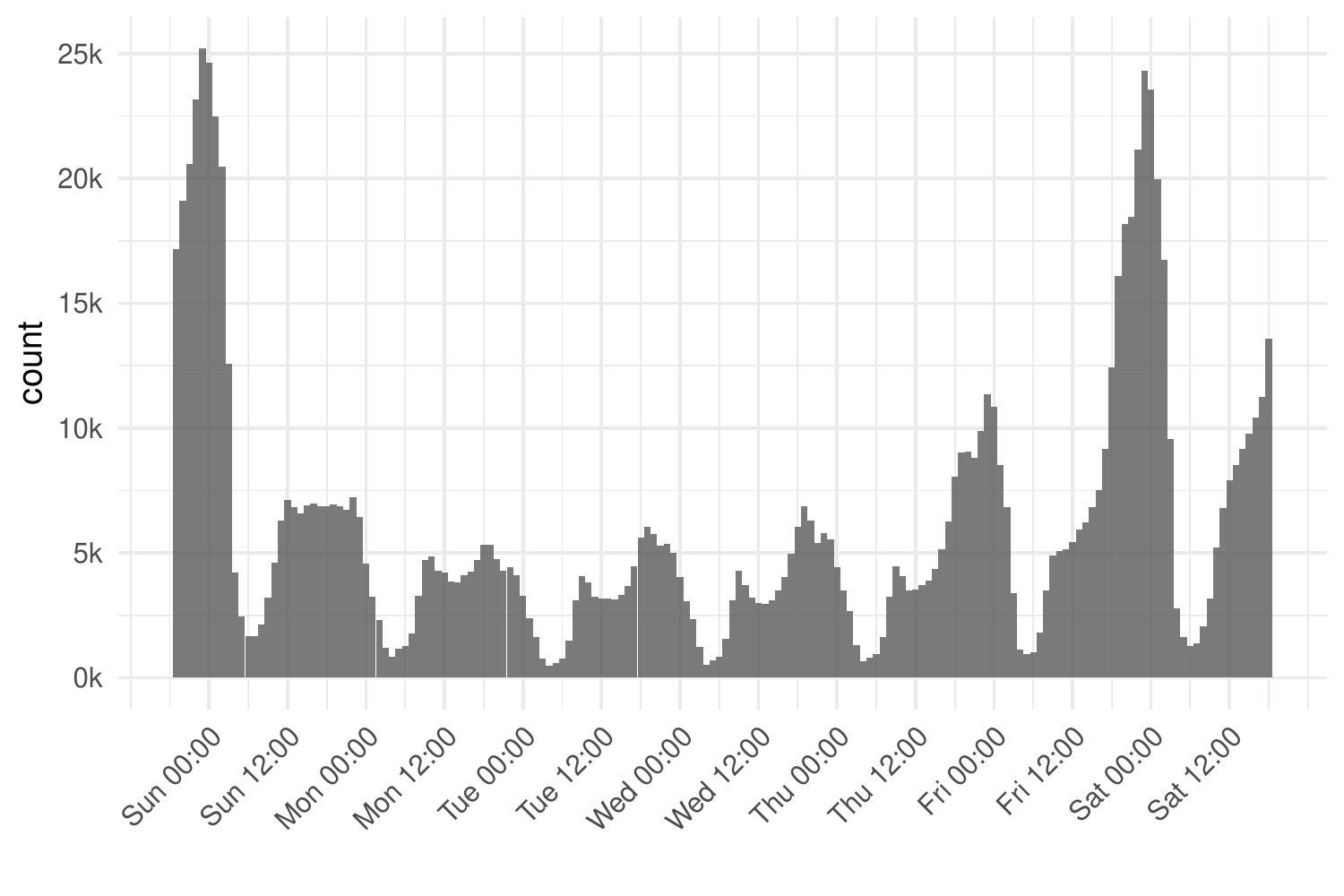}
        \caption{Counts aggregated by time units}
        \label{fig:prod:timely}
    \end{subfigure}
    \begin{subfigure}[h]{0.495\linewidth}
        \centering
        \includegraphics[width=.8\linewidth]{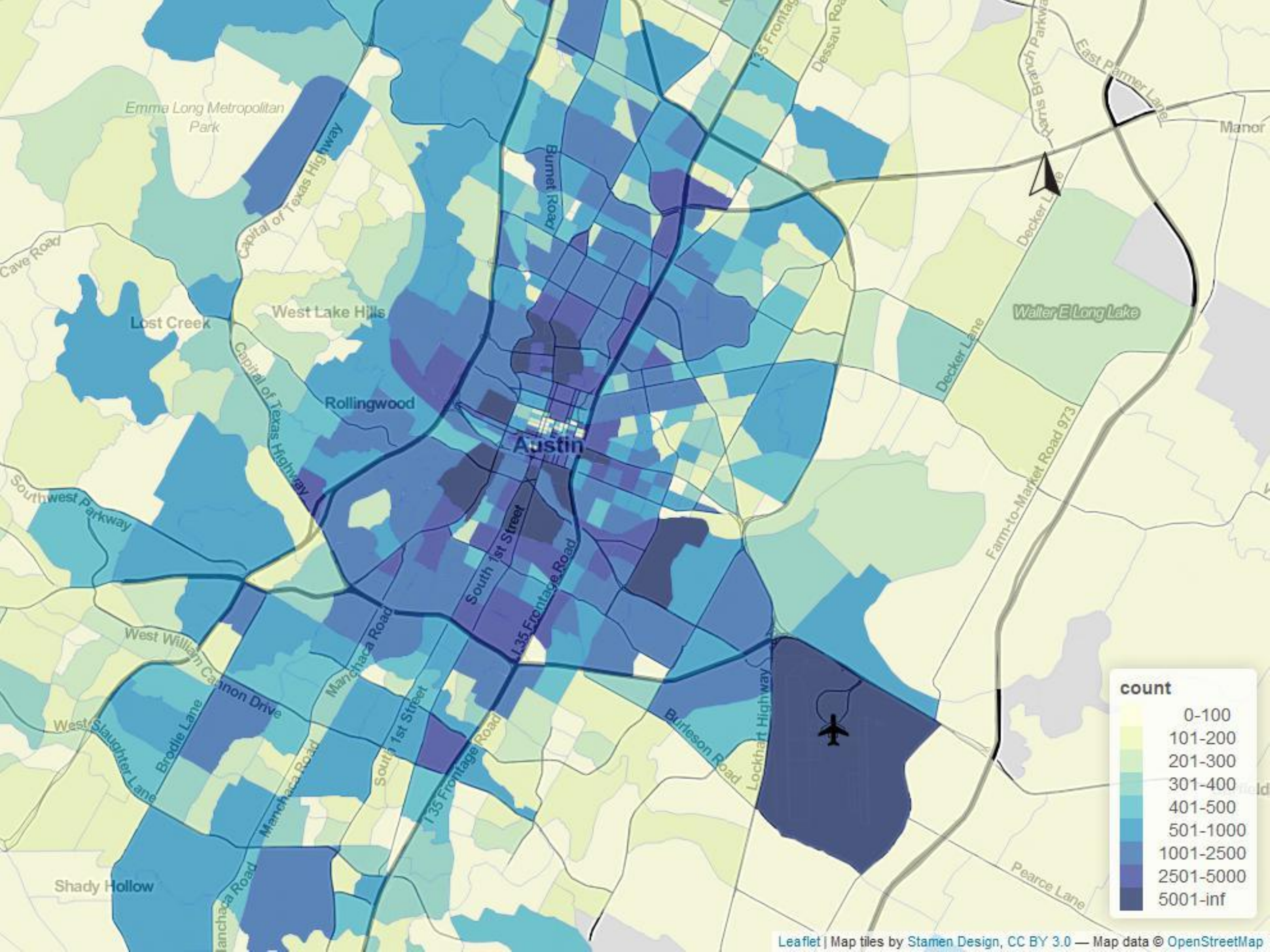}
        \caption{Counts aggregated by space units}
        \label{fig:prod:spatial}
    \end{subfigure}%
    \caption{Description of the estimated productivity metrics} 
    \label{fig:prod_spacetime}
\end{figure}

Figure \ref{fig:quantile-splitting} shows the distribution of the productivity variable defined in Section \ref{subsection:productivity} aggregated globally from all TAZs. Figure \ref{fig:prod:idlehist} and \ref{fig:prod:idlemap} show the distribution of the idle time variable (also defined in Section \ref{subsection:productivity}) aggregated globally and aggregated by TAZ. We can observe in these figures how TAZs that are more distant from the central business district on average present larger idle times. Finally, figures \ref{fig:prod:timely} \ref{fig:prod:spatial} show the total counts of trips aggregated by time and aggregated by TAZ. These figures show that there is a strong weekly periodicity with a strong distinction between weekdays and weekends; and show that most of the trips concentrate around central Austin and the airport area.

\section{Tail Probabilities of Not Exceeding Living Wage}
\label{appendix:maps_livingwages}
\setcounter{figure}{0}
\setcounter{table}{0}

\begin{figure}[H]
    \centering
    \begin{minipage}[t]{0.48\linewidth}
        \centering
        \includegraphics[width=0.9\linewidth]{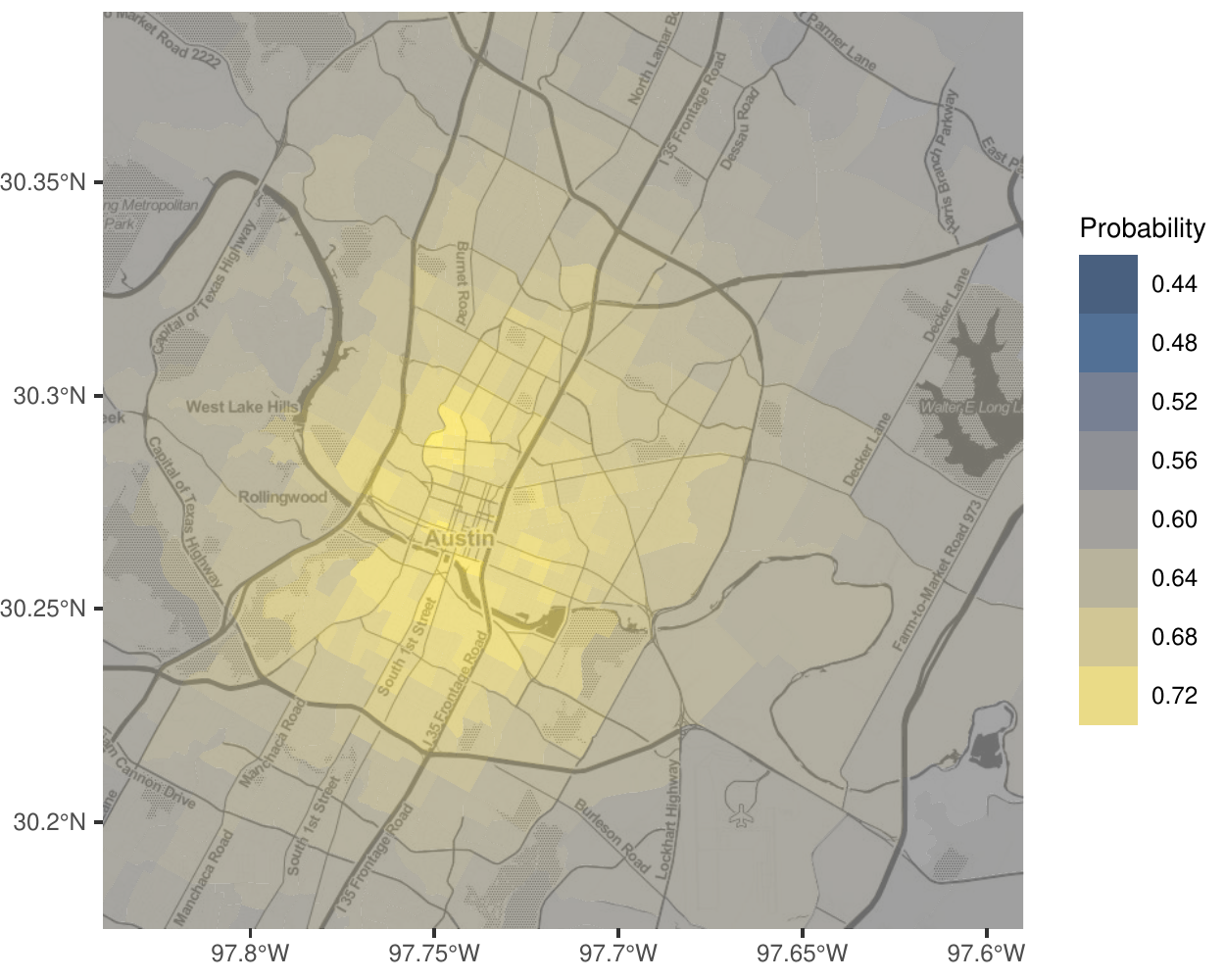}
        \subcaption{Sunday 12 AM}
        \label{fig:wages:appendix1:a}
    \end{minipage}
    \begin{minipage}[t]{0.48\linewidth}
        \centering
        \includegraphics[width=0.9\linewidth]{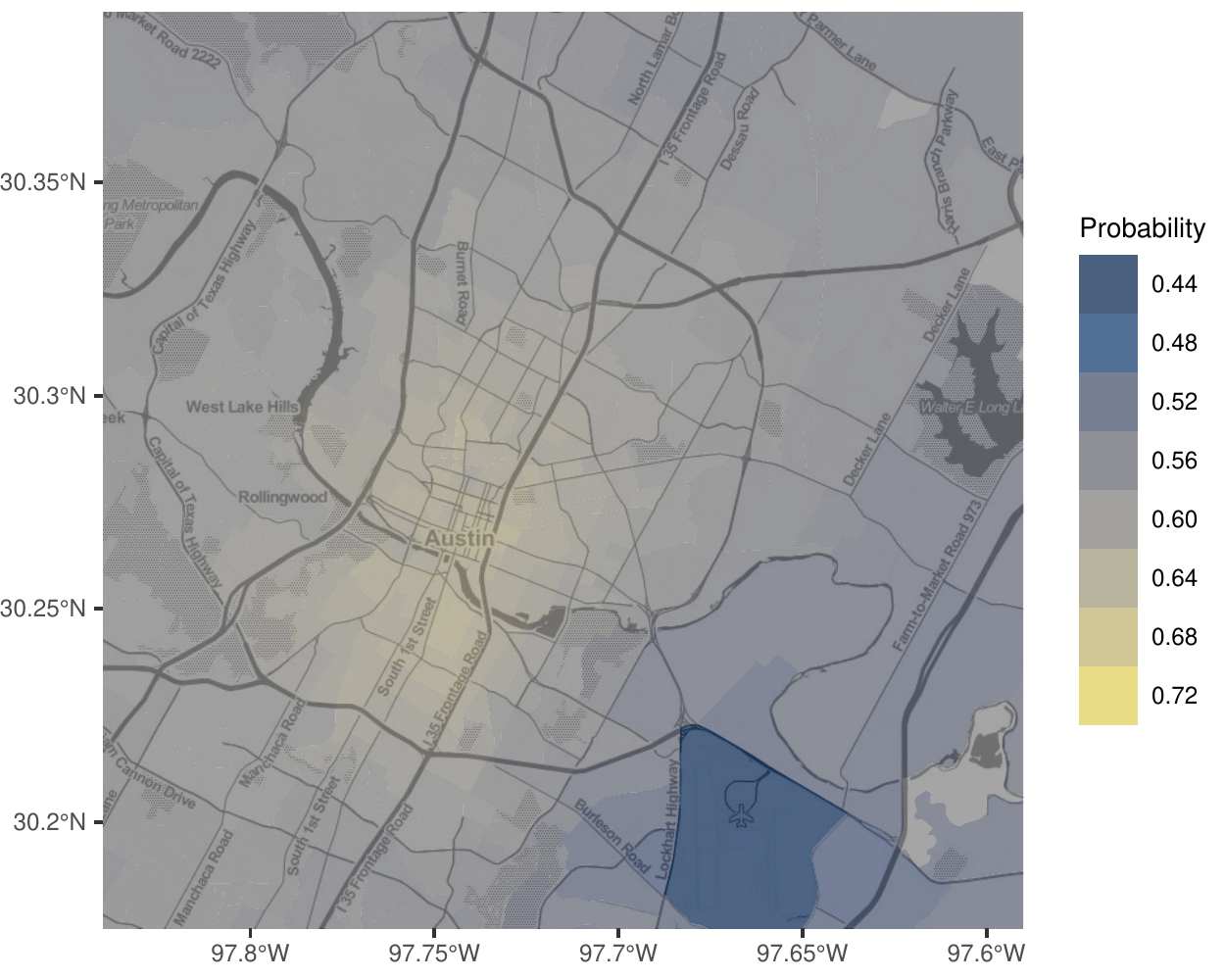}
        \subcaption{Monday 12 PM}
        \label{fig:wages:appendix1:b}
    \end{minipage}
    \caption{Probability of exceeding \$18.56 in the next hour given a current location (living wage with costs for one single working adult with no children).}
    \label{fig:wages:appendix1}
\end{figure}

\begin{figure}[H]
    \centering
    \begin{minipage}[t]{0.48\linewidth}
        \centering
        \includegraphics[width=0.9\linewidth]{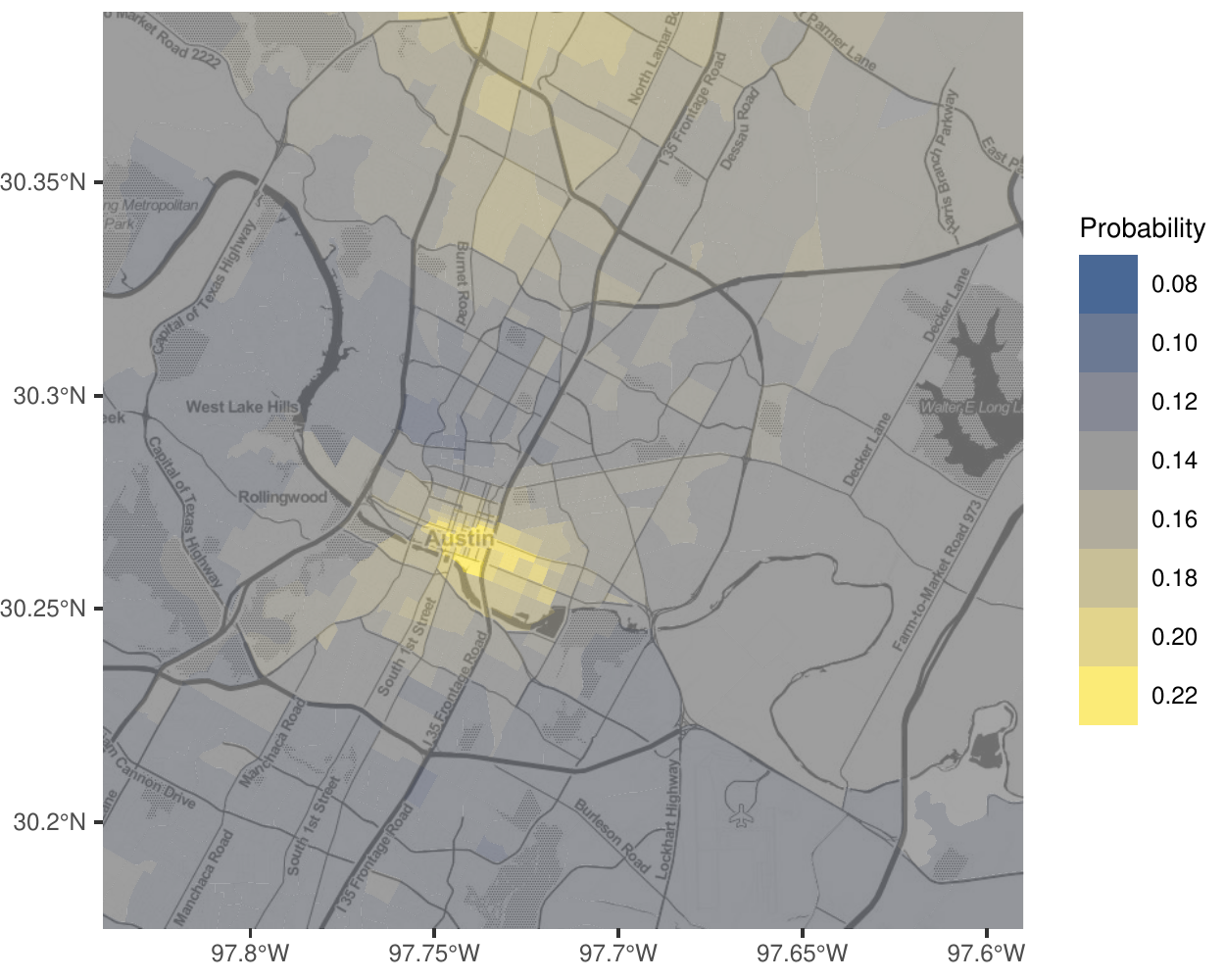}
        \subcaption{Sunday 12 AM}
        \label{fig:wages:appendix2:a}
    \end{minipage}
    \begin{minipage}[t]{0.48\linewidth}
        \centering
        \includegraphics[width=0.9\linewidth]{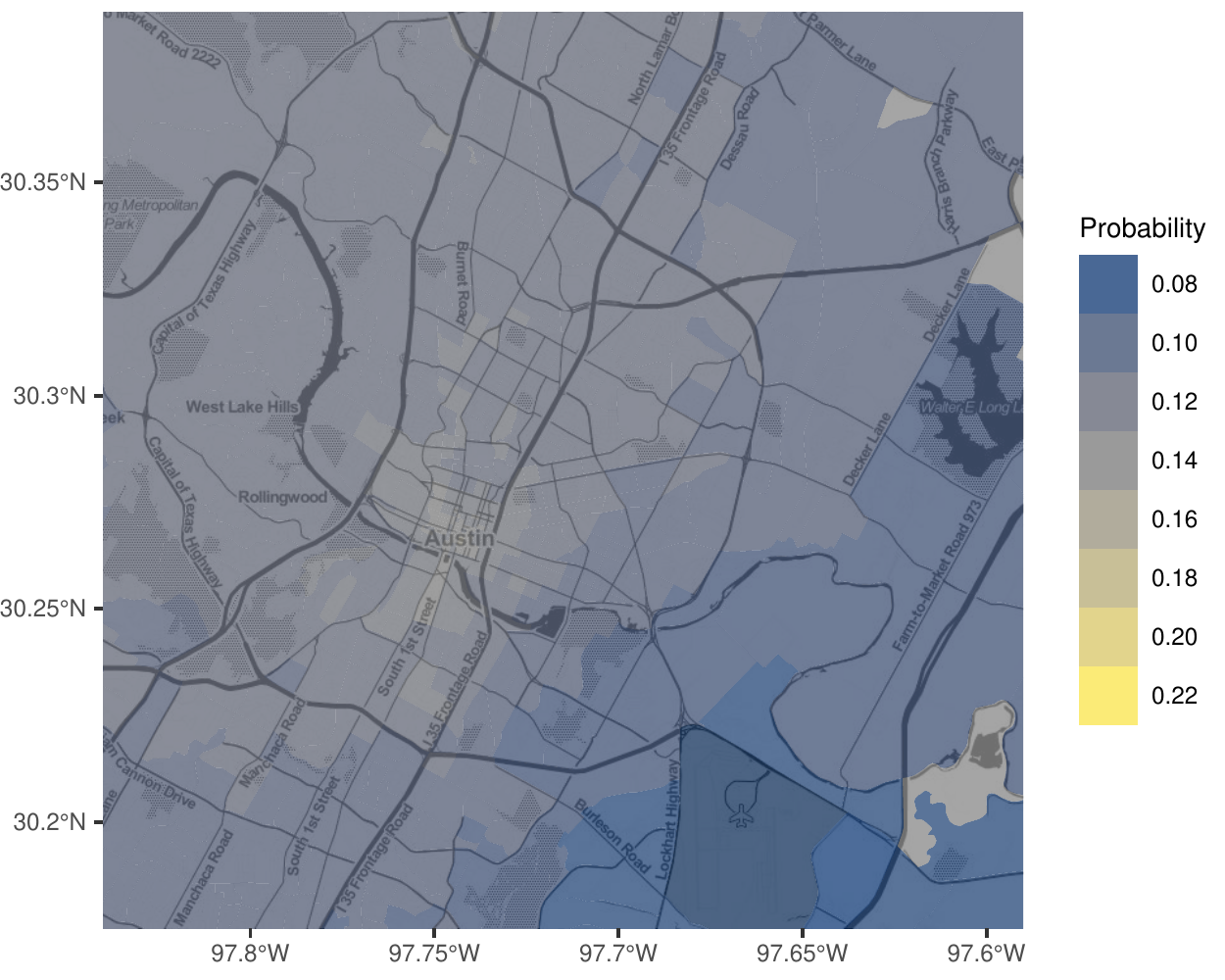}
        \subcaption{Monday 12 PM}
        \label{fig:wages:appendix2:b}
    \end{minipage}
    \caption{Probability of exceeding \$32.73 in the next hour given a current location (living wage with costs for two adults, one working, and two children).}
    \label{fig:wages:appendix2}
\end{figure}

\begin{figure}[H]
    \centering
    \begin{minipage}[t]{0.48\linewidth}
        \centering
        \includegraphics[width=0.9\linewidth]{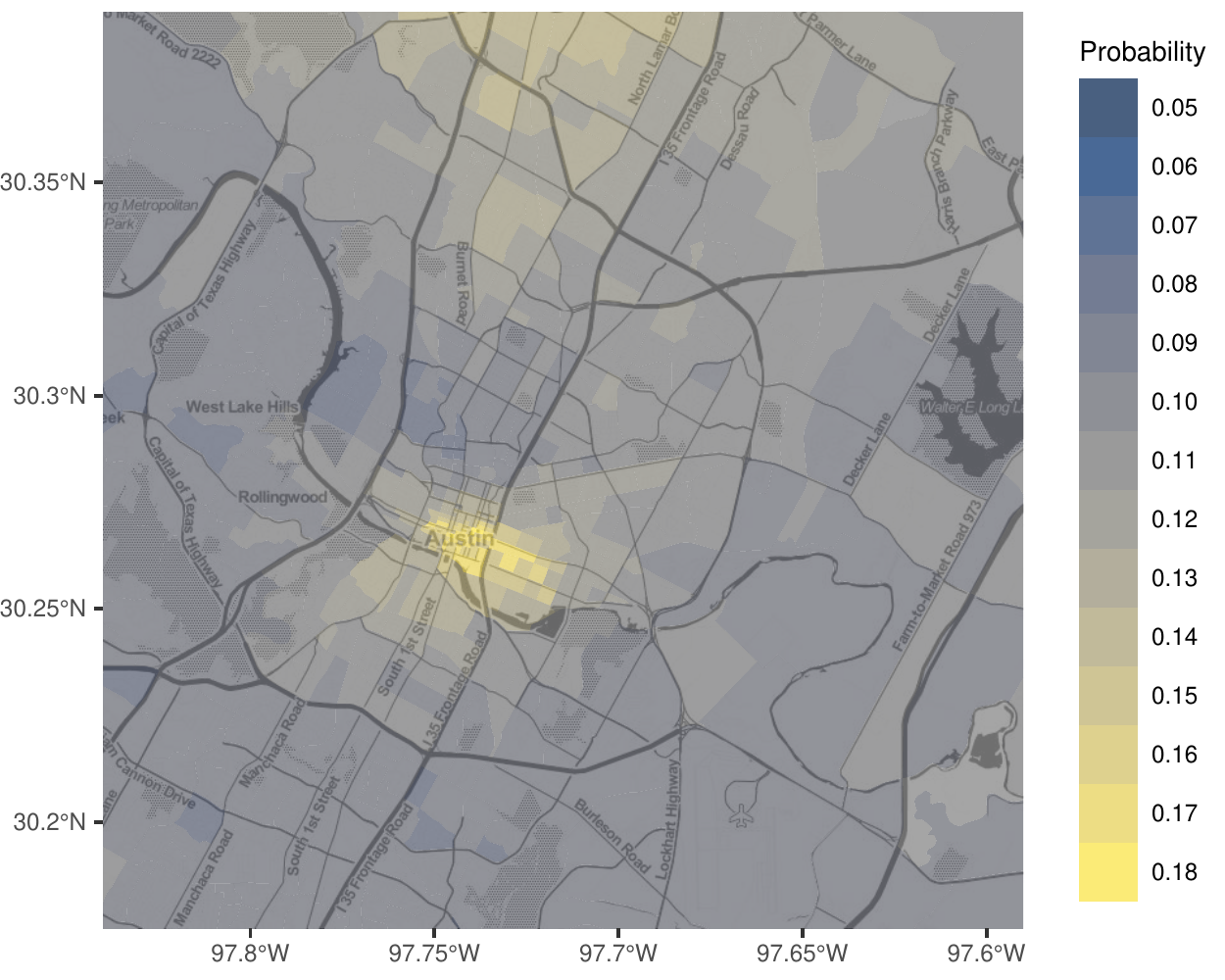}
        \subcaption{Sunday 12 AM}
        \label{fig:wages:appendix3:a}
    \end{minipage}
    \begin{minipage}[t]{0.48\linewidth}
        \centering
        \includegraphics[width=0.9\linewidth]{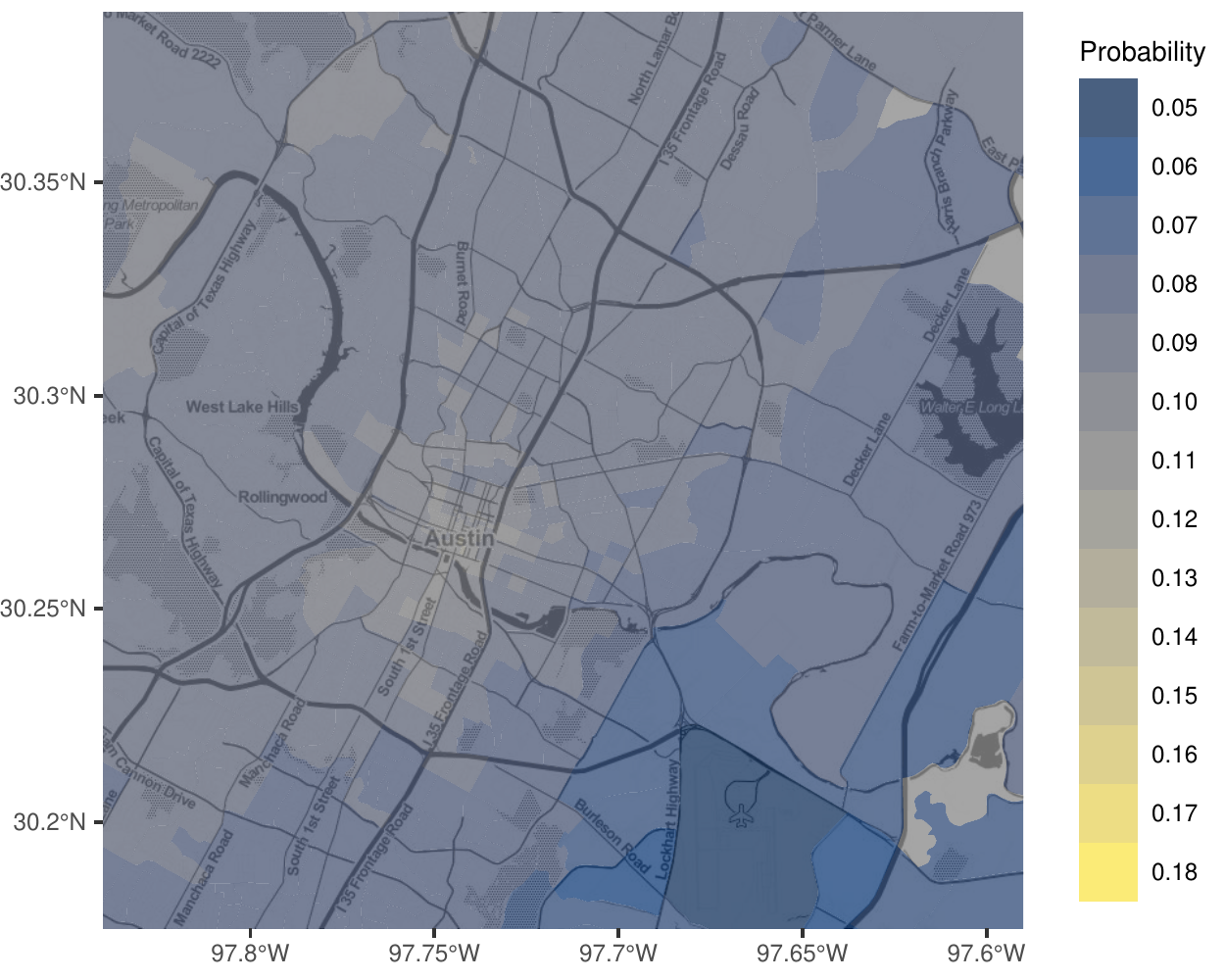}
        \subcaption{Monday 12 PM}
        \label{fig:wages:appendix3:b}
    \end{minipage}
    \caption{Probability of exceeding \$34.74 in the next hour given a current location (living wage with costs for one adult with two children).}
    \label{fig:wages:appendix3}
\end{figure}

\section{Quantiles}
\label{appendix:quantiles}
\setcounter{figure}{0}
\setcounter{table}{0}

\begin{figure}[H]
    \centering
    \begin{minipage}[t]{0.48\linewidth}
        \centering
        \includegraphics[width=0.9\linewidth]{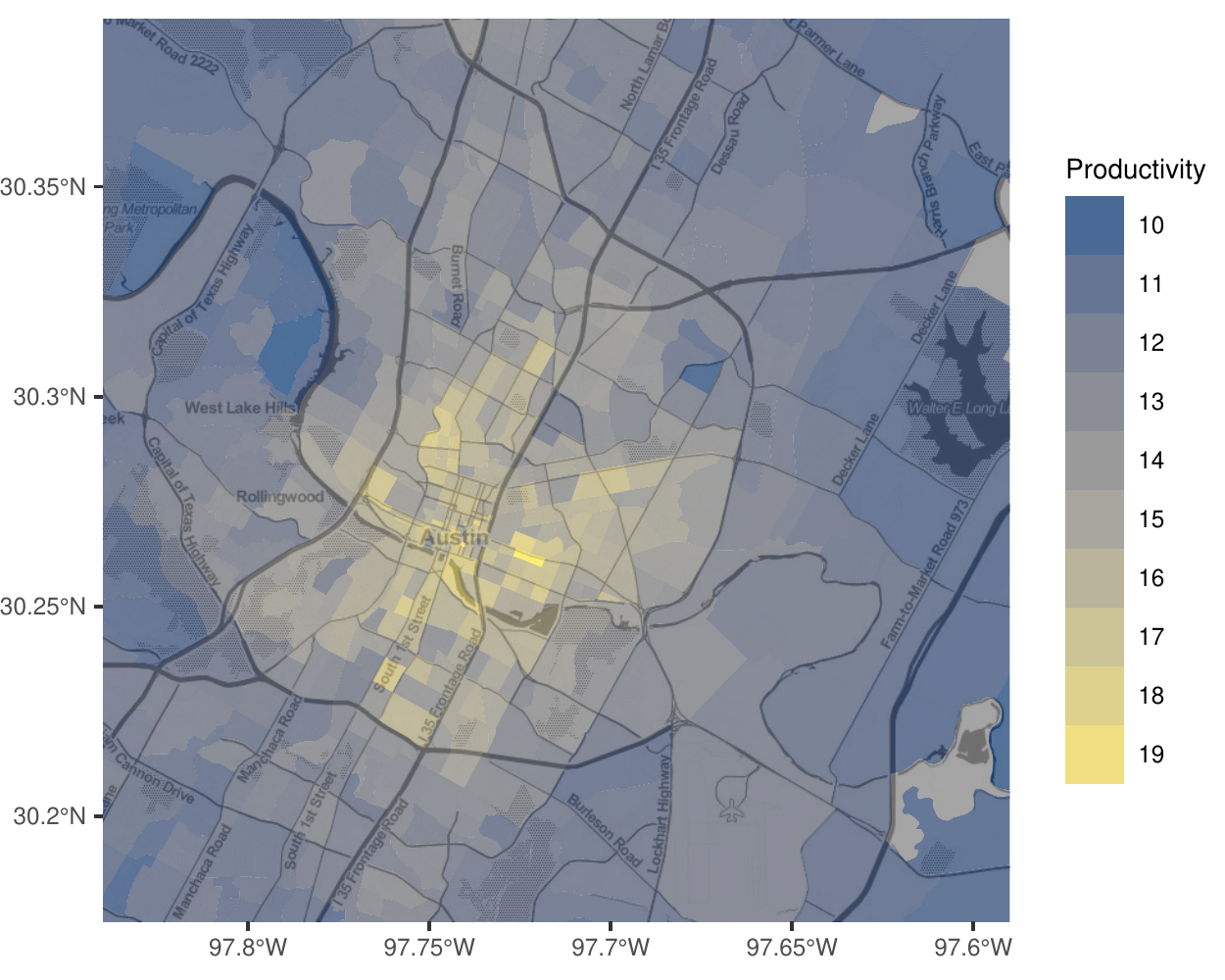}
        \subcaption{Sunday 12 AM}
        \label{fig:quantiles:0.25:a}
    \end{minipage}
    \begin{minipage}[t]{0.48\linewidth}
        \centering
        \includegraphics[width=0.9\linewidth]{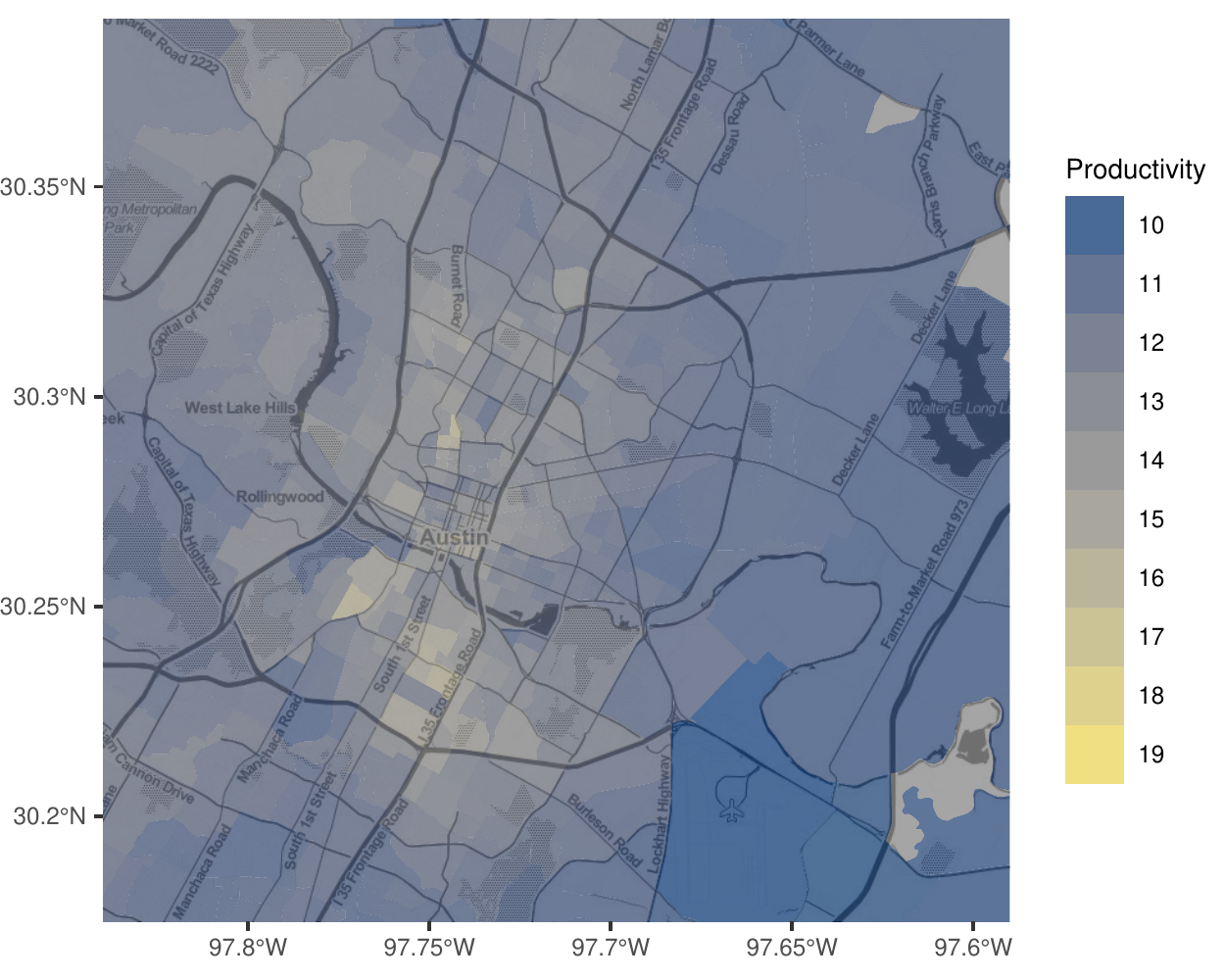}
        \subcaption{Monday 12 PM}
        \label{fig:quantiles:0.25:b}
    \end{minipage}
    \caption{Lower 25\% quantile of productivity for different times and locations.}
    \label{fig:quantiles:0.25}
\end{figure}

\begin{figure}[H]
    \centering
    \begin{minipage}[t]{0.48\linewidth}
        \centering
        \includegraphics[width=0.9\linewidth]{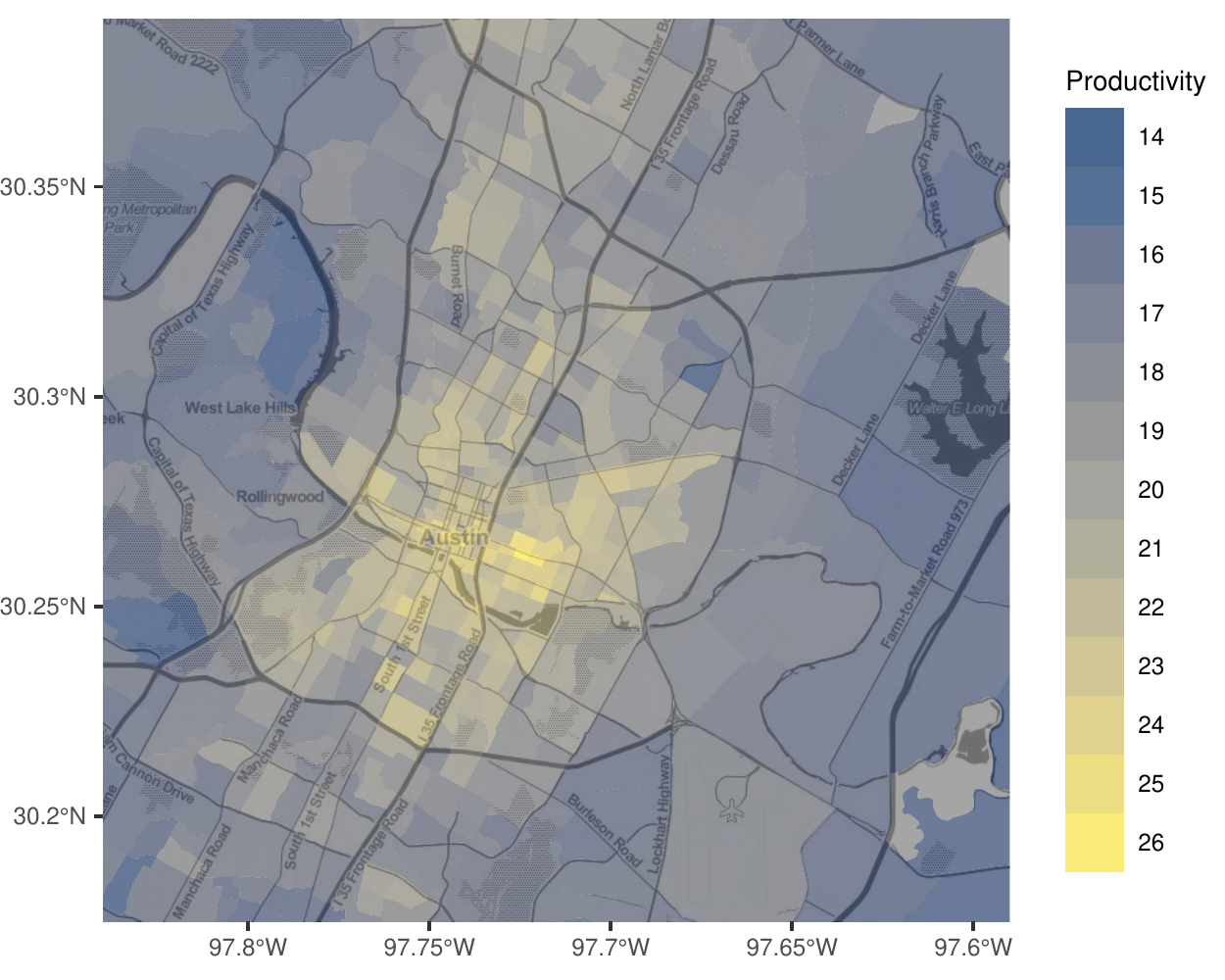}
        \subcaption{Sunday 12 AM}
        \label{fig:quantiles:0.50:a}
    \end{minipage}
    \begin{minipage}[t]{0.48\linewidth}
        \centering
        \includegraphics[width=0.9\linewidth]{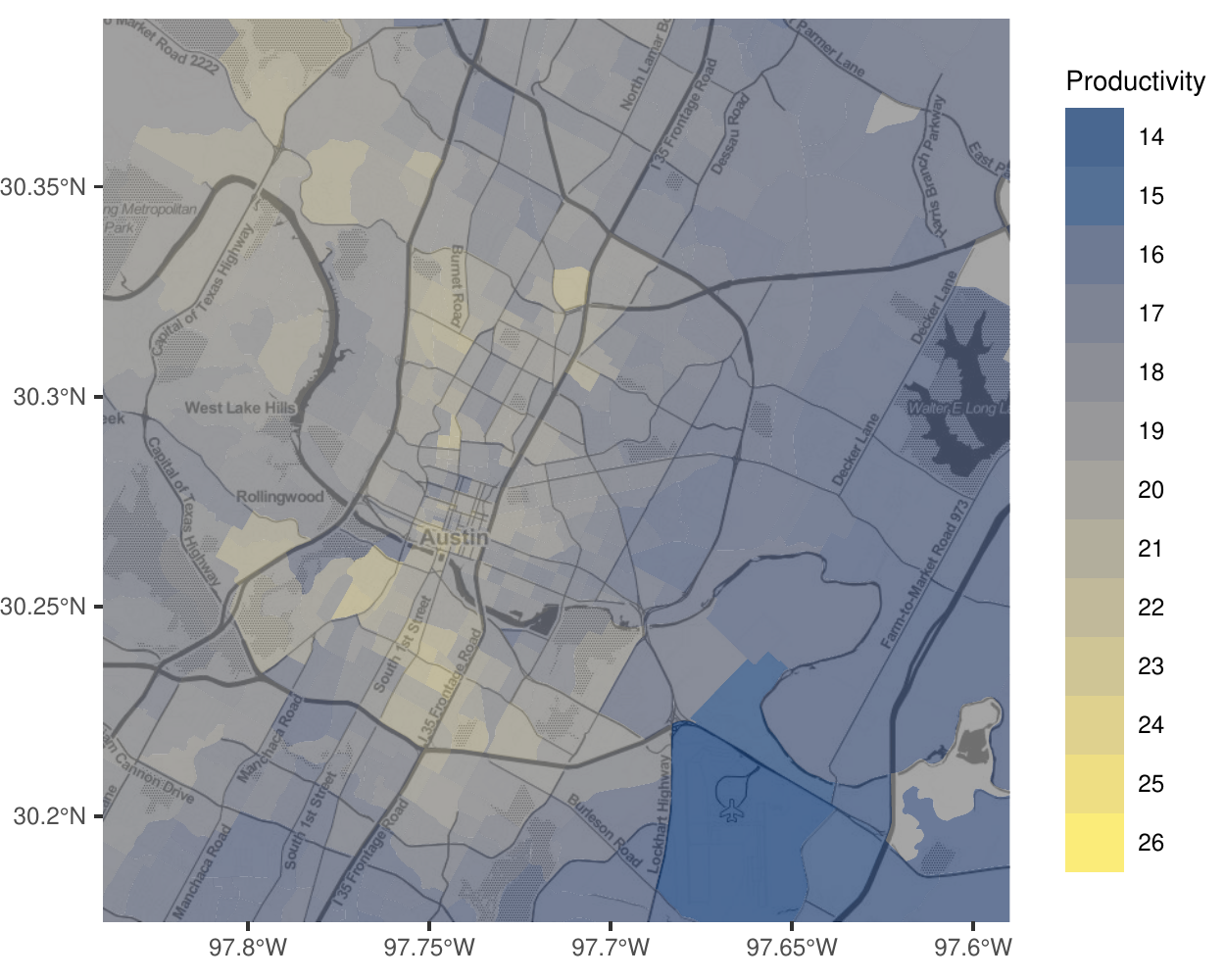}
        \subcaption{Monday 12 PM}
        \label{fig:quantiles:0.50:b}
    \end{minipage}
    \caption{Median of productivity for different times and locations.}
    \label{fig:quantiles:0.5}
\end{figure}
\end{appendix}

\bibliographystyleA{agsm.bst}
\bibliographyA{ref}

\end{document}